\documentclass[11pt]{article}
\usepackage[a4paper,bottom=4.2cm,top=1cm,head=3cm,width=18cm,dvipdfm]{geometry}
\evensidemargin=\oddsidemargin

\usepackage[dotinlabels]{titletoc}
\usepackage{titlesec}

\usepackage[linktocpage=true,bookmarks=true,bookmarksnumbered=true,breaklinks=true,pdfpagemode=Fullscreen,pdfstartview=FitBH]{hyperref}

\usepackage{graphicx}
\usepackage{dcolumn}
\usepackage{bm}
\usepackage{amssymb}
\usepackage{mathrsfs}
\usepackage{amsmath}
\usepackage{epsfig}
\usepackage{here}
\usepackage[dvips]{color}
\usepackage{hhline}
\usepackage{srcltx}

\allowdisplaybreaks[4]

\renewcommand{\thefootnote}{\fnsymbol{footnote}}
\newcommand{\beq}{\begin{equation}}
\newcommand{\eeq}{\end{equation}}
\newcommand{\bq}{\begin{equation}}
\newcommand{\eq}{\end{equation}}
\newcommand{\ba}{\begin{array}}
\newcommand{\ea}{\end{array}}
\newcommand{\beqa}{\begin{eqnarray}}
\newcommand{\eeqa}{\end{eqnarray}}
\newcommand{\beqs}{\begin{subequations}}
\newcommand{\eeqs}{\end{subequations}}

\def\nn{\nonumber}
\def\non{\nonumber}

\newcommand{\GG}{\mathbb{G}}

\def\bd{\boldsymbol}
\def\hf{\frac{1}{2}}
\def\pp{\prime}
\def\ot{\otimes}

\def\om{\omega}

\def\ZZ{\mathbb{Z}}
\def\ZZs{\mathbb{Z}_2^s}
\def\ZZmt{\mathbb{Z}_2^{\mu\tau}}
\def\ZZsp{\mathbb{Z}_2^{\prime s}}
\def\ZZmtp{\mathbb{Z}_2^{\prime\mu\tau}}
\def\O{{\cal O}}

\def\II{\mathcal{I}}

\def\End{\end{document}}

\def\to{\rightarrow}
\def\To{\Rightarrow}

\def\dis{\displaystyle}
\def\f{\frac}
\def\ov{\overline}

\def\[{\left[}
\def\]{\right]}
\def\({\left(}
\def\){\right)}

\def\under{\underline}

\def\a{\alpha}

\def\ab{\bar{\alpha}}

\def\U1EM{U(1)_{\rm em}}
\def\O{\mathcal O}

\def\leqq{\leqslant}
\def\geqq{\geqslant}

\def\N{{\cal N}}
\def\O{\mathcal{O}}
\def\mD{m_D^{}}
\def\mDT{m_D^T}
\def\mbD{\overline{m}_D^{}}
\def\mbDT{\overline{m}_D^T}

\def\Dt{\widetilde{D}}

\def\[{\left[}
\def\]{\right]}
\def\dis{\displaystyle}

\def\sia{\sigma_1^{}}
\def\sib{\sigma_2^{}}

\def\N{{\cal N}}
\def\d{\delta}
\def\da{\delta_a}
\def\dx{\delta_x}
\def\ds{\delta_s}

\def\gaa{\gamma_1^{}}
\def\gab{\gamma_2^{}}
\def\gac{\gamma_3^{}}

\def\dlj{d_\ell^{(j)}}
\def\dnu{d_\nu^{}}
\def\dnuj{d_\nu^{(j)}}

\def\mh{\widehat{m}}

\def\a{\alpha}

\def\ep{\epsilon}
\def\deg{\circ}

\def\d{\delta}

 \def\tab{\theta_{12}^{~}}
 \def\tbc{\theta_{23}^{~}}
 \def\tac{\theta_{13}^{~}}
 \def\ts{\theta_{s}}
 \def\ta{\theta_{a}}
 \def\tx{\theta_{x}}
 \def\ma{m_1^{}}
 \def\mb{m_2^{}}
 \def\mc{m_3^{}}
 \def\mutau{\mu\!-\!\tau}
 \def\mmutau{\mu\tau}
 
 \def\nuL{\nu_L^{}}
 
 \def\mh{\widehat{m}}
 
 \def\tm{{\widetilde{m}}}

 \def\ahat{\widehat{\alpha}}
 \def\att{\tilde{a}}

 \def\bta{\tilde{b}_1^{}}
 \def\btb{\tilde{b}_2^{}}
 \def\cta{\tilde{c}_1^{}}
 \def\ctb{\tilde{c}_2^{}}
 \def\mtt{\widetilde{m}}
 \def\Mtt{\widetilde{M}}

\numberwithin{equation}{section}

\begin{document}
 \thispagestyle{empty}
 \setcounter{footnote}{0}
 \titlelabel{\thetitle.\quad \hspace{-0.8em}}
\titlecontents{section}
              [1.5em]
              {\vspace{4mm} \large \bf}
              {\contentslabel{1em}}
              {\hspace*{-1em}}
              {\titlerule*[.5pc]{.}\contentspage}
\titlecontents{subsection}
              [3.5em]
              {\vspace{2mm}}
              {\contentslabel{1.8em}}
              {\hspace*{.3em}}
              {\titlerule*[.5pc]{.}\contentspage}
\titlecontents{subsubsection}
              [5.5em]
              {\vspace{2mm}}
              {\contentslabel{2.5em}}
              {\hspace*{.3em}}
              {\titlerule*[.5pc]{.}\contentspage}


 \begin{center}
 {\bf {\Large
  Common Origin of $\boldsymbol{\mu\!-\!\tau}$ and CP Breaking
  in Neutrino Seesaw, \\[2mm]
  Baryon Asymmetry, and Hidden Flavor Symmetry}}

 \vspace*{8mm}

 {\sc Hong-Jian He}\,$^{a,b,c}$\,\footnote{hjhe@tsinghua.edu.cn} ~~and~~
 {\sc Fu-Rong Yin}\,$^a$\,\footnote{yfr@tsinghua.edu.cn}

\vspace*{3mm}

$^a$\,Institute of Modern Physics and Center for High Energy Physics,
\\
Tsinghua University, Beijing 100084, China
\\[1mm]
$^b$\,Center for High Energy Physics, Peking University, Beijing 100871, China
\\[1mm]
$^c$\,Kavli Institute for Theoretical Physics China, \\
Chinese Academy of Sciences, Beijing 100190, China\\

\vspace{25mm}
\end{center}

 \vspace*{3mm}
 \begin{abstract}
 \baselineskip 17pt
 \noindent
 We conjecture that all CP violations (both Dirac and Majorana types)
 arise from a common origin in neutrino seesaw.
 With this conceptually attractive and simple conjecture,
 we deduce that {\it $\mutau$ breaking shares the
 common origin with all CP violations.}
 We study the common origin of $\mutau$ and CP breaking
 in the Dirac mass matrix of seesaw Lagrangian
 (with right-handed neutrinos being $\mutau$ blind), which uniquely leads to
 inverted mass-ordering of light neutrinos.
 We then predict a very different correlation between the two small $\mutau$ breaking
 observables  $\,\theta_{13}^{}\!-0^\circ\,$ and $\,\theta_{23}^{} \!-45^\circ\,$,\,
{\it  which can saturate the present experimental upper limit on $\theta_{13}^{}$}.
 This will be tested against our previous normal mass-ordering scheme
 by the on-going oscillation experiments.
 We also analyze the correlations of $\,\theta_{13}^{}\!$ with
 Jarlskog invariant and neutrinoless $\beta\beta$-decay observable.
 From the common origin of CP and $\mutau$ breaking in the neutrino seesaw,
 we establish a direct link between the low energy CP violations and
 the cosmological CP violation for baryon asymmetry.
 With these we further predict a {\it lower bound} on $\,\theta_{13}$\,,\,
 supporting the on-going probes of $\,\theta_{13}$\,
 at Daya Bay, Double Chooz and RENO experiments.
 Finally, we analyze the general model-independent $\ZZ_2\otimes\ZZ_2$ symmetry structure
 of the light neutrino sector, and map it into the seesaw sector,
 where one of the $\ZZ_2$'s corresponds to the $\mutau$ symmetry $\ZZ_2^{\mmutau}$
 and another the hidden symmetry $\ZZ_2^s$
 (revealed in our previous work) which dictates the solar mixing angle $\,\theta_{12}\,$.\,
 We derive the physical consequences of this $\,\ZZ_2^s$\, and its possible partial violation
 in the presence of $\mutau$ breaking {\it (without or with neutrino seesaw),}
 regarding the $\,\theta_{12}\,$ determination and the correlation between
 $\mutau$ breaking observables.
\\[4mm]
\\[2mm]
{\tt Phys.\,Rev.\,D\,84\,(2011)\,033009} ~and~ {\tt arXiv:1104.2654}
%
\end{abstract}

 \newpage
 \setcounter{page}{2}

 \tableofcontents

 \newpage
 \setcounter{footnote}{0}
 \renewcommand{\thefootnote}{\arabic{footnote}}
 \baselineskip 18.5pt

\section{\large Introduction}
 \label{sec:introduction}

 We conjecture that all CP violations (both Dirac and Majorana types)
 arise from a common origin in neutrino seesaw.
 With this conceptually attractive and simple conjecture,
 we deduce that {\it $\mutau$ breaking shares the
 common origin with all CP violations,}
 since the $\mutau$ symmetric limit enforces vanishing
 mixing angle $\theta_{13}$ and
 thus Dirac CP conservation.

 In a recent work\,\cite{GHY}, we studied the common origin of soft
 $\mutau$ and CP breaking in the neutrino seesaw, which is uniquely formulated
 in the dimension-3 Majorana mass term of singlet right-handed neutrinos.
 This formulation predicts the normal mass ordering (NMO) for light neutrinos.
 In this work, we study in parallel a different realization of the
 common origin of $\mutau$ and CP breaking in the ``$\mutau$ blind seesaw",
 where the right-handed neutrinos are singlet under the $\mutau$ transformation.
 We then find {\it the Dirac mass-matrix to be the unique place
 for the common origin of $\mutau$ and CP breaking} in the $\mutau$ blind seesaw.
 Since the Dirac mass-matrix arises from Yukawa interactions with Higgs boson(s),
 this can also provide an interesting possibility of realizing
 spontaneous CP violation with CP phases originating from
 the vacuum expectation values of Higgs fields.
 Different from our previous construction\,\cite{GHY}, we reveal that
 the common origin of $\mutau$ and CP breaking in the Dirac mass-matrix uniquely
 leads to the inverted mass-ordering (IMO) of light neutrinos and thus different
 neutrino phenomenology. Hence, the present mechanism can be distinguished from
 the previous one\,\cite{GHY} by the on-going and upcoming experiments on the
 neutrino oscillations\,\cite{review}
 and neutrinoless double-beta decays\,\cite{0nu2beta}.

 The oscillation data from solar and atmospheric neutrinos, and from the
 terrestrial neutrino beams produced in the reactor and accelerator experiments,
 have measured two mass-squared differences
 $\left( \Delta m^2_{31},\,\Delta m^2_{21} \right)$ and two
 large mixing angles $(\theta_{12},\,\theta_{23})$
 to good accuracy\,\cite{Fit2010}\cite{Fogli-08}.
 The two compelling features are\,\cite{Fit2010}\cite{Fogli-08}:
 (i) the atmospheric neutrino mixing angle $\theta_{23}$
 has only small deviations from its maximal value of $\,\theta_{23}=45^\deg$\,;\,
 (ii) the reactor neutrino mixing angle $\theta_{13}$ is found to be small,
 having its allowed range still consistent with $\,\theta_{13}=0^\deg$\, at $90\%$C.L.
 Hence, the pattern of \,$(\theta_{23},\,\theta_{13})=(45^\deg,\,0^\deg)$\,
 is strongly supported by the experimental data as {\it a good zeroth order approximation.}
 It is important to note that
 this pattern corresponds to the $\mutau$ symmetry and Dirac CP conservation
 in the neutrino sector, where the $\mutau$ symmetry is determined by both values
 of  \,$(\theta_{23},\,\theta_{13})=(45^\deg,\,0^\deg)$\, and the
 Dirac CP conservation is due to $\,\theta_{13}=0^\deg$\,.\,
 On the theory ground, it is natural and tempting to expect a common origin for all
 CP-violations, although the Dirac and Majorana CP-violations appear
 differently in the light neutrino mass-matrix of the low energy effective theory.
 Given such a common origin for two kinds of CP-violations, then they must vanish
 together in the $\mutau$ symmetric limit.
 For the $\mutau$ blind seesaw, we can uniquely formulate this common breaking
 in the Dirac mass matrix, leading to distinct neutrino phenomenology.

 With such a conceptually attractive and simple construction of the common breaking of
 two discrete symmetries, we can predict the $\mutau$ breaking at low energies and derive
 {\it quantitative correlations between the two small deviations,}
 $\,\tbc \!-45^\circ$\, and $\,\tac\!-0^\circ$,\,
 very different from that of the previous NMO scheme\,\cite{GHY}.
 Our predicted range of $\,\tac\,$ can saturate its present experimental upper limit.
 The improved measurements of $\,\theta_{23}\,$ will come from
 the Minos\,\cite{MINOS} and T2K\,\cite{T2K} experiments, etc,
 while $\,\theta_{13}\,$ will be more accurately probed by the
 on-going reactor experiments,
 Daya Bay\,\cite{DayaBay}\cite{footnote-1},
 Double Chooz\,\cite{2CHOOZ}, and RENO\,\cite{RENO}, as well as the accelerator
 experiments T2K\,\cite{T2K}, NO$\nu$A\,\cite{NOvA} and LENA\,\cite{LENA}, etc.
 We further derive the observed baryon asymmetry via leptogenesis
 at seesaw scale, and analyze the correlation between
 the leptogenesis and the low energy neutrino observables in the present IMO scheme.
 Especially, we deduce a lower bound on the reactor neutrino mixing angle
 $\,\theta_{13}\gtrsim 1^\deg\,$,\, and demonstrate that most of the
 predicted parameter space will be probed by the on-going
 Double Chooz, Daya Bay, and RENO reactor experiments.

 Finally, we will analyze the most general $\ZZ_2\otimes\ZZ_2$ symmetry structure of the
 light neutrino sector, and map it into the seesaw sector, where one of the $\ZZ_2$'s
 is the $\mutau$ symmetry $\ZZ_2^{\mmutau}$ and another the hidden symmetry $\ZZ_2^s$
 (revealed in our recent work\,\cite{GHY} for the NMO scheme),
 which dictates the solar mixing angle $\,\theta_{12}\,$.\,
 We derive the physical consequences of the $\,\ZZ_2^s$\, for the most general
 light neutrino mass-matrix (without seesaw) and for the seesaw models (with different
 $\mutau$ breaking mechanisms).
 In particular, we analyze the partial violation of $\,\ZZ_2^s$\,
 in the presence of $\mutau$ breaking for the $\mutau$ blind seesaw, which leads to
 {\it a modified new correlation between the $\mutau$ breaking observables,}
 very different from that of Ref.\,\cite{GHY}.
 The determination of $\,\theta_{12}\,$ is systematically studied for the current
 IMO scheme and the partial violation of $\,\ZZ_2^s\,$ will be clarified.

We organize this paper as follows.
In Sec.\,2 we present a unique construction for the common origin
of the $\mutau$ and CP breakings
in the neutrino seesaw with $\mutau$ blind right-handed neutrinos.
Then, we give in Sec.\,3 a model-independent
reconstruction of light neutrino mass-matrix under inverted mass-ordering
and with small $\mutau$ and CP violations at low energies.
In Sec.\,4.\,1, we explicitly derive
the low energy $\mutau$ and CP violation observables
from the common breaking in the Dirac mass-matrix of the $\mutau$ blind seesaw.
These include the two small deviations for the mixing angles
$\,\tbc \!-45^\circ$\, and $\,\tac\!-0^\circ$,  the Jarlskog invariant for
CP-violations, and the $M_{ee}$ element for neutrinoless double-beta decays.
In Sec.\,4.\,2 we study the cosmological CP violation via
leptogenesis in our model, this can generate
the observed baryon asymmetry of the universe.
Using all the existing data from neutrino oscillations
and the observed baryon asymmetry\,\cite{WMAP08,PDG},
we derive the direct link between the cosmological CP-violation
and the low energy Jarlskog invariant $J$\,.\,
We further predict a lower bound on the reactor mixing angle $\,\theta_{13}\,$,\,
and deduce a nonzero Jarlskog invariant $J$ with negative range.
We also establish a lower limit on the leptogenesis scale for
producing the observed baryon asymmetry.
In Sec.\,5, we analyze the determination of solar mixing angle
$\,\theta_{12}\,$ and its relation to the hidden symmetry $\ZZ_2^s$
in the light neutrino sector (without seesaw) and in the seesaw sector
(with two different realizations of $\mutau$ breaking).
Finally, conclusions are summarized in the last section\,6.

\section{\large
 Common Origin of $\bd{\mutau}$ and CP Breaking from Neutrino Seesaw \\
 with Inverted Ordering}
\label{sec:seesaw}

The current global fit of neutrino data\,\cite{Fit2010} for the three mixing angles and
two mass-squared differences is summarized in Table-1.
We note a striking pattern of the mixing angles, where the atmospheric angle $\theta_{23}$
has its central value slightly below
the maximal mixing\,\cite{footnote-2} of $45^\deg$ and the reactor angle
$\theta_{13}$ slightly above $0^\deg$.  So the neutrino data support two small deviations
$\,\theta_{23}-45^\deg$\, and $\,\theta_{23}-0^\deg$\, of the same order,
\beqa
\label{eq:da-dx-exp}
-7.0^\deg < (\theta_{23}-45^\deg) < 5.5^\deg\,,~~~~~~~~~~
0^\deg \leqq (\theta_{13}-0^\deg)<9.5^\deg \,,
\eeqa
at 90\%C.L., with the best fitted values,
$\,(\theta_{23}-45^\deg)=-2.2^\deg$\, and \,$(\theta_{13}-0^\deg)=5.1^\deg$.\,
This justifies a fairly good {\it zeroth order approximation,}
\,$\theta_{23}=45^\deg$\, and \,$\theta_{13}=0^\deg$,\,
under which two exact discrete symmetries emerge,
i.e., the $\mutau$ symmetry\,\cite{mutauRev}
and the Dirac CP conservation in the neutrino sector.
It is clear that the $\mutau$ symmetry and the associated Dirac CP-invariance
are well supported by all neutrino data as a good {\it zeroth order approximation},
and have to appear in any viable theory for neutrino mass-generation.
We also note that the $\,\theta_{13}=0^\deg$\, limit does not remove the possible
low energy Majorana CP-phases, but since the Majorana CP-violation comes from a common
origin with the Dirac CP-violation in our theory construction (cf.\ below),
it has to vanish as the Dirac CP-violation goes to zero in the $\mutau$ symmetric limit.
\begin{table*}[h]
\begin{center}
\label{tab:1} {\small
\begin{tabular}{c||ccccc}
\hline\hline
& & & \\
{\tt Parameters} & {\tt Best Fit}
 & 90\%\,{\tt C.L.}  & 99\%\,{\tt C.L.}
 & ${1\sigma}$ {\tt Limits} & ${3\sigma}$ {\tt Limits}
\\
& & & & & \\[-2mm]
\hline
& & & & & \\[-2mm]
 $ \Delta m^2_{21}(10^{-5}{\rm eV}^2)$ &
$7.59 $  &$7.26-7.92$&$7.00-8.11$& $7.39-7.79$&$6.90-8.20$
\\[1mm]
\hline
& & & & & \\[-2mm]
$ \Delta m^2_{31}(10^{-3}{\rm eV}^2)({\tt NMO})$ &
$2.46 $ &$2.26-2.66$&$2.14-2.78$& $2.34-2.58$ &$2.09-2.83$
\\[1mm]
\hline& & & & & \\[-2mm]
$ \Delta m^2_{13}(10^{-3}{\rm eV}^2)({\tt IMO})$ &
$2.36 $  &$2.18-2.54$&$2.04-2.68$& $2.25-2.47$&$1.99-2.73$
\\[1mm]
\hline
& & & & & \\[-2mm]
$\theta_{12}$ & $34.5^\circ$  &$32.8^\circ-36.0^\circ$ &
$32.1^\circ-37.2^\circ$&  $33.5^\circ-35.5^\circ$ & $31.7^\circ-37.7^\circ$\\[1mm]
\hline
& & & & & \\[-2mm]
$\theta_{23}$ & $42.8^\circ$ &
$38.0^\circ-50.5^\circ$ & $36.5^\circ-52.0^\circ$ & $39.9^\circ-47.5^\circ$ & $35.5^\circ-53.5^\circ$\\[1mm]
\hline
& & & & & \\[-2mm]
$\theta_{13}$ & $5.1^\circ$  &
$0^\circ \!-9.5^\circ$ & $0^\circ \!-  11.3^\circ$& $1.8^\circ-8.1^\circ$ & $0^\circ \!-12.0^\circ$\\
\hline\hline
\end{tabular}
}
\vspace*{-2mm}
\caption{Updated global analysis\,\cite{Fit2010} of solar, atmospheric, reactor and
   accelerator neutrino data for three-neutrino oscillations, where the AGSS09 solar
   fluxes and the modified Gallium capture cross-section\,\cite{Abdurashitov:2009tn} are used.}
\end{center}
\end{table*}
\vspace*{-6mm}

 In our theory construction, we conjecture that all CP violations (both Dirac and Majorana types)
 have a common origin and thus they must share the common origin with the $\mutau$ breaking.
 For the neutrino seesaw with heavy right-handed neutrinos blind to the $\mutau$ symmetry,
 this common origin can only come from the Dirac mass-term.
 In the following, we first consider the minimal neutrino
 seesaw Lagrangian with exact $\mutau$ and CP invariance, from which we will derive
 the seesaw mass-matrix for the light neutrinos.
 Diagonalizing this zeroth order mass-matrix we predict the inverted mass-ordering of
 light neutrinos and deduce the mixing angles,
 \,$(\theta_{23},\,\theta_{13})_0=(45^\deg,\,0^\deg)$\,,\,
 as well as a formula for the solar angle $\,\theta_{12}\,$.
 Then we will construct the common origin for the $\mutau$ and CP breaking
 in the Dirac mass-matrix.
 Finally, we systematically expand the small $\mutau$ and CP breaking effects
 in the seesaw mass-matrix to the first nontrivial order.

\vspace*{3mm}
 \subsection{%
 $\bd{\mutau}$ and CP Symmetries of Neutrino Seesaw with Inverted Ordering}
 \label{sec:zeroth-seesaw}

 The right-handed neutrinos are singlets under the standard model gauge group,
 and thus can be Majorana fields with large masses. This naturally realizes the
 seesaw mechanism\,\cite{Mseesaw} which provides the simplest explanation for the small
 masses of light neutrinos.  For simplicity,
 we consider the Lagrangian for the minimal neutrino seesaw\,\cite{MSS,He2003},
 with two right-handed singlet Majorana neutrinos
 besides the standard model (SM) particle content,
 \beqa
 {\cal L}_{\rm ss}
 & = &
     - \;\ov{L}\;Y_\ell\;\Phi\ell_R^{}
  \; - \;\ov{L}\;Y_\nu\widetilde\Phi\;\N + \f{1}{2}\N^TM_R\widehat{C}\N + {\rm h.c.}
 \nn\\
 & = &
 - \;\ov{\ell_L}\;M_\ell\;\ell_R
              \;-\;\ov{\nuL}\;\mD\;\N + \f{1}{2}\N^TM_R\widehat{C}\N
 + {\rm h.c.} + (\textrm{interactions})
 \,,
 \label{eq:L-seesaw}
 \eeqa
 where $\,L\,$ represents three left-handed neutrino-lepton weak doublets,
 $\,\ell=(e,\,\mu,\,\tau)^T\,$ denotes charged leptons,
 $\,\nuL =(\nu_e^{},\,\nu_\mu^{},\,\nu_\tau^{})^T\,$ is
 the light flavor neutrinos, and $\,\N =(N_1,\,N_2)^T$\, contains two heavy
 right-handed singlet neutrinos.
 The lepton Dirac-mass-matrix $\,M_\ell = v\,Y_\ell /\sqrt{2}\,$
 and the neutrino Dirac-mass-matrix $\,\mD = \f{v}{\sqrt{2}}\,Y_\nu\,$ arise from the
 Yukawa interactions after spontaneous electroweak symmetry breaking,
 $\,\left<\Phi\right> = (0,\,\f{v}{\sqrt{2}}\,)^T\neq 0\,$,
 and the Majorana mass-term for $\,M_R\,$ is a gauge-singlet.
 We can regard this
 minimal seesaw Lagrangian in Eq.\,(\ref{eq:L-seesaw}) as
 an effective theory of the general three-neutrino seesaw where the right-handed
 singlet $N_3$ is much heavier than the other two $(N_1,\,N_2)$ and
 thus can be integrated out at the mass-scales of $(N_1,\,N_2)$,
 leading to Eq.\,(\ref{eq:L-seesaw}).
 As a result, the minimal seesaw generically predicts
 a massless light neutrino\,\cite{MSS};
 this is always a good approximation as long as
 one of the light neutrinos has a negligible mass
 in comparison with the other two (even if not exactly massless).
 Extension to the three-neutrino seesaw will be discussed in Sec.\,4.\,3.

 Let us integrate out the heavy neutrinos $(N_1,\,N_2)$
 in (\ref{eq:L-seesaw}) and derive the seesaw formula for the
 $3\times 3$ symmetric Majorana mass-matrix of the light neutrinos,
 \begin{eqnarray}
  M_\nu ~\simeq~
  \mD M_R^{-1} \mDT \,,
  \label{eq:seesaw-formula}
 \end{eqnarray}
 where $\mD$ is the $3\times 2$ Dirac mass-matrix, and $M_R$ is the $2\times 2$
 Majorana mass-matrix. The diagonalization of $M_\nu$ is achieved by unitary rotation
 matrix $U_{\nu}^{}$ via
 $\,U_{\nu}^T M_\nu U_{\nu}^{} = D_\nu\,$
 with $\,D_\nu =\textrm{diag}(\ma,\,\mb,\,\mc)\,$.

 The Lagrangian (\ref{eq:L-seesaw}) is defined to
 respect both the $\mutau$ and CP symmetries.
 Under the $\mutau$ symmetry $\,\mathbb{Z}_2^{\mu\tau}$, we have the transformation,
 \,$\nu_\mu \leftrightarrow p\nu_\tau$\,,\, where
 \,$p=\pm$\, denotes the even/odd parity assignments of the light neutrinos
 under $\,\mathbb{Z}_2^{\mu\tau}$\,.\, Since the $\mutau$ symmetry has been tested at
 low energy via mixing angles of light neutrinos, it is logically possible that the
 right-handed heavy Majorana neutrinos in the seesaw Lagrangian (\ref{eq:L-seesaw})
 are singlets under $\,\mathbb{Z}_2^{\mu\tau}$ (called ``$\mutau$ blind"), which is
 actually the simplest realization of $\mutau$ symmetry in the neutrino seesaw.
 In this work we consider that the right-handed Majorana neutrinos $\N$ to be $\mutau$ blind,
 i.e., both $(N_1,\,N_2)$ are the singlets under $\mathbb{Z}_2^{\mu\tau}$, and thus can
 be first rotated into their mass-eigenbasis without affecting the $\mutau$ symmetric
 structure of the Dirac mass-matrix $\,\mD$\,.\,
 So, in the mass-eigenbasis of $(N_1,\,N_2)$,
 we have $\,M_R=\textrm{diag}(M_1,\,M_2)\,$.\,
 Under the $\mutau$ and CP symmetries, the Dirac mass-matrix $\mD$ is real and obeys the
 invariance equation,
 \begin{eqnarray}
 \label{eq:mdtrans}
 G_\nu^T m_D^{} ~=~ m_D^{} \,,
 \end{eqnarray}
with
 \begin{eqnarray}
 \label{eq:T3T2}
 G_\nu^{} ~=~ \!\!\(
  \ba{ccc}
   1 & 0 & 0 \\
   0 & 0 & p \\
   0 & p & 0
  \ea \) .
 \end{eqnarray}
 Next, we note that due to the large mass-splitting of $\mu$ and $\tau$ leptons,
 the lepton sector can exhibit, in general, a different
 flavor symmetry $\mathbb{G}_\ell$  from the
 $\mutau$ symmetry $\mathbb{Z}_2^{\mu\tau}$ in the neutrino sector.
 The two symmetries $\mathbb{Z}_2^{\mu\tau}$ and $\mathbb{G}_\ell$ could
 originate from spontaneous breaking of a larger flavor symmetry
 $\mathbb{G}_F$\,\cite{Gf}. Under the transformation of left-handed leptons
 $\,F_{\ell}^{}\,\in\mathbb{G}_\ell^{}\,$,\, we have the invariance equation
 of lepton mass-matrix,
 $\,F_{\ell}^\dag M_\ell^{} M_\ell^\dag F_{\ell}^{} = M_\ell^{} M_\ell^\dag\,$.\,
 As we will show in Sec.\,4.\,2, we are free to choose an equivalent representation
 $\,d_{\ell}=U_\ell^\dag F_\ell U_\ell^{}\,$ of $\mathbb{G}_\ell$ from the start under which
 the left-handed leptons are in their mass-eigenbasis, where $U_\ell$ is the transformation
 matrix diagonalizing the lepton mass-matrix,
 $\,U_{\ell}^\dag M_\ell^{} M_\ell^\dag U_{\ell}^{} =D_{\ell}^2\,$ with
 $\,D_{\ell}^{}=\textrm{diag}(m_e^{},\,m_\mu^{},\,m_\tau^{})\,$.\,
 This means that in the lepton mass-eigenbasis,
 the conventional Pontecorvo-Maki-Nakagawa-Sakata (PMNS) mixing
 matrix $V$ \cite{PMNS} in the leptonic charged current
 (an analog of the CKM matrix\,\cite{CKM} in the quark sector) is fixed by
 the transformation $U_{\nu}^{}$ of neutrino mass-diagonalization,
 \,$V = U_{\nu}^{}$\,.\,
 We can further rotate the right-handed leptons into their mass-eigenbasis,
 without affecting the PMNS matrix, except making the lepton-mass-term diagonal
 in the seesaw Lagrangian (\ref{eq:L-seesaw}), i.e.,
 $\,M_\ell = \textrm{diag}(m_e^{},\,m_\mu^{},\,m_\tau^{})\,$\,.\,

 Under the $\mutau$ and CP symmetries, we find the Dirac mass-matrix $m_D^{}$ to
 have the following form,
 \begin{eqnarray}
 \label{eq:mD}
 m_D^{} ~= \(
  \ba{ll}
    \bar a  &~~ \bar a' \\[2mm]
    \bar b  &~~ \bar c \\[2mm]
    \bar b  &~~ \bar c
  \ea \)
  \,=\,
  \(
  \ba{ll}
  \sia \,a &~~  \sib \,a' \\[2mm]
  \sia \,b &~~  \sib \,c \\[2mm]
  \sia \,b &~~  \sib \,c
  \ea \) \!,
 \end{eqnarray}
 with all elements being real, and
 $\,\sia\equiv\sqrt{\widehat{m}_0^{}M_1}\,$,~
 $\,\sib\equiv\sqrt{\widehat{m}_0^{}M_2}\,$.
 As will be shown shortly, the parameter $\,\widehat{m}_0^{}\,$ is defined at the seesaw
 scale and equals the nonzero mass-eigenvalue of the light neutrinos
 at zeroth-order under the $\mutau$ symmetric limit.
 In (\ref{eq:mD}) we have also defined four dimensionless parameters,
 \begin{eqnarray}
 \label{eq:abc-m0}
 ( a,\, b)~\equiv\, \f{(\bar a,\, \bar b)}{\sqrt{\widehat m_0M_1}\,},
 &~~~&
 ( a',\, c)~\equiv\, \f{(\bar a',\, \bar c)}{\sqrt{\widehat m_0M_2}\,}\,.
 \end{eqnarray}
 Then, we find it convenient to define a dimensionless Dirac matrix,
 \beqa
 \label{eq:mD-bar}
 \ov{m}_D^{} ~\equiv~ m_D^{} (\mh_0^{}M_R^{})^{-\f{1}{2}}
 ~=~   \(
  \ba{ll}
  a &~~  a' \\[2mm]
  b &~~  c \\[2mm]
  b &~~  c
  \ea \) \!.
 \eeqa
 Substituting the above into the seesaw equation (\ref{eq:seesaw-formula}),
 we derive the $\mutau$ and CP symmetric mass-matrix for light neutrinos,
 \begin{eqnarray}
  M_\nu^{} ~\simeq~
  \mD M_R^{-1} \mDT
  &=& \mh_0^{}\(\mbD\mbDT\)
 ~=~ \widehat m_0
 \begin{pmatrix}
     a^2 +a'^2  &a b+ a' c  &  a b+ a'c
  \\[5mm]
  &   b^2+ c^2  & b^2+ c^2
  \\[3mm]
  & &b^2+ c^2
  \end{pmatrix} \!,~~~~~~
  \label{eq:Mnu-0}
 \end{eqnarray}
 which we call the {\it zeroth order mass-matrix.}
 In the next subsection we will further include the small $\mutau$ and CP breaking effect.
 Note that from (\ref{eq:Mnu-0}), we have $\,\det(M_\nu)=0\,$,\,
 which generally holds in any minimal seesaw.

 Diagonalizing the mass-matrix (\ref{eq:Mnu-0}),
 we derive the mass-eigenvalues and mixing angles at zeroth order,
 \begin{subequations}
 \label{eq:LO-Mass-Angle}
 \beqa
 \label{eq:m12}
 && \widehat{m}_{1,2}^{}
 ~=~ \f{\widehat m_0}{2}\[ (a^2+a'^2+2b^2+2c^2)\mp
 \sqrt{[(a^2+a'^2)-2(b^2+c^2)]^2+8(ab+a'c)^2}\]\!,~~~~~~~~~~
 \\[2mm]
 &&
 \label{eq:m3}
 \widehat m_3^{} ~~~=~ 0\,,
 \\[2mm]
 &&
 \tan 2\theta_{12} \,=\, \f{2\sqrt{2}|ab+a'c|}{|a^2+a'^2-2(b^2+c^2)|}\,, ~~~~~
   \theta_{23} \,=\, 45^\deg\,, ~~~~~
   \theta_{13} \,=\, 0^\deg \,,
 \label{eq:theta-123}
 \eeqa
 \end{subequations}
where we have made all mass-eigenvalues positive and
the mixing angles $\,(\theta_{12},\,\theta_{13},\,\theta_{23})\,$
within the range $\[0,\,\f{\pi}{2}\]$ by properly defining the rotation matrix.
(As shown in Table-1, the solar angle $\theta_{12}$ is most precisely measured and
its $3\sigma$ range is below $37.7^\deg$, so we always have $\,2\theta_{12}<\f{\pi}{2}\,$
and $\,\tan 2\theta_{12}>0\,$.)
The mixing angles $\,(\theta_{23},\,\theta_{13})=(45^\deg,\,0^\deg)\,$ are direct consequence
of the $\mutau$ symmetry, but this symmetry does not fix $\,\theta_{12}$\,.\,
Eqs.\,(\ref{eq:LO-Mass-Angle}a)-(\ref{eq:LO-Mass-Angle}b)
show that the mass-spectrum of light neutrinos falls into the
``inverted mass-ordering'' (IMO), $\,\mh_2\gtrsim \mh_1 \gg \mh_3\,$.

Table-1 shows that the ratio of two mass-squared differences,
$\,\frac{\Delta m_{21}^2}{|\Delta m_{31}^2|}\ll 1$\,.\,
Since for the minimal seesaw model with IMO, the equation $\,\det(M_\nu)=0\,$ leads to
$\,\mh_3^{}=0\,$,\,  so the above ratio requires the approximate degeneracy
$\,\mh_1^{}\simeq \mh_2^{}\,$  to be a good zeroth order approximation as enforced by
the neutrino oscillation data. So, we will realize the exact degeneracy
$\,\widehat{m}_{1}^{}=\widehat{m}_{2}^{}\,$ for the $\mutau$ and CP symmetric mass-matrix
(\ref{eq:Mnu-0}),  by imposing the relations for Eq.\,(\ref{eq:m12}),
\beqa
\label{eq:abc-1}
(a^2+a'^2)-2(b^2+c^2)\,=\,0\,,
&~~~&
ab+a'c \,=\, 0\,.
\eeqa
As will be shown in the next subsection,
including the common origin of $\mutau$ and CP breaking in the neutrino
seesaw can produce small non-degeneracy between $\mh_1^{}$ and $\mh_2^{}$ at the next-to-leading
order (NLO).
Since the mass-parameter $\mh_0^{}$ is introduced in (\ref{eq:abc-m0})
for defining the dimensionless parameters $(a,\,b,\,c)$,
we can now fix $\,\mh_0^{}$ by defining
\beqa
\label{eq:m0=m1=m2}
\mh_0^{}\,\equiv\, \widehat{m}_{1}^{} \,=\, \widehat{m}_{2}^{} \,,
\eeqa
as the zeroth order mass-eigenvalue of light neutrinos, under the normalization condition,
\beqa
\label{eq:m0=m1=m2-cond}
(a^2+a'^2) + 2(b^2+c^2) ~=~ 2 \,.
\eeqa
Combining this relation to Eq.\,(\ref{eq:abc-1}), we can deduce,
\beqa
\label{eq:abc-2}
a^2 \,=\, 2c^2 \,=\, 1-2b^2 \,,~~~~
a'^2 \,=\, 2b^2\,,~~~~
c^2 \,=\, \f{1}{2}-b^2 \,,~~~~
a'c \,=\, - ab \,,
\eeqa
where we see that three of the four parameters, $(a,\,a',\,c)$, can all be solved in terms
of $\,b$\,.\, The last equation in (\ref{eq:abc-2}) is not independent, but it helps to
fix a relative sign.
We note that in (\ref{eq:Mnu-0}) the $\mutau$ symmetric seesaw mass-matrix $\,M_\nu\,$
contains five parameters, the mass-parameter $\,\widehat{m}_0^{}\,$ and the four dimensionless
parameters $(a,\,b,\,c,\,a')$. The inverted mass-spectrum have imposed a LO condition
$\,\widehat{m}_1^{}=\widehat{m}_2^{}\,$,\, which results in two constraints in (\ref{eq:abc-1}),
and the normalization condition $\,\widehat{m}_0^{}\equiv \widehat{m}_1^{}\,$
in (\ref{eq:m0=m1=m2}) leads to the third constraint (\ref{eq:m0=m1=m2-cond}).
In consequence, we end up with only two independent parameters,
$\,\widehat{m}_0^{}\,$ and $\,b\,$.

We note that under the condition of (\ref{eq:abc-1}), the mixing angle $\theta_{12}$ given by
(\ref{eq:theta-123}) has no definition at the zeroth order (the $\mutau$ symmetric limit)
due to the vanishing numerator and denominator in the formula of
$\,\tan 2\theta_{12}\,$.\, But including the small $\mutau$ breaking effect will generate the
nonzero expression of $\theta_{12}$ at the NLO even though its final formula does not depend on
the $\mutau$ breaking parameter (cf.\ Sec.\,2.\,2).
As we will show in Sec.\,2.\,2,  the $\mutau$ breaking arises from deviation
in the element $c$ of $\mD$, so we can apply the l\,$'$H\^{o}pital rule to the expression of
$\tan 2\theta_{12}$ by taking the first-order derivatives on its numerator/denominator respect to
$\,c$\, and deduce,
\beqa
 \tan 2\theta_{12} \,=\,\frac{|a'|}{\sqrt{2}\,|c|\,}
                   \,=\, \frac{|\bar a'|}{\sqrt{2}\,|\bar c|\,} \,,
 \label{eq:t12}
\eeqa
which is consistent with (\ref{eq:t12-NLO}) of Sec.\,4.\,1
from the explicit NLO analysis. For the case with $\mutau$ breaking arising from deviation
in the element $b$ of $\mD$, we can apply the l\,$'$H\^{o}pital rule again to infer the
formula,
\beqa
 \tan 2\theta_{12} \,=\,\frac{|a|}{\sqrt{2}\,|b|\,}
 \,=\, \f{\sqrt{2}|c|}{|a'|} \,,
 \label{eq:t12-b}
\eeqa
which is {\it the inverse of (\ref{eq:t12}).}  As will be shown in Sec.\,5.\,2, the different
forms of $\mutau$ breaking will affect the determination of the solar mixing angle
$\,\theta_{12}$.\, But it is worth to note that the expression of $\,\theta_{12}$\,
is fixed by the $\mutau$ symmetric $\mD$ as in (\ref{eq:t12}) or (\ref{eq:t12-b}),
and does not explicitly depend on the $\mutau$ breaking parameter.
We will systematically analyze these features in Sec.\,5 and clarify the difference
from our previous construction\,\cite{GHY}.

\vspace*{2mm}
\subsection{%
 Common Origin of $\bd{\mutau}$ and CP Breaking in the $\bd{\mutau}$  Blind Seesaw}
\label{sec:unique}

 In this subsection, we will construct a unique breaking term providing a common origin
 for both $\mutau$ and CP breaking. From this we will further derive predictions of
 the common $\mutau$ and CP breaking for the low energy light neutrino mass-matrix,
 by treating the small breaking as perturbation up to the first nontrivial order
 (Sec.\,4). We will analyze the seesaw-scale leptogenesis and
 its correlations with the low energy observables in Sec.\,4.\,2.

 As we have explained, the $\mutau$ symmetry serves as a good
 zeroth order flavor symmetry of the neutrino sector, which predicts
 $\,\theta_{13}=0\,$ and thus the Dirac CP-conservation.
 Hence, the $\mutau$ symmetry breaking is generically small,
 and must generate all Dirac CP-violations at the same time.
 On the theory ground, it is natural and tempting to expect a {\it common origin for all
 CP-violations,} even though the Dirac and Majorana CP-violations appear
 differently in the light neutrino mass-matrix of the low energy effective theory.
 For the two kinds of CP-violations arising from a common origin,
 then they must vanish together in the $\mutau$ symmetric limit.

 Different from our previous study\,\cite{GHY}, we consider the heavy right-handed neutrinos
 to be $\mutau$ blind in the neutrino seesaw. Thus the Majorana mass-matrix $M_R^{}$ of the
 right-handed neutrinos must be $\mutau$ singlet.
 Hence, we deduce that {\it the unique common origin of the $\mutau$ and CP breaking
 must arise from the Dirac mass-matrix of the seesaw Lagrangian (\ref{eq:L-seesaw}).}
 For the minimal seesaw, the most general form of $\,\mD$\, is
 \begin{eqnarray}
 \label{eq:mD-2}
  \mD  ~= \(
  \ba{ll}
    \overline a  &~~ \overline a' \\[2mm]
    \overline b_1^{}  &~~ \overline c_1^{} \\[2mm]
    \overline b_2^{}  &~~ \overline c_2^{}
  \ea \)
  =
    \(
  \ba{ll}
  \sia \,a      &~~  \sib \,a'     \\[2mm]
  \sia \,b_1^{} &~~  \sib \,c_1^{} \\[2mm]
  \sia \,b_2^{} &~~  \sib \,c_2^{}
  \ea \) \!,
\end{eqnarray}
 where the scaling factors
 $\,\sia\equiv\sqrt{\widehat{m}_0^{}M_1}\,$ and
 $\,\sib\equiv\sqrt{\widehat{m}_0^{}M_2}\,$
 are real mass-parameters as defined in Eq.\,(\ref{eq:mD}).
 The six elements of $\,\mD$\, can be complex in general.
 But there are three rephasing degrees of freedom for the left-handed lepton-doublets.
 So we can always rotate the three elements in the first column of $\mD$
 to be all real, hence the remaining CP phases
 (associated with the $\mutau$ breaking) have to appear in
 the elements $\,c_1^{}\,$ and $\,c_2^{}\,$
 because  $\,a'\,$ cannot break $\mutau$ symmetry and thus should be real.
 We have conjectured that all CP violations arise from a {\it common origin},
 which then must originate from the $\mutau$ breaking;
 so we can formulate such a common origin as a single phase in either
 $\,c_1^{}\,$ or $\,c_2^{}\,$ in the minimal construction, where the
 other two elements in the second column of $\mD$ should be real.
 Hence, we present a unique minimal construction
 to formulate the common origin of $\mutau$ and CP breaking
 in the Dirac mass-matrix $\mD$ as follows,
 \begin{eqnarray}
  \mD ~=
      \(
  \ba{ll}
  \sia \,a      &~~  \sib \,a'     \\[2mm]
  \sia \,b      &~~  \sib \,c(1-\zeta') \\[2mm]
  \sia \,b      &~~  \sib \,c(1-\zeta e^{i\omega})
  \ea \!\) \!,
 \label{eq:mD-B-1}
 \end{eqnarray}
 where the dimensionless parameters
 $\,-1<\zeta' <1$,\,\, $\,0\leqq \zeta <1\,$,\,
 and the CP-phase angle $\,\omega\in[0,2\pi)$\,.\,
 Here we have set $\,b_1^{}=b_2^{}\equiv b\,$
 since $(b_1^{},\,b_2^{})$ are already made real
 and thus cannot serves as the common source of the $\mutau$ and CP breaking.
 Inspecting (\ref{eq:mD-B-1}) we see that, {\it for any nonzero $\zeta$ and $\omega$,
 the $\mutau$ and CP symmetries are broken by the common source of
 $\,\zeta e^{i\omega}\,$.\,}
 We could also absorb the real parameter $\zeta'$ into $c$ by defining
 $\,c'\equiv c(1-\zeta')\,$. Thus we have,
 \begin{eqnarray}
  \mD ~=
      \(
  \ba{ll}
  \sia \,a      &~~  \sib \,a'     \\[2mm]
  \sia \,b      &~~  \sib \,c' \\[2mm]
  \sia \,b      &~~  \sib \,c'(1-\zeta'' e^{i\omega'})
  \ea \!\) \!,
 \label{eq:mD-B-1a}
 \end{eqnarray}
  with
  \beqa
  \zeta''e^{i\omega'}
  ~=~ \f{\,\zeta e^{i\omega}-\zeta'\,}{1-\zeta'} \,.
  \eeqa
  Given the ranges of $(\zeta,\,\zeta')$ as defined above, we see that
  the corresponding new parameter $\,\zeta''\,$ of the $\mutau$ breaking
  has a much larger range, including values within
  $\,1\lesssim |\zeta''|\lesssim 3\,$
  (when $\,|\zeta|,|\zeta'|\leqq 0.6\,$ for instance),
  which are beyond the perturbative expansion.
  We find that if enforce $\,|\zeta''|<1\,$,\,  the parameter-space of
  (\ref{eq:mD-B-1a}) becomes smaller than (\ref{eq:mD-B-1}) and insufficient for
  making the model fully viable. This means that our formulation of
  (\ref{eq:mD-B-1}) is more general and has larger parameter-space for
  making theoretical predictions. Hence, we will apply
  (\ref{eq:mD-B-1}) for the physical analyses below.

 We note another formulation of such a breaking in the Dirac mass-matrix $\,\mD$\,,
 \begin{eqnarray}
 \label{eq:mD-B-2}
  \widehat{m}_D^{} ~=
      \(
  \ba{ll}
  \sia \,a      &~~  \sib \,a'     \\[2mm]
  \sia \,b      &~~  \sib \,c(1-\zeta e^{i\omega}) \\[2mm]
  \sia \,b      &~~  \sib \,c(1-\zeta')
  \ea \!\) \!,
 \end{eqnarray}
 which is connected to (\ref{eq:mD-B-1})
 by a $\mutau$ transformation for the light neutrinos
 $\nu=(\nu_e^{},\,\nu_\mu^{},\,\nu_\tau^{})\,$ into
 $\nu'=(\nu_e^{},\,\nu_\tau^{},\,\nu_\mu^{})\,$,\,
 via $\,\nu =  G_\nu^{} \nu'\,$,\, with $\,G_\nu [p=1]\,$ defined
 in Eq.\,(\ref{eq:T3T2}).
 Accordingly, the mass-matrix (\ref{eq:mD-B-2}) transforms as,
 \beqa
 \widehat{m}_D^{}~~~\rightarrow~~~
 \widehat{m}_D' \,=\, G_\nu^T \widehat{m}_D^{} \,=\, \mD \,,
 \eeqa
 which goes back to (\ref{eq:mD-B-1}).
 So the two different formulations (\ref{eq:mD-B-1}) and (\ref{eq:mD-B-2})
 just cause the $\mutau$ asymmetric parts in the seesaw mass-matrix
 $\,M_\nu = \mD M_R^{-1} \mDT$\, to differ by an overall minus sign.
 As we will comment further in Sec.\,4.\,1, this does not affect our predictions
 for the physical observables and their correlations. So we only need to focus
 on the formulation (\ref{eq:mD-B-1}) for the rest of our analysis.

 We may also first rotate the three elements in the second column of (\ref{eq:mD-2})
 to be real and then formulate the common origin of $\mutau$ and CP breaking as follows,
 \begin{eqnarray}
  \mD ~=
      \(
  \ba{ll}
  \sia \,a                           &~~  \sib \,a'     \\[2mm]
  \sia \,b(1-\zeta')                 &~~  \sib \,c \\[2mm]
  \sia \,b(1-\zeta e^{i\omega})      &~~  \sib \,c
  \ea \!\) \!.
 \label{eq:mD-B-3}
 \end{eqnarray}
 As will be clarified in Sec.\,5, this will lead to the determination of
 solar mixing angle $\,\theta_{12}\,$ as in (\ref{eq:t12-b}),
 in contrast to (\ref{eq:mD-B-1}) which predicts a different $\,\theta_{12}\,$
 as in (\ref{eq:t12}). Here $\,\theta_{12}\,$ is explicitly fixed by the $\mutau$
 and CP symmetric parameters of $\mD$ in either case.
 But, we find the predictions for all other $\mutau$ and CP breaking observables and
 their correlations to remain the same as those from the construction in
 (\ref{eq:mD-B-1}).

 Finally, it is interesting to note that for an extended Higgs sector
 (consisting of two Higgs doublets or more)
 we can generate all CP-phases in the Dirac mass-matrix $\mD$ via spontaneous
 CP violation\,\cite{TDLee74}, which is beyond the current scope and
 will be elaborated elsewhere\,\cite{HeYin3}.

\vspace*{3mm}
\subsection{Perturbative Expansion for $\bd{\mutau}$ and CP Breaking}
\label{sec:common-origin}

 Let us first consider the $3\times 3$ mass-matrix $M_\nu$ light neutrinos,
 which can be generally presented as,
 \beqa
 M_\nu &\!\!=\!\!&
 \(\ba{lll}
 A & B_1 & B_2 \\[1mm]
   & C_1 & D   \\[1mm]
   &     & C_2
 \ea \)
 ~\equiv~
 \(\ba{lll}
 A_0 & B_0 & B_0 \\[1mm]
     & C_0 & D_0   \\[1mm]
     &     & C_0
 \ea \) +
 \(\ba{lll}
 \d A & \d B_1 & \d B_2 \\[1mm]
      & \d C_1 & \d D   \\[1mm]
   &           & \d C_2
 \ea \)
\nn \\[5mm]
 &\!\!\equiv\!\!&
 M_\nu^{(0)} + \d M_\nu
 ~=~ M_\nu^{(0)} + \d M_\nu^{(1)} + \O(\zeta_i^2) \,,
 \label{eq:Mu=Mu0+dMu}
 \eeqa
 where the zeroth order matrix $M_\nu^{(0)}$ corresponds to vanishing $\mutau$ breaking
 with $\,\zeta_i^{}=0\,$,\, and the NLO mass-matrix $\d M_\nu^{(1)}$ includes the
 $\mutau$ breaking to the first nontrivial order.
 We find it useful to further decompose $\,\d M_\nu^{(1)}\,$  into the
 $\mutau$ symmetric and anti-symmetric parts,
 \beqa
 \d M_\nu^{(1)} ~\equiv~ \d M_\nu^{s} + \d M_\nu^{a}
 &\equiv&
   \(\ba{lll}
 \d A & \d B_s & \d B_s   \\[1mm]
      & \d C_s & \d D   \\[1mm]
      &      & \d C_s
 \ea \) +
    \(\ba{llc}
    0 & \d B_a & -\d B_a   \\[1mm]
      & \d C_a & 0         \\[1mm]
      &        & -\d C_a
 \ea \)  \!,~~~~~
 \label{eq:dMu=s+a}
 \eeqa
 with
 \beqs
 \beqa
 \d B_s  ~\equiv~ \f{1}{2}\({\d B_1 + \d B_2}\) \,, &&
 \d B_a  ~\equiv~ \f{1}{2}\({\d B_1 - \d B_2}\) \,,
 \\[2.5mm]
 \d C_s  ~\equiv~ \f{1}{2}\({\d C_1 + \d C_2}\) \,, &&
 \d C_a  ~\equiv~ \f{1}{2}\({\d C_1 - \d C_2}\) \,.
 \eeqa
 \label{eq:delta-BC-sa}
 \eeqs
 This decomposition is actually unique.

 From our construction in the previous subsection, the $\mutau$ and CP breaking
 Dirac mass-matrix $\mD$ as well as the Majorana mass-matrix $M_R^{}$
 is uniquely parameterized as follows,
 \begin{eqnarray}
 \label{eq:mDMR}
  \mD  ~=
    \(
  \ba{ll}
  \sia \,a      &~~  \sib \,a'     \\[2mm]
  \sia \,b      &~~  \sib \,c_1^{} \\[2mm]
  \sia \,b      &~~  \sib \,c_2^{}
  \ea \)  \!,
  \label{eq:MD-B-I}
 &~~~&
  M_R^{} ~=~ \textrm{diag}(M_1,\,M_2) \,,
 \end{eqnarray}
 with $\,\sigma_{1,2}^{} \equiv \sqrt{\widehat{m}_0^{}M_{1,2}\,}\,$\, and
 \begin{eqnarray}
 \label{eq:c1-c2}
 c_1^{} ~=~ c\(1-\zeta'\) ,
 &~~~&
 c_2^{} ~=~ c\(1-\zeta e^{i\omega}\) .
 \end{eqnarray}
 Thus, we can explicitly derive the seesaw mass-matrix for light neutrinos,
 \begin{eqnarray}
 M_\nu^{} ~=~ \widehat m_0
 \begin{pmatrix}
     a^2 +a'^2  & a b+ a' c_1^{}  &  a b+ a'c_2^{}
  \\[3mm]
  &   b^2+ c_1^2  & b^2+ c_1^{}c_2^{}
  \\[3mm]
  & & b^2+ c_2^2
  \end{pmatrix} \!.
  \label{eq:Mnu-all-I}
   \end{eqnarray}
 Since the neutrino data require the $\mutau$ breaking to be small, we
 can further expand $M_\nu$ in terms of small breaking parameter $\,\zeta\,$ as,
 \begin{eqnarray}
 M_\nu^{} ~\equiv~ M_\nu^{(0)}+\delta M_\nu
 ~=~ M_\nu^{(0)}+\delta M_\nu^{(1)}+O(\zeta^2) \,,
 \end{eqnarray}
 with
 \begin{subequations}
 \begin{eqnarray}
 M_\nu^{(0)}
 & \!\!=\! & \widehat m_0^{}
 \begin{pmatrix}\,
     a^2 +a'^2  &a b+ a' c  &  a b+ a'c
  \\[3mm]
  &   b^2+ c^2  & b^2+ c^2
  \\[3mm]
  & &b^2+ c^2 \,
  \end{pmatrix}
  =~  \widehat m_0^{}
 \begin{pmatrix}\,
  1  & 0  &  0 \,~
  \\[1.5mm]
  &   \hf  & \hf \,~
  \\[1.5mm]
  & &  \hf \,~
  \end{pmatrix}  ,
  \label{eq:Mnu0}
 \\[2mm]
 \delta M_\nu^{(1)}
  & \!\!=\! & \widehat m_0^{}
 \begin{pmatrix}\,
    0
  & -a'c ~\zeta'
  & -a'c ~\zeta e^{i\omega}
  \\[3mm]
&-2~c^2 \zeta'
  & -c^2 (\zeta' +\zeta e^{i\omega})
  \\[3mm]
  & &-2c^2 \zeta e^{i\omega}
  \end{pmatrix} ,
  \label{eq:Mnu1}
   \end{eqnarray}
 \end{subequations}
 where we have used the solution (\ref{eq:abc-2}) for the second step of (\ref{eq:Mnu0})
 and the $\mutau$ breaking expression (\ref{eq:c1-c2}) for deriving (\ref{eq:Mnu1}).
 For our current model with the expansion up to $\O(\zeta,\,\zeta')$,
 we deduce from (\ref{eq:Mnu0})-(\ref{eq:Mnu1}) and
 (\ref{eq:dMu=s+a})-(\ref{eq:delta-BC-sa}),
 \beqs
  \begin{eqnarray}
  && A_0~=~ \mh_0^{}(a^2+a'^2) ~=~ \mh_0^{} \,,
  \\[1.5mm]
  && B_0~=~ \mh_0^{}(ab+a'c) ~=~ 0 \,,
  \\[1.5mm]
  && C_0~=~D_0 ~=~ \mh_0^{}(b^2+c^2) ~=~ \hf \mh_0^{}\,,
  \end{eqnarray}
 \eeqs
 and
 \beq
 \label{eq:seesaw-Delta-m}
 \ba{ll}
 \dis\d A   ~=\, 0\,,
 &~~~~~
 \dis\d D   ~=\,  -\widehat m_0 c^2 (\zeta' +\zeta e^{i\omega})\,,
 \\[2.4mm]
 \dis\d B_s ~=\, -\hf{\widehat m_0 a' c(\zeta'+\zeta e^{i\omega})} \,,
 &~~~~~
 \dis\d C_s ~=\, -\widehat m_0 c^2 (\zeta' +\zeta e^{i\omega})\,,
 \\[3mm]
 \dis\d B_a ~=\, -\hf{\widehat m_0 a' c(\zeta' -\zeta e^{i\omega})} \,,
 &~~~~~
 \dis\d C_a ~=\, -\widehat m_0 c^2 (\zeta' -\zeta e^{i\omega}) \,.
 \ea
 \eeq
Note that from (\ref{eq:seesaw-Delta-m}) we can compute the ratio,
\beqa
\label{eq:dBa/dCa-2}
\f{\d B_a}{\d C_a} ~=~ \f{a'}{2c} ~=\, -\f{b}{a} \,,
\eeqa
 where in the last step we have used the resolution (\ref{eq:abc-2}).
 It is interesting to note that the ratio (\ref{eq:dBa/dCa-2})
 of the $\mutau$ asymmetric parts in the light neutrino mass-matrix $M_\nu$
 only depends on the $\mutau$ symmetric elements of the Dirac mass-matrix
 $\mD$\,.\,  This ratio just corresponds to the determination of the solar angle
 $\theta_{12}$ in (\ref{eq:t12}) and
 will be further confirmed later by the full NLO analysis of Sec.\,4.\,1.

 \vspace{3mm}
 \section{\large Inverted Ordering: Reconstructing Light Neutrino Mass Matrix \\
           with $\bd{\mutau}$ and CP Violations at Low Energy}
 \label{sec:general}

 In this section, we give the model-independent reconstruction
 of the Majorana mass-matrix for light neutrinos under inverted mass-ordering (IMO),
 in terms of the low energy observables (mass-eigenvalues, mixings angles and CP
 phases). We expand this reconstruction by experimentally well-justified
 small parameters up to the next-to-leading order (NLO).
 Applying this reconstruction formulation to our model will allow us to
 systematically derive the physical predictions for the correlations among
 the low energy observables as well as for the link to the baryon asymmetry
 via leptogensis at the seesaw scale.

\vspace*{3mm}
\subsection{Notation Setup and Model-Independent Reconstruction}
\label{sec:model-independent-reconstruction-formalism}

 Let us consider the general $3\times3$ symmetric and complex
 Majorana mass-matrix for the light neutrinos,
 \beqa
 \label{eq:Mnu33}
  M_\nu
 &\equiv&
  \begin{pmatrix}
    m_{ee} & m_{e \mu} & m_{e \tau} \\[2mm]
    & m_{\mu \mu} & m_{\mu \tau} \\[2mm]
    && m_{\tau \tau}
  \end{pmatrix}
 ~\equiv~
  \begin{pmatrix}
    A & B_1 & B_2 \\[2mm]
      & C_1 & D   \\[2mm]
      &     & C_2
  \end{pmatrix} .
 \eeqa
 In the mass-eigenbasis of charged leptons, the neutrino mass-matrix $M_\nu$
 can be diagonalized by a unitary transformation \,$V(=U_{\nu}^{})$\,, i.e.,
 $\,
  V^T M_\nu V = D_\nu \equiv \textrm{diag}(\ma,\,\mb,\,\mc)
 $\,,\,
 and thus we can write the reconstruction equation,
 \beqa
  M_\nu ~=~ V^* D_\nu V^\dagger \,.
 \label{eq:Mnu-V-D}
 \eeqa
 The mixing matrix $V$ can be generally expressed as a
 product of three unitary matrices including a CKM-type
 mixing matrix $U$ plus two diagonal rephasing matrices $U'$ and $U''$,
 \begin{subequations}
 \label{eq:V-def}
 \beqa
 \label{eq:V}
 && V ~\equiv~ U'' U U' \,,
 \\[2mm]
 && U ~\equiv~
  \begin{pmatrix}
    c_s c_x & - s_s c_x & - s_x e^{i \delta_D}
  \\[1.5mm]
    s_s c_a - c_s s_a s_x e^{-i\delta_D}
  & c_s c_a + s_s s_a s_x e^{-i\delta_D}
  & - s_a c_x
  \\[1.5mm]
    s_s s_a + c_s c_a s_x e^{-i\delta_D}
  & c_s s_a - s_s c_a s_x e^{-i\delta_D}
  & c_a c_x
  \end{pmatrix} \!,
  \label{eq:U}
  \\[3mm]
  &&
  U'  ~\equiv~ \textrm{diag}(e^{i \phi_1^{}},\,e^{i \phi_2^{}},\,e^{i \phi_3^{}})
   \,,  ~~~~~~
  U'' ~\equiv~ \textrm{diag}(e^{i \alpha_1^{}},\,e^{i \alpha_2^{}},\,e^{i \alpha_3^{}})\,,
 \label{eq:U'U''}
 \eeqa
 \end{subequations}
 where $\d_D$ is the Dirac CP-phase. For notational convenience,
 we have denoted the three neutrino mixing angles of the PMNS matrix as,
 $\, (\theta_{12},\,\theta_{23},\,\theta_{13})
      \equiv (\theta_s,\,\theta_a,\,\theta_x) \,,
 $\,
 by following Ref.\,\cite{He2003}.  We will further use the notations,
 $\, (s_s,\,s_a,\,s_x) \equiv
     (\sin\ts ,\,\sin\ta ,\,\sin\tx ) \,$ and
 $\, (c_s,\,c_a,\,c_x) \equiv
     (\cos\ts ,\,\cos\ta ,\,\cos\tx ) \,.
 $\,
 For the diagonal rephasing matrix $U'$, only two of its three Majorana phases are
 measurable (such as $\phi_3-\phi_1$ and $\phi_2-\phi_1$) after extracting
 an overall phase factor. The matrix $U''$ contains another three phases which
 associate with the flavor-eigenbasis of light neutrinos and are needed for
 the consistency of diagonalizing a given mass-matrix \,$M_\nu$\,.

 For convenience we define the rephased mass-eigenvalues
 $\,\widetilde{D}_\nu \equiv U^{\prime *}D_\nu U^{\prime\dag}
    \equiv (\tm_1,\,\tm_2,\,\tm_3) =
    (m_1^{}e^{-i2\phi_1^{}},\,m_2^{}e^{-i2\phi_2^{}},\\\,m_3^{}e^{-i2\phi_3^{}})\,$,\,
 so the reconstruction equation (\ref{eq:Mnu-V-D}) becomes,
 \beqa
 \label{eq:Mnu-V'-Dt}
 M_\nu \,=\, V^{\pp *} \widetilde{D}_\nu V^{\pp\dag}\,,
 &~~~~& (\,V'\equiv U''U\,)\,.
 \eeqa
 Thus, we can fully reconstruct all elements of $M_\nu$
 in terms of the rephased mass-eigenvalues $(\tm_1,\,\tm_2,\,\tm_3)$,
 the mixing angles $(\ts ,\,\ta ,\,\tx )$, the Dirac phase $\d_D$,
 and the rephasing phases $\alpha_i$ (which do not appear in physical
 PMNS mixing matrix),
 \begin{subequations}
  \begin{eqnarray}
  \hspace*{-2mm}
    m_{ee}^{}
  &~ =~ &
    e^{-i2\alpha_1^{}}
  \[
    c^2_s c^2_x \widetilde m_1^{}
  + s^2_s c^2_x \widetilde m_2^{}
  + s^2_x e^{- 2 i \delta_D} \widetilde m_3^{} \] \!,
  \label{eq:Mnu-Reconstruct-ee}
  \\[1.5mm]
  \hspace*{-2mm}
    m_{\mu \mu}^{}
  & =&
    e^{-i2\alpha_2^{}}
  \[
    ( s_s c_a  - c_s s_a s_x e^{i \delta_D} )^2 \widetilde m_1^{}
  + ( c_s c_a  + s_s s_a s_x e^{i \delta_D} )^2 \widetilde m_2^{}
  + s^2_a c^2_x \widetilde m_3^{}
  \] \!,
  \label{eq:Mnu-Reconstruct-mm}
  \\[1.5mm]
  \hspace*{-2mm}
    m_{\tau\tau}^{}
  & \!\!\!\!=\!\!\! &
    e^{-i2\alpha_3^{}}
  \[
    ( s_s s_a + c_s c_a s_x e^{i \delta_D} )^2 \widetilde m_1^{}
  + ( c_s s_a - s_s c_a s_x e^{i \delta_D} )^2 \widetilde m_2^{}
  +   c^2_a c^2_x \widetilde m_3^{}
  \] \!,
  \label{eq:Mnu-Reconstruct-tt}
 \\[1.5mm]
 \hspace*{-2mm}
    m_{e\mu}^{}
  & \!\!\!\!=\!\!\! &
    e^{-i(\alpha_1^{}+\alpha_2^{})}
  \!\[
    c_s c_x (s_s c_a \!-\! c_s s_a s_x e^{i \delta_D}) \widetilde m_1^{}
  \!-\! s_s c_x (c_s c_a + s_s s_a s_x e^{i \delta_D}) \widetilde m_2^{}
  \!+\! s_a s_x c_x e^{- i\delta_D} \widetilde m_3^{}
  \] \!,~~~~~~~~
  \label{eq:Mnu-Reconstruct-em}
  \\[1.5mm]
  \hspace*{-2mm}
    m_{e\tau}^{}
  & \!\!\!\!=\!\!\! &
    e^{-i(\alpha_1^{}+\alpha_3^{})}
  \!\[
    c_s c_x (s_s s_a \!+\! c_s c_a s_x e^{i \delta_D}) \widetilde m_1^{}
  \!-\! s_s c_x (c_s s_a \!-\! s_s c_a s_x e^{i \delta_D}) \widetilde m_2^{}
  \!-\! c_a s_x c_x e^{- i \delta_D} \widetilde m_3^{}
  \] \!,~~~~~~~~
  \label{eq:Mnu-Reconstruct-et}
  \\[1.5mm]
  \hspace*{-2mm}
    m_{\mu\tau}^{}
  & \!\!\!\!=\!\!\! &
    e^{-i(\alpha_2^{}+\alpha_3^{})}
  \!\[
    ( s_s c_a - c_s s_a s_x e^{i \delta_D} )
    ( s_s s_a + c_s c_a s_x e^{i \delta_D} ) \widetilde m_1^{}
  \right.
    \nonumber
  \\
  \hspace*{-2mm}
  & \!\!\!\!\!\!\! &
  \hspace{16mm}
  \left.
  ~~ + ( c_s c_a + s_s s_a s_x e^{i \delta_D} )
    ( c_s s_a - s_s c_a s_x e^{i \delta_D} ) \widetilde m_2^{}
  -   s_a c_a c^2_x \widetilde m_3^{}
  \] \!,
  \label{eq:Mnu-Reconstruct-mt}
  \end{eqnarray}
  \label{eq:Mnu-Reconstruct}
 \end{subequations}
 where among the Majorana phases $\phi_{1,2,3}^{}$
 (hidden in the mass-parameters $\tm_{1,2,3}^{}$) only two are
 independent because an overall phase factor of $U'$ can be
 absorbed into the diagonal rephasing-matrix $U''$.\,
 For the case with a vanishing mass-eigenvalue (such as $m_3^{}=0$ in our present model),
 only one independent phase combination, say $e^{i(\phi_2^{}-\phi_1^{})}$, will survive.
 If we impose $\mutau$ symmetry on the light neutrino mass-matrix $M_\nu$,
 we can deduce\,\cite{GHY},
 \beqa
  \(\ta ,\,\tx \)_0
 \,=\, (45^\deg ,\, 0^\deg ) \,,
 &~~~~~& \a_{20}^{} \,=\, \a_{30}^{} \,.
  \label{eq:mutau-solution}
 \eeqa
 The solar mixing angle $\ts$ is independent of the $\mutau$ symmetry
 and is thus left undetermined. To predict $\ts$, we will uncover a new flavor
 symmetry beyond the $\ZZ_2^{\mu\tau}$ (cf.\ Sec.\,5).

\vspace*{3mm}
\subsection{Reconstruction of Light Neutrino Mass Matrix with Inverted Ordering }
\label{sec:Reconstruct-IMO}

 Now we are ready to apply the above general reconstruction formalism to
 the inverted mass-ordering (IMO), $\,\mb\gtrsim \ma \gg \mc\,$, with $\,\mc=0\,$
 (as predicted by the present minimal seesaw model), in contrast to our previous model
 which predicts the normal mass-ordering (NMO)\,\cite{GHY}.
 We introduce a small mass-ratio for light neutrinos,
 \begin{equation}
  y' ~\equiv~\frac{m_2^2-m_1^2}{m_1^2} ~=~
 \f{\Delta  m^2_{21}}{\Delta m^2_{13}} ~=~ 0.029-0.036
   \,\ll\, 1 \,,
  \label{eq:definition-y}
 \end{equation}
 as constrained by the neutrino data at $90\%$\,C.L.\,(Table-1).
 So it is sufficient to make perturbative expansion in $y'$ up to its
 linear order. Thus, at the zeroth order of $y'$, we have equal mass-eigenvalues,
 $\,m_{10}^{}=m_{20}^{}=m_0^{}\,$.\,
 Under the $y'$-expansion up to next-to-leading
 order (NLO),  $\,m_{i}^{}=m_{0}^{}+\d m_i^{}\,$,\, we have
\beqa
\dis y' ~\simeq~ \f{\,2(\d m_2^{}-\d m_1^{})\,}{m_1^{}}\,
~=~ \frac{\,2(m_2^{}-m_1^{})\,}{m_1^{}} \,.
\eeqa
We can define another small ratio $\dis\,z\equiv\f{\d m_1^{}}{m_{1}^{}}=\O(y')\,$,\,
and deduce,
\beqa
\label{eq:z-y}
\d m_1^{} ~=~ z\,m_{1}\,, &~~~&
\d m_2^{} ~=~ \(z+\frac{y'}{2}\)m_1^{} \,,
\eeqa
where $\,\ma = \sqrt{\Delta m^2_{13}}\,$ is fixed by the neutrino data, and
$\,m_0^{}=\ma - \d\ma = (1-z)\ma \simeq \sqrt{\Delta m^2_{13}}\,$.\,

Next, we consider the mixing angles and CP-phases. Since the neutrino oscillation data
strongly support the $\mutau$ symmetry as a good approximate symmetry
(\ref{eq:mutau-solution}), we can define the small deviations from the general
$\mutau$ symmetric solution (\ref{eq:mutau-solution}),
\beqa
\da ~\equiv~ \ta -\f{\pi}{4} \,,
&~~~~~&
\dx ~\equiv~ \tx - 0\,,
\eeqa
which characterize the $\mutau$ symmetry breaking.
From the data in Table-1, we can infer the constrained $90\%$\,C.L. ranges,
\beqa
0 ~\leqq ~\d_x^2~ \leqq~ 0.027\,, &~~~&
0 ~\leqq ~\d_a^2~ \leqq~ 0.015\,.
\eeqa
For our analysis we will systematically expand the small parameters
$\,(\d_a,\,\d_x,\,y',\, z)\,$ up to their linear order.
For the Majorana CP-phases,  $\phi_3^{}$ drops due to \,$\mc=0\,$;\,
we also remove an overall redundant Majorana phase $\phi_1^{}$ (from $U'$)
into the redefinition of $\,\a_j^{}$ (in $U''$).\,
So, the remaining independent Majorana phase is only $\,\phi\,$,
\beqs
\beqa
\ab_j^{} &\equiv & \a_j^{} + \phi_1^{} \,, ~~~~~(j=1,2,3)\,,
\\[0.9mm]
\phi^{} &\equiv&  \phi_2^{}-\phi_1^{} ~=~ \phi_0^{} +\d\phi \,.
\eeqa
\eeqs
The expansion up to the NLO for our current reconstruction analysis
will include $\,(\d \ab_1^{},\,\d\ab_2^{},\,\d\ab_3^{},\,\delta\phi)$\,.\,
The solar angle $\,\theta_s\,(\equiv\theta_{12})$\,
is independent of the $\mutau$ breaking and thus receives no NLO correction.
Furthermore, we note that the Dirac phase $e^{i\d_D}$ is always associated
with the small mixing parameter $\,s_x\,(\simeq \d_x)\,$,\,
so it only appears at the NLO and
thus receive no more correction at this order of expansion.

Finally, we give a summary of all relevant
NLO parameters in our reconstruction analysis,
\beqa
 (\, y',\,
     z,\,
     \d_a,\,
     \d_x,\,
     \d\ab_1^{},\,
     \d\ab_2^{},\,
     \d\ab_3^{}, \,\delta\phi)\,,
  \label{eq:NLO-parameters}
\eeqa
Each of them is defined as the difference between its full value and zeroth-order
value under the $\mutau$ symmetric limit.
In Sec.\,4 we will derive these deviations from
our seesaw model for the common origin of $\mutau$ and CP breaking,
and analyze their correlations.

Making the perturbative expansion of (\ref{eq:NLO-parameters}) under the
inverted mass-ordering,
we first deduce the LO form of the light neutrino mass-matrix
(\ref{eq:Mnu33}),
\begin{subequations}
  \beqa
m_{ee}^{(0)}\equiv  A_0
  &=&
    m_0^{} e^{-2i \bar\alpha_{10}}
  \left(
    c^2_{s}
  + s^2_{s}  e^{- i2\phi_{0}}
  \right) ,
  \\[2mm]
m_{e \mu}^{(0)}= m_{e \tau}^{(0)}\equiv   B_0
  & = &
    \f{1}{\sqrt{2}\,} m_0^{} s_{s} c_{s}
    e^{-i(\bar\alpha_{10}+\bar\alpha_{20})}
  \left(1  -  e^{-i2\phi_{0}}
  \right) ,
  \\[2mm]
 m_{\mu\mu}^{(0)}=m_{\tau\tau}^{(0)} \equiv  C_0
  & = &
    \f{1}{2} m_0^{} e^{-2i\bar\alpha_{20}}
  \left(
    s^2_{s}
  + c^2_{s} e^{- 2 i \phi_{0}}
  \right) \,=~ D_0 \,,
  \eeqa
 \label{eq:dMnu-0}
\end{subequations}
 where we have also matched to our notation of
 $\,M_\nu^{(0)}\,$ in (\ref{eq:Mu=Mu0+dMu}).
 Then, we derive elements of the NLO mass-matrix
 $\,\d M_\nu^{(1)}\,$ from (\ref{eq:Mnu-Reconstruct}),
\begin{subequations}
  \begin{eqnarray}
    \d m_{ee}^{(1)}  & \,\equiv\, &  \d A ~=~
   m_0e^{- i2\ab_{10}^{}}
   \[z+\frac{s_s^2}{2}y'- i2(s_s^2\,\delta\phi+\delta\bar\alpha_1)\]\,,
\label{eq:dMnu-1-ee}
  \\[2mm]
    \d m_{e\mu}^{(1)}  & \,\equiv\, &  \d B_1 ~=~
    \frac{m_{0}}{\sqrt{2}}e^{-  i(\ab_{10}^{}+\ab_{20}^{})}
    \[-\frac{c_ss_s}{2}y'-e^{i\delta_D}\delta_x+ i2c_ss_s\delta\phi\],
\label{eq:dMnu-1-em}
  \\[2mm]
    \d m_{e\tau}^{(1)}  & \,\equiv\, &  \d B_2 ~=~
   \frac{m_{0}}{\sqrt{2}}e^{- i(\ab_{10}^{}+\ab_{20}^{})} \[-\frac{c_ss_s}{2}y'+e^{i\delta_D}\delta_x+ i2c_ss_s\delta\phi\] ,
\label{eq:dMnu-1-et}
  \\[2mm]
    \d m_{\mu\mu}^{(1)}  & \,\equiv\, &  \d C_1 ~=~
    m_0e^{- i2\ab_{20}^{}}
    \[\frac{z}{2}+\frac{c_s^2}{4}y'-\delta_a
      -i(c_s^2\,\delta\phi+\delta\bar\alpha_2)\],
\label{eq:dMnu-1-mm}
  \\[2mm]
\d m_{\tau\tau}^{(1)}  & \,\equiv\, &  \d C_2 ~=~
   m_0e^{- i2\ab_{20}^{}}
   \[\frac{z}{2}+\frac{c_s^2}{4}y'+\delta_a
     -i(c_s^2\,\delta\phi+\delta\bar\alpha_3)\],
\label{eq:dMnu-1-tt}
  \\[2mm]
\d m_{\mu\tau}^{(1)}  & \,\equiv\, &  \d D ~=~
    m_0e^{- i2\ab_{20}^{}}
    \[\frac{z}{2}+\frac{c_s^2}{4}y'-\frac{i}{2}(2c_s^2\,\delta\phi
      +\delta\bar\alpha_2+\delta\bar\alpha_3)\],
\label{eq:dMnu-1-mt}
  \end{eqnarray}
  \label{eq:dMnu-1}
\end{subequations}
where we have matched to our notation of $\d M_\nu^{(1)}$
as defined in (\ref{eq:Mu=Mu0+dMu}).
In the above formulas, we have used the $\mutau$ symmetric relations for the
LO parameters, $\,(\theta_{a0},\,\theta_{x0})=(\f{\pi}{4},\,0)\,$
and $\,\ab_{20}^{}=\ab_{30}^{}\,$, as well as $\,m_3\equiv 0\,$\,.

From (\ref{eq:dMu=s+a}),
we can uniquely decompose the elements of $\d M_\nu^{(1)}$
in (\ref{eq:dMnu-1}) as the $\mutau$ symmetric and anti-symmetric parts,
$\,\d M_\nu^{(1)} \equiv \d M_\nu^{s}+\d M_\nu^{a}\,$,\,
with their elements given by,
\beq
\label{eq:Reconstruct-dMnu-sa}
\ba{l}
\d B_s ~\equiv~ \dis\f{\d B_1+\d B_2}{2}
~=~ \frac{m_{0}}{\sqrt{2}}e^{- i(\ab_{10}^{}+\ab_{20}^{})}
\[-\frac{c_ss_s}{2}y'+ i2c_ss_s\delta\phi\] \,,
\\[5mm]
\d B_a ~\equiv~ \dis\f{\d B_1-\d B_2}{2}
~=~ -\frac{m_{0}}{\sqrt{2}}e^{- i(\ab_{10}^{}+\ab_{20}^{})}e^{i\delta_D}\delta_x\,,
\\[5mm]
\d C_s ~\equiv~ \dis\f{\d C_1+\d C_2}{2}
~=~ m_0e^{- i2\ab_{20}^{}}
\[\frac{z}{2}+\frac{c_s^2}{4}y'
  -\frac{i}{2}(2c_s^2~\delta\phi+\delta\overline\alpha_2
  +\delta\overline\alpha_3)\] \,=~ \d D \,,
\\[5mm]
\d C_a ~\equiv~ \dis\f{\d C_1-\d C_2}{2}
~=~ -m_{0}e^{- i2\ab_{20}^{}}
\[\delta_a+\frac{i}{2}(\delta\overline\alpha_2
  -\delta\overline\alpha_3)\] .
\ea
\eeq
With these, we will be ready to apply the above reconstruction formulas
(\ref{eq:dMnu-0}), (\ref{eq:dMnu-1}) and (\ref{eq:Reconstruct-dMnu-sa})
to match with (\ref{eq:Mu=Mu0+dMu})
in our seesaw model at the LO and NLO, respectively.
We will systematically solve these matching conditions in the next section,
which allows us to connect the seesaw parameters to the low energy neutrino
observables and deduce our theoretical predictions.

For matching the seesaw predictions to our reconstruction formalism, we note that
the latter was presented at the low energy scale so far. We need to connect the
low energy neutrino parameters to the model predictions at the seesaw scale,
where the possible renormalization group (RG) running effects should be
taken into account in principle.
Such RG effects were extensively discussed in the literature\,\cite{nuRG},
and can be straightforwardly applied to the present analysis.
Below the seesaw scale, heavy right-handed neutrinos can be integrated out from the
effective theory and the seesaw mass-eigenvalues $\,m_j^{}\,$($j=1,2,3$)
for light neutrinos obey the approximate one-loop RG equation (RGE)\,\cite{nuRG},
\beqa
\label{eq:RGE-mj}
\f{dm_j^{}}{dt} &\,=\,& \f{\ahat}{16\pi^2}m_j^{} \,,
\eeqa
to good accuracy\,\cite{footnote-3},\,
where $\,t=\ln(\mu/\mu_0^{})\,$ with $\,\mu\,$ the renormalization scale.
For the SM, the coupling-parameter
$\,\ahat \simeq -3g_2^2 +6y_t^2 +\lambda\,$,\,
with $\,(g_2^{},\,y_t^{},\,\lambda)\,$ denoting the $SU(2)_L$ weak gauge coupling,
the top Yukawa coupling and Higgs self-coupling, respectively.
Hence, we can deduce the running mass-parameter
$\,m_j^{}\,$ from scale $\mu_0^{}$ to $\,\mu\,$,
%
\beqa
\label{eq:RG-Run-mj}
m_j^{}(\mu) ~=~ \chi (\mu,\mu_0^{})\,m_j^{}(\mu_0^{})
~\simeq~ \exp\[\f{1}{16\pi^2}\int_{0}^t \ahat(t')\,dt'\]m_j^{}(\mu_0^{})\,,
\eeqa
%
with $\,t=\ln(\mu/\mu_0^{})\,$.\,
In the present analysis we will choose,
$\,(\mu_0^{},\,\mu) = (M_Z,\,M_1)\,$,\,
with $Z$ boson mass $M_Z$ representing the weak scale and
the heavy neutrino-mass $M_1$ characterizing the seesaw scale.

Consider the minimal neutrino seesaw with inverted mass-spectrum,
$\,\mb\gtrsim \ma \gg \mc\,=0\,$.\,
We note that the zero-eigenvalue $\,\mc\,$ and the mass ratio $\,y'$
do not depend on the RG running scale $\,\mu\,$.\,  So we can derive the running
of the two nonzero mass-parameters from weak scale to seesaw scale,
\beqs
\label{eq:RG-Run-m123}
\beqa
&& \mh_1^{} ~\equiv~ m_1^{}(M_1) ~=~ \chi_1^{}\, m_1^{}(M_Z) ,
\\[1mm]
&& \mh_2^{} ~\equiv~ m_2^{}(M_1) ~=~ \chi_1^{}\,m_2^{}(M_Z)
            ~=~ \sqrt{1+y'}\,\,\mh_1^{}\,\,,
\eeqa
\eeqs
with $\,\chi_1^{} ~\equiv~ \chi (M_1,M_Z)\,$.\,
In Sec.\,4, we will compute the RG running factor
$\,\chi_1^{}\equiv \chi (M_1,M_Z)\,$ numerically,
which depends on the inputs of initial values for
$\,\alpha_2^{}=g_2^2/(4\pi)\,$,\, $\,y_t^{}\,$ and the Higgs boson mass $\,M_H\,$,
via the combination $\,\ahat\,$ defined above.
Using the electroweak precision data\,\cite{PDG,mH-fit},
$\,\a_2^{-1}(M_Z)=29.57\pm 0.02\,$,\, $\,m_t^{}=173.1\pm 1.4\,$GeV,
and the Higgs-mass range
$\,115\leqq M_H^{} \leqq 149\,$GeV [90\%\,C.L.] for the SM,
we find the running factor $\,\chi (M_1,M_Z)\simeq 1.3-1.4\,$
for $\,M_1 = 10^{13}-10^{16}\,$GeV.
Other running effects due to the leptonic mixing angles and CP-phases
are all negligible for the present study since their RGEs contain only
flavor-dependent terms and are all suppressed by
$\,y_\tau^2 =\O(10^{-4})\,$ at least\,\cite{nuRG}.
For the analyses below (Sec.\,4), we will first evolve the
mass-parameters from the seesaw scale $\,M_1\,$ down to the
low energy scale for neutrino oscillations, and then match them with
those in our reconstruction formalism. Including such RG effects just requires to
replace the light mass-eigenvalues $\,(\mh_1^{},\,\mh_2^{})\,$
at seesaw scale $\,M_1\,$ by the corresponding $\,(m_1^{},\,m_2^{})\,$
at low energy, and vice versa.

\section{\large
Predictions of Common $\bd{\mutau}$ and CP Breaking with Inverted Ordering}
\label{sec:diagonalization}

 In this section we apply the reconstruction formalism (including the RG running effects)
 in Sec.\,3.\,2 to our common $\mutau$ and CP breaking seesaw
 in Sec.\,2.\,3.
 Then, we systematically derive the predictions for the low energy
 neutrino observables.  This includes the nontrivial correlation between
 two small $\mutau$ breaking parameters
 $\,\d_x \(\equiv \theta_{13} - 0\)\,$ and $\,\d_a \(\equiv\theta_{23}-\f{\pi}{4}\)\,$.\,
 Furthermore, we study the correlations of $\,\theta_{23}-45^\deg\,$ and $\,\theta_{13}\,$
 with Jarlskog invariant $\,J\,$ and neutrinoless $\beta\beta$-decay observable
 \,$M_{ee}$\,.\,  Finally, we study the matter-antimatter asymmetry
 (baryon asymmetry) via leptogenesis in the $\mutau$ blind seesaw, and establish the
 direct link with low energy neutrino observables.
 Furthermore, we will derive a nontrivial lower bound on the reactor
 mixing angle, $\,\theta_{13}\gtrsim 1^\deg\,$,\, and restrict the
 Jarlskog invariant into a negative range, $\,-0.037\lesssim J \lesssim -0.0035\,$.

\vspace{3mm}
\subsection{Predicting Correlations of Low Energy Neutrino Observables}
\label{sec:correlation-1}

 Both $\mutau$ and CP violations arise from
 a common origin in the seesaw Lagrangian of our model,
 which is characterized by the breaking parameter
 $\,\zeta e^{i\om}\,$ and shows up at the NLO of our perturbative expansion.
  Hence, in the light neutrino mass-matrix,
  the small $\mutau$ breaking parameters $(\d_a,\,\d_x)$ together with all CP-phases
  are controlled by $\,\zeta\,$ and $\,\om\,$.\,
  In the following, we will use the reconstruction formalism (Sec.\,3.\,2)
  under IMO for diagonalizing the light neutrino mass-matrix at the NLO.
  Then, we will further derive quantitative predictions for these
  low energy observables and their correlations.

 We first inspect the reconstructed LO mass-matrix $M_\nu^{(0)}$
 in (\ref{eq:dMnu-0}).  Matching (\ref{eq:dMnu-0}) with our model prediction
 (\ref{eq:Mnu0}) at the same order, we find the solutions,
\begin{subequations}
\beqa
\label{eq:alpha10-20-phi0}
&& \bar\alpha_{10}^{} \,=\, \bar\alpha_{20}^{} \,=\, \phi_0^{}
   \,=\, 0\,, 
\\
\label{eq:m10-20-m3}
&& m_{10}^{}\,=\, m_{20}^{} \,=\, m_0 \,, ~~~ m_3=0 \,,
\\
\label{eq:abca'}
&& a^2 \,=\, 2c^2 \,=\, 1-2b^2 \,,~~~
a'^2 \,=\, 2b^2\,,~~~
c^2 \,=\, \f{1}{2}-b^2 \,,~~~
a'c \,=\, - ab \,,
\label{eq:alpha-20}
\eeqa
\label{eq:LO-sol}
\end{subequations}
which is also consistent with Eq.\,(\ref{eq:abc-2}).
Here all the LO CP-phases
$\,(\bar\alpha_{10}^{},\, \bar\alpha_{20}^{},\, \phi_0^{})=0\,$
because the original CP-violation in the seesaw Lagrangian vanishes in the
$\,\zeta =0\,$ limit (Sec.\,2.\,2).

Then, we analyze the NLO light neutrino mass-matrix $\d M_\nu^{(1)}$, as given by
(\ref{eq:dMu=s+a}) of our model and by the reconstruction formula (\ref{eq:dMnu-1}).
We match the two sets of equations at the low energy
for the $\mutau$ symmetric elements,
\begin{subequations}
  \begin{eqnarray}
   \d A ~=~ 0  & = & m_0^{}
   \[z +\frac{s_s^2}{2}y'-i2(s_s^2~\delta\phi+\delta\bar\alpha_1)\]\,,
    \label{eq:re-delta-equality-A}
  \\[2mm]
   \d B_s ~=~-\frac{m_0^{}}{2} a' c(\zeta' +\zeta e^{i\omega})  & = &
  \frac{m_0}{\sqrt{2}\,}
  \[-\frac{c_ss_s}{2}y'+ i2c_ss_s\delta\phi\] \,,
    \label{eq:re-delta-equality-Bs}
  \\[2mm]
  \d C_s ~=~ -m_0^{} c^2 (\zeta' +\zeta e^{i\omega})
  & = & \frac{m_0^{}}{2}
        \[z +\f{c_s^2}{2}y'-i(2c_s^2\,\delta\phi + \delta\bar\alpha_2
          +\delta\bar\alpha_3)\] \,=\delta D,
    \label{eq:re-delta-equality-Es}
  \end{eqnarray}
  \label{eq:re-delta-equality-s}
\end{subequations}
and for $\mutau$ anti-symmetric elements,
\beqs
\beqa
 \dis\d B_a  ~=-\frac{m_0^{}}{2} a' c(\zeta'-\zeta e^{i\omega})
 &=&
 -\frac{m_0^{}}{\sqrt{2}\,}\,e^{i\delta_D}\delta_x \,,
 \label{eq:dBa}
 \\[2mm]
 \dis\d C_a  ~= - m_0^{} c^2 (\zeta'-\zeta e^{i\omega})
 &=& - m_0^{}\[\delta_a+\frac{i}{2}(\delta\bar\alpha_2
                      -\delta\bar\alpha_3)\],
 \label{eq:dCa}
\eeqa
\label{eq:dBa-dCa}
\eeqs
where using Eq.\,(\ref{eq:RG-Run-m123}) we have run the mass-parameter
$\,\mh_{0}^{}\,$ from the seesaw scale down to the corresponding $\,m_{0}^{}\,$
at low energy for the left-hand-sides of
Eqs.\,(\ref{eq:re-delta-equality-s}) and (\ref{eq:dBa-dCa}).

From the $\mutau$ symmetric
 Eqs.\,(\ref{eq:re-delta-equality-s}a)-(\ref{eq:re-delta-equality-s}b),
 we can infer six independent conditions
 for the real and imaginary parts of
 $\,(\delta A,\,\delta B_s,\,\delta C_s)\,$,\, respectively,
 \begin{subequations}
 \begin{eqnarray}
 z &~=~& -\frac{s_s^2}{2}y' \,,
 \label{eq:zy}
 \\[1mm]
\delta\bar\alpha_1^{} &=& -s_s^2\,\delta\phi \,,
\\[1.5mm]
\frac{c_ss_s}{\sqrt{2}\,}\,y' &=& a'c\(\zeta' +\zeta \cos\omega\) \,,
\label{eq:ab}
\\[1.5mm]
2\sqrt{2}~c_ss_s\delta\phi &=& -a'c \,\zeta\sin\omega \,,
\\[1mm]
\frac{z}{2}+\frac{c_s^2}{4}\,y' &=& -c^2\(\zeta' +\zeta \cos\omega\) \,,
\label{eq:bb}\\
-\frac{1}{2}(2c_s^2~\delta\phi+\delta\bar\alpha_2^{}
+\delta\bar\alpha_3^{}) &=& -c^2 \,\zeta\sin\omega \,.
\end{eqnarray}
\label{eq:solve-s}
\end{subequations}
Thus, with the aid of (\ref{eq:zy}) we take the ratio of
(\ref{eq:ab}) and (\ref{eq:bb}),  and derive
\begin{equation}
  \tan2\ts ~=~ -\frac{a'}{\sqrt{2}\,c\,} ~=~ \frac{\sqrt{2}\,b}{a} \,,
  \label{eq:t12-NLO}
\end{equation}
which coincides with (\ref{eq:t12}) in Sec.\,2.\,1.
Using Eq.\,(\ref{eq:t12-NLO}), we deduce from Eq.\,(\ref{eq:alpha-20}),
\begin{subequations}
\begin{eqnarray}
 a \,=\, p_a^{}\cos2\theta_s \,, ~~~&&~~~
 b \,=\, p_a^{}\frac{1}{\sqrt{2}\,}\sin 2\theta_s \,,
 \\
 a' =\, p_{a'}^{}\sin 2\theta_s \,, ~~~&&~~~
 c \,= -p_{a'}^{}\frac{1}{\sqrt{2}\,}\cos 2\theta_s \,,
\end{eqnarray}
\label{eq:expression-abc-new}
\end{subequations}
with $\,p_a^{},\,p_{a'}^{}=\pm$\, denoting the signs of $(a,\,a')$.
Here we see that the four dimensionless LO parameters
$(a,\,a',\,b,\,c)$ in the Dirac mass-matrix (\ref{eq:mD-B-1}) are fixed
by the solar mixing angles $\,\ts$\,,\, since the conditions in (\ref{eq:abca'})
make three of them non-independent.
Finally, we further resolve (\ref{eq:solve-s}) and derive the NLO parameters,
\begin{subequations}
\begin{eqnarray}
y' &=& -2\cos2\theta_s\(\zeta'+\zeta \cos\omega\) ,
\label{eq:solve-y'}
\\[2mm]
z &=& s_s^2\cos 2\theta_s \(\zeta' +\zeta\cos\omega\) ,
\label{eq:solve-z}
\\[2mm]
\delta\bar\alpha_1^{}
&=&
-\f{1}{2} s_s^2(c_s^2-s_s^2) \,\zeta\,\sin\omega \,,
\label{eq:solve-dalpha1}
\\[2mm]
\delta\phi &=&
\f{1}{2}(c_s^2-s_s^2) \,\zeta\,\sin\omega \,,
\label{eq:dphi}
\\[2mm]
\delta\bar\alpha_2^{} + \delta\bar\alpha_3^{}
&=&
s_s^2(s_s^2-c_s^2) \,\zeta\,\sin\omega \,.
\label{eq:solve-dalpha-23}
\end{eqnarray}
\label{eq:solve-II-s}
\end{subequations}
It is interesting to note that the present model predicts a generically small
Majorana CP-phase angle at low energy, $\,\phi =\d\phi = O(\zeta)\,$,\,
in contrast to our soft breaking model\,\cite{GHY}
where the low energy Majorana CP-phase angle ($\phi_{23}^{}$) is not suppressed.

Next, we analyze the $\mutau$ anti-symmetric equations (\ref{eq:dBa})-(\ref{eq:dCa})
for $\,\d M_\nu^{(1)}$.\,  With (\ref{eq:expression-abc-new}),
we can deduce from (\ref{eq:dBa})-(\ref{eq:dCa}),
\beqs
\beqa
 \frac{1}{2}\sin 2\theta_s\cos 2\theta_s(\zeta' -\zeta e^{i\omega})
 &~=~& -e^{i\delta_D}\,\delta_x \,,
 \label{eq:dBa-1}
 \\[2mm]
 \frac{1}{2}\cos^22\theta_s\(\zeta' -\zeta e^{i\omega}\)
 &=& \delta_a + \frac{i}{2}(\delta\bar\alpha_2^{}
 -\delta\bar\alpha_3^{}) \,,
 \label{eq:dCa-1}
\eeqa
\label{eq:dBa-dCa-1}
\eeqs
which decompose into
\begin{subequations}
\begin{eqnarray}
\cos\delta_D~\delta_x
&~=~&
-\frac{1}{2}\sin2\theta_s\cos2\theta_s\(\zeta' -\zeta \cos\omega\) ,
\label{eq:solve-II-a-a}
\\[2mm]
\sin\delta_D~\delta_x
&=&
\frac{1}{2}\sin2\theta_s\cos2\theta_s \(\zeta\,\sin\omega\) ,
\label{eq:solve-II-a-b}
\\[2mm]
\delta_a
&=&
\frac{1}{2}\cos^22\theta_s
\(\zeta' -\zeta \cos\omega\) ,
\label{eq:solve-II-a-c}
\\[2mm]
\delta\bar\alpha_2 -\delta\bar\alpha_3
&=&
-\cos^22\theta_s \(\zeta\,\sin\omega\) \,.
 \end{eqnarray}
 \label{eq:solve-II-a}
 \end{subequations}
Thus the Dirac CP-phase angle $\delta_D$ can be derived from the ratio
of (\ref{eq:solve-II-a-a}) and (\ref{eq:solve-II-a-b}),
\begin{eqnarray}
\tan \delta_D ~=~
\frac{\zeta\,\sin\omega}{\zeta\cos\omega -\zeta'}
~=~ \frac{\,\delta\bar\alpha_2-\delta\bar\alpha_3\,}{2\delta_a} \,.
\label{eq:delta-D}
\end{eqnarray}
With Eqs.\,(\ref{eq:solve-y'}), (\ref{eq:delta-D}) and (\ref{eq:solve-II-a}),
we finally deduce,
%
\begin{eqnarray}
\zeta' +\zeta \cos\omega \,=\, -\frac{1}{2\cos2\theta_s}\,y'\,,
&~~~~~~&
-\zeta \sin\omega \,=\, \frac{2\tan\delta_D}{\cos^22\theta_s}\,\delta_a \,,
\end{eqnarray}
%
and thus
\begin{subequations}
  \begin{eqnarray}
     \cos\delta_D \dx
  &\!= \!\!&  -\frac{\sin2\theta_s}{4}\(y'+4\cos2\theta_s~\zeta' \)
  \,=~\frac{\sin 2\theta_s}{4}\(y'+4\cos2\theta_s\cos\omega\,\zeta\) ,
\label{eq:sol-deltax}
\\[2mm]
    \da &\!= \!\!&
  \frac{\cos2\theta_s}{4}\(y'+4\cos2\theta_s~\zeta'\)
 \,=\, -\frac{\cos2\theta_s}{4}\(y'+4\cos2\theta_s\cos\omega\,\zeta\) ,
\label{eq:sol-deltaa}
\\[2mm]
    \delta\bar\alpha_2^{} -\delta\bar\alpha_3^{}
    &=& 2\tan\delta_D\,\da  \,.
  \end{eqnarray}
\label{eq:NLO-Solution}
\end{subequations}
From Eqs.\,(\ref{eq:sol-deltax}) and (\ref{eq:sol-deltaa}),
we derive a nontrivial correlation between the low energy
$\mutau$ breaking observables $\,\da\,$ and $\,\dx$\,,
\begin{eqnarray}
{\da} ~=\, -\cot 2\theta_s \cos\delta_D \,\dx \,.
\label{eq:da-dx-1}
\end{eqnarray}
This shows that at the NLO the two small $\mutau$ breaking parameters
are {\it proportional to each other,} $\,\dx \propto \da\,$.\,
Because of \,$|\cos\d_D| \leqq 1$\,,\,
we can infer from Eq.\,(\ref{eq:da-dx-1}) a generic {\it lower bound}
on $\,\delta_x$\,,\, for any nonzero $\,\da$\,,
\beqa
\label{eq:Lbound-dx}
  \delta_x &\geqq&
  |\delta_a|\tan2\theta_s  \,,
\eeqa
where we have $\,\dx \equiv \theta_{13}\in [0,\f{\pi}{2}]\,$ in our convention.
It is worth to note that our previous soft breaking model\,\cite{GHY}
also predicted a correlation and a lower bound,
\beqs
\begin{eqnarray}
\label{eq:NMO-da-dx-1}
&&
{\da} ~=\, -\cot \theta_s \cos\delta_D \,\dx \,,
~~~~~~ (\textrm{Prediction~of~Ref.\,\cite{GHY}}),~~~~
\\[1mm]
&&\To~~~ \delta_x ~\geqq~ |\delta_a|\tan\theta_s  \,,
\end{eqnarray}
\eeqs
where the quantitative difference from the present predictions is that
we have the coefficient $\,\cot 2\theta_s\,$ in Eq.\,(\ref{eq:da-dx-1})
as compared to $\,\cot\theta_s\,$ in Eq.\,(\ref{eq:NMO-da-dx-1}).
In fact, this is a {\it profound difference.}
From the present oscillation data in Table-1, we observe that the deviation
of the solar angle $\theta_s\,(\equiv\theta_{12})$ from its maximal mixing value
is relatively small,
\beqa
\label{eq:45-t12}
9.0^\deg ~<~ 45^\deg - \theta_{12} ~<~ 12.2^\deg\,,
&~~~~~~& \textrm{(at 90\%\,C.L.),}
\eeqa
and this limit only relaxes slightly at 99\%\,C.L.,
$\,7.8^\deg < 45^\deg - \theta_{12} < 12.9^\deg\,$.\,
Hence, we see that the range of the deviation $\,45^\deg - \theta_{12}\,$ is
at the same level as the two other small deviations $\,\theta_{23}-45^\deg\,$
and $\,\theta_{13}-0^\deg\,$ shown in Eq.\,(\ref{eq:da-dx-exp}).
So, we can define a new naturally small quantity,
\beqa
\ds ~\equiv~ \f{\pi}{4} - \theta_s \,,
\eeqa
and make expansion for $\,\ds\,$ as well. Then, we immediately observe a
{\it qualitative difference} between $\,\cot 2\ts\simeq 2\ds\ll 1\,$ in (\ref{eq:da-dx-1})
and $\,\cot \ts\simeq 1+2\ds \gtrsim 1\,$ in (\ref{eq:NMO-da-dx-1}).
Hence, we can rewrite the two correlations (\ref{eq:da-dx-1}) and (\ref{eq:NMO-da-dx-1})
in the well expanded form,
\beqs
\label{eq:da-dx-expand-ds}
\beqa
\label{eq:IMO-da-dx-expand-ds}
{\da} &\,\simeq& - 2\cos\delta_D \,(\ds\dx ) ~\ll~ \dx \,,
~~~~~~~\, (\textrm{Current Prediction}),
\\[1.5mm]
\label{eq:NMO-da-dx-expand-ds}
{\da} &\,\simeq& - \cos\delta_D \,\dx ~=~ O(\dx ) \,,
~~~~~~~~~~ (\textrm{Prediction~of~Ref.\,\cite{GHY}}).
\eeqa
\eeqs
Two comments are in order. First, we deduce from (\ref{eq:da-dx-expand-ds}) the following
patterns of the three mixing angles,
\beqs
\label{eq:t12-t23-t13-pattern}
\beqa
\label{eq:t12-t23-t13-IMO}
(\theta_{12},\,\theta_{23},\,\theta_{13})
&\,=&  \(\f{\pi}{4}-\ds,\, \f{\pi}{4}-O(\ds\dx),\, \dx\) ,
~~~~~~ (\textrm{in the current model}),
\\[1.5mm]
\label{eq:t12-t23-t13-NMO}
(\theta_{12},\,\theta_{23},\,\theta_{13})
&\,=&  \(\f{\pi}{4}-\ds,\, \f{\pi}{4}-\da,\, \dx\) ,
~~~~~~~~~~~~~~ (\textrm{in the model of Ref.\,\cite{GHY}}),
\eeqa
\eeqs
where for the current model Eq.\,(\ref{eq:t12-t23-t13-IMO}) predicts a nearly maximal
atmospheric angle  \,$\theta_{23}\simeq \f{\pi}{4}$\,;\,
while for the soft-breaking model\,\cite{GHY}, Eq.\,(\ref{eq:t12-t23-t13-NMO})
allows all three deviations to be comparable.
Second, for each given nonzero $\,\da =\theta_{23}-\f{\pi}{4}\,$,\, we can deduce
the lower limits on $\,\dx =\theta_{13}\,$ from (\ref{eq:da-dx-expand-ds}),
\beqs
\label{eq:t13-lowerB}
\beqa
\label{eq:t13-lowerB-IMO}
\dx &~\geqq~& \f{|\da |}{\,2\ds\,} ~\gg~ |\da |\,,
~~~~~~ (\textrm{Current Prediction}),
\\[2mm]
\label{eq:t13-lowerB-NMO}
\dx &~\geqq~& |\da | \,,
~~~~~~~~~~~~~~~~~~\, (\textrm{Prediction~of~Ref.\,\cite{GHY}}).
\eeqa
\eeqs
Given the 99\%\,C.L. range of $\,7.8^\deg < \ds <12.2^\deg\,$,\,
we derive the lower limit from (\ref{eq:Lbound-dx}) or (\ref{eq:t13-lowerB-IMO})
for the present model,
\beqa
\theta_{13} ~\geqq~ (3.6\sim 2.1)|\theta_{23}-45^\deg| \,,
\eeqa
which allows $\,\theta_{13}$ to easily saturate its current upper limit.
As another illustration, taking the current ``best fit" values
$\,(\theta_{12},\,\theta_{23})=(34.5^\deg,\,42.8^\deg)\,$
as in Table-1, we derive from (\ref{eq:Lbound-dx}) or (\ref{eq:t13-lowerB-IMO})
the lower limits $\,\theta_{13} \geqq 6^\deg\,$ for
the present model, and $\,\theta_{13} \geqq 1.5^\deg\,$ for Ref.\,\cite{GHY}.
Hence, in contrast with Ref.\,\cite{GHY}, the present model favors a
larger $\,\theta_{13}$, and can saturate its current upper limit,
as will be demonstrated in Fig.\,2 below.

\begin{figure}[t]
  \centering
  \hspace*{-3mm}
  \includegraphics[width=8.5cm,height=6.5cm,clip=true]{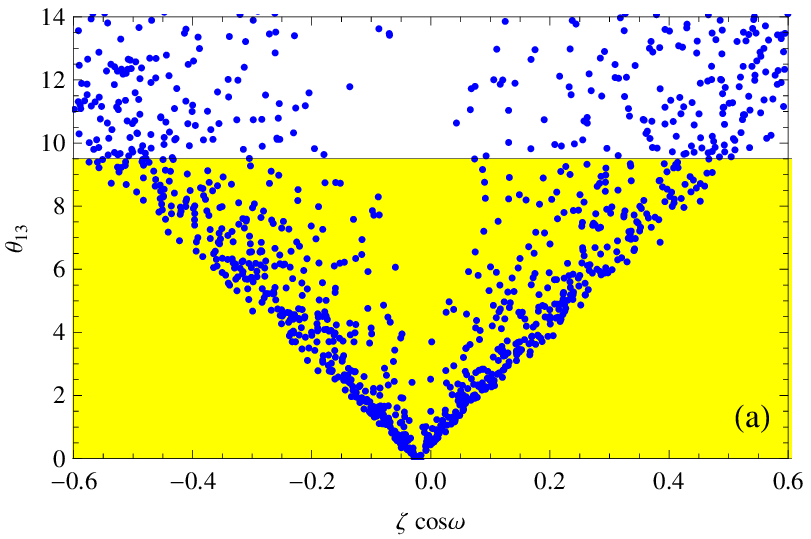}
  \hspace*{-4mm}
  \includegraphics[width=8.5cm,height=6.5cm,clip=true]{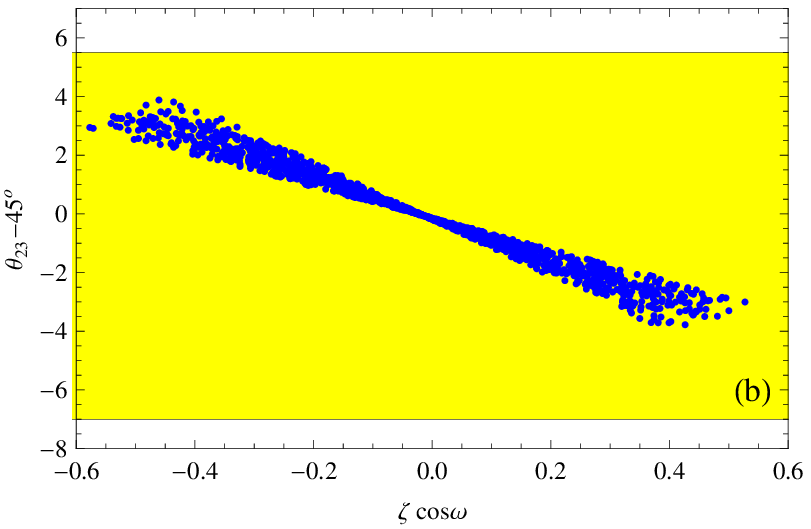}
   \hspace*{-8mm}
  \includegraphics[width=8.0cm,height=6.0cm,clip=true]{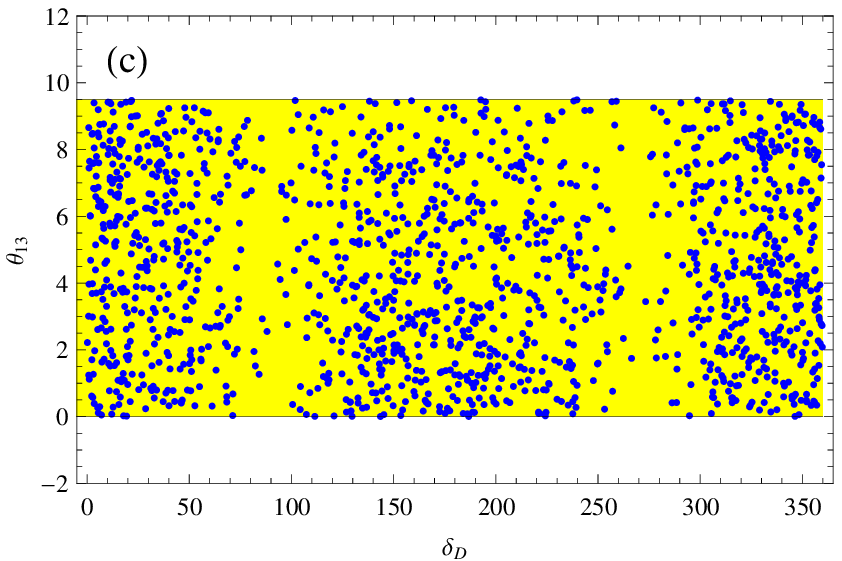}
  \includegraphics[width=8.0cm,height=5.8cm,clip=true]{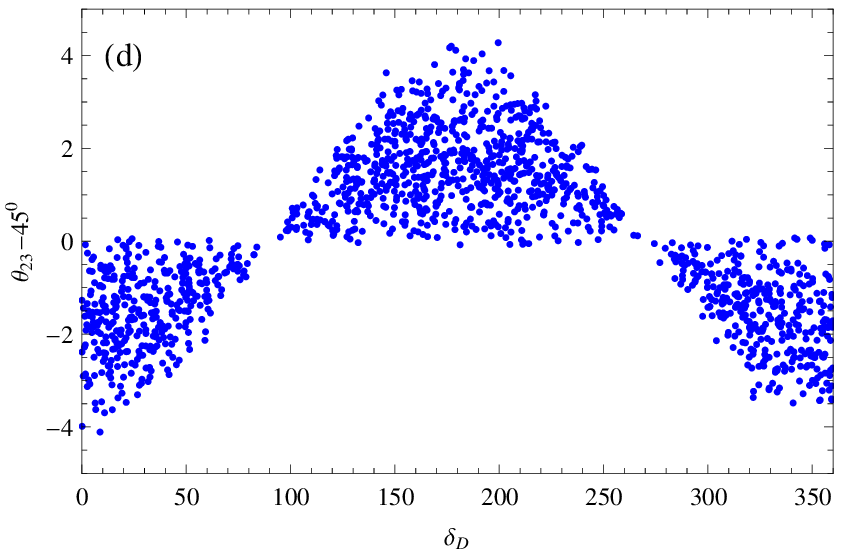}
  \caption{Predictions of $\,\theta_{13}\,$ and $\,\theta_{23}-45^\deg\,$
  as functions of the $\mutau$ breaking parameter $\,\zeta\cos\omega\,$
  and CP breaking parameter $\,\delta_D^{}$\,.\,
  The experimental inputs are scanned within 90\%\,C.L. ranges
  and the Dirac phase angle $\,\d_D^{}\in [0,\,2\pi)$\,,\, with 1500 samples.
  The shaded region (yellow) denotes the $90\%$\,C.L. limits on
  $\,\theta_{13}\,$ and $\,\theta_{23}-45^\deg$,\, from Table-1.}
  \label{fig:dx-ta-zeta}
\end{figure}
\begin{figure}[t]
\centering
\includegraphics[width=14cm,height=10.5cm]{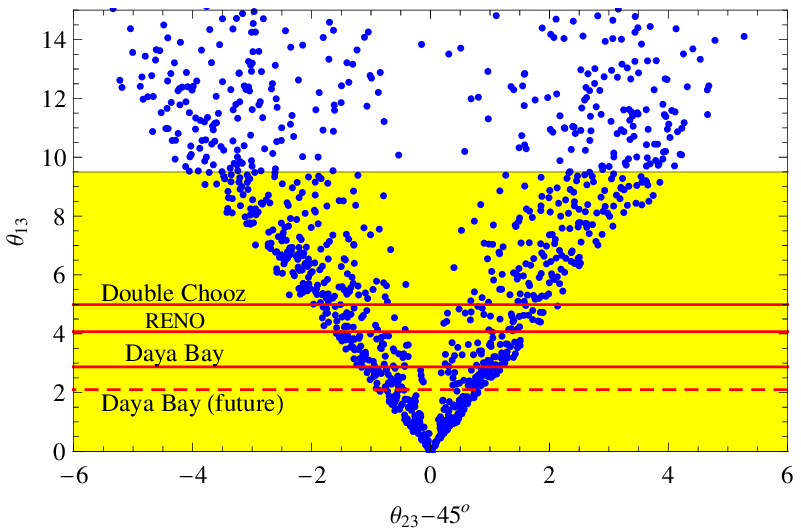}
  \caption{Correlation between $\,\theta_{13}\,$ and
  $\,\theta_{23}-45^\deg\,$,\, based on Eqs.\,(\ref{eq:sol-deltax})-(\ref{eq:sol-deltaa}),
  where the experimental inputs are scanned within 90\%\,C.L. ranges
  and the Dirac phase angle $\,\d_D^{}\in [0^\deg,\,360^\deg)$\,,\, with 1500 samples.
  The sensitivities of
  Double Chooz\,\cite{2CHOOZ}, RENO\,\cite{RENO} and Daya Bay\,\cite{DayaBay} experiments
  to $\,\theta_{13}$\, are shown by the three horizontal (red) solid lines at 90\%\,C.L.,
  as $5.0^\deg$, $4.1^\deg$ and $2.9^\deg$ (from top to bottom).
  The Daya Bay's future sensitivity ($2.15^\deg$) is shown by the horizontal
  dashed (red) line.
  }
  \label{fig:dx-da}
\end{figure}
In the following, we systematically analyze the predicted parameter space and
correlations in the present model (with inverted mass-ordering).
We will find these to be {\it very different} from that in our soft breaking
model (with normal mass-ordering)\,\cite{GHY}. So, the present model can be
tested against that in Ref.\,\cite{GHY} by the on-going and upcoming neutrino experiments.

Using the neutrino data for $\ts$ and $(\Delta m^2_{21},\,\Delta m^2_{13})$
(Table-1), and scanning the Dirac CP phase-angle
$\,\d_D^{} \in [0^\deg ,\, 360^\deg )$\,,\,
we can plot the two $\mutau$ breaking mixing angles,
$\,\theta_{13}\,(\equiv\dx)$\, and \,$\theta_{23} -45^\deg\,(\equiv\da)$,\,
from (\ref{eq:sol-deltax})-(\ref{eq:sol-deltaa}) and (\ref{eq:da-dx-1}),
as functions of the theory parameter $\,\zeta\cos\omega\,$ and $\,\delta_D\,$.\,
Our findings are depicted in Fig.\,\ref{fig:dx-ta-zeta}(a)-(d)
with the experimental inputs varied within 90\%\,C.L. ranges
and with $\,\zeta\cos\omega\in [-0.6,\,0.6]$\, in the natural perturbative region.
Here we find that the theory prediction of
\,$\theta_{23} -45^\deg$ lies in the range,
\beqa
\label{eq:t23-45-limit}
-4^\deg ~\leqq~ \theta_{23} -45^\deg ~\leqq~ 4^\deg \,,
\eeqa
which is within the current experimental bounds.
On the other hand the predicted $\,\theta_{13}\,$ can saturate the current
experimental limits, and has distinct distributions.

From the theory relations (\ref{eq:sol-deltax})-(\ref{eq:sol-deltaa}),
we can further explore the {\it correlation}
between the two $\mutau$ breaking mixing angles
$\,\theta_{13}\,$ and $\,\theta_{23} -45^\deg\,$.\,
This is displayed in Fig.\,\ref{fig:dx-da},
where we have varied the measured parameters within
their 90\%\,C.L.\ ranges, and input the Dirac-phase angle $\,\d_D^{}\in [0,\,2\pi)$\,
as well as $\,|\zeta'|\leqq 0.6\,$.\,
The current 90\%\,C.L.\ limits on $\,\theta_{13}\,$ are shown by the shaded region (yellow),
while the $\,\theta_{13}\,$ sensitivities of the on-going
Double Chooz\,\cite{2CHOOZ}, RENO\,\cite{RENO} and Daya Bay\,\cite{DayaBay} experiments
are depicted by the three horizontal (red) lines at 90\%\,C.L.,
as $5.0^\deg$, $4.1^\deg$ and $2.9^\deg$
(from top to bottom), based on three years of data-taking.
The horizontal dashed (red) line represents Daya Bay's future sensitivity ($2.15^\deg$)
with six years of running\,\cite{Lindner}.

Inspecting Fig.\,\ref{fig:dx-da}, we find
that the sharp edges on the two sides of the allowed parameter space
are essentially determined by the lower bound given in
(\ref{eq:Lbound-dx}), $\,\dx \geqq |\da|\tan2\ts\,$,\,
where the current data require,
$\,2.2 \leqq \tan 2\ts \leqq 3.1\,$ at 90\%\,C.L.\ (Table-1)
and the lower limit $\,\tan 2\ts=2.2\,$ just corresponds to the slopes
of the sharp edges which are nearly straight lines.
Hence, for {\it any measured nonzero value of
$\,\theta_{23}-45^\deg \neq 0\,$,\, the Fig.\,\ref{fig:dx-da}
imposes a lower bound on $\,\theta_{13}\,$,} which will be tested
by the reactor experiments such as Daya Bay, RENO and Double Chooz.
The current oscillation data favor the central value of $\theta_{23}$
to be smaller than $45^\deg$ (Table-1) and this feature
is quite robust\,\cite{D23-Smirnov}.
From Fig.\,\ref{fig:dx-da}, we see that taking the current central value
of $\,\theta_{23}-45^\deg = -2.2^\deg\,$ (Table-1), the lower bound on
$\,\theta_{13}\,$ is already very close to the sensitivity of Double Chooz
experiment; and a minor deviation of $\,\theta_{23}-45^\deg = -1.4^\deg\,$
will push $\,\theta_{13}\,$ up to the sensitivity of Daya Bay experiment.
Hence, {\it the Daya Bay, RENO and Double Chooz reactor experiments hold great
potential to discover a nonzero $\,\theta_{13}\,$.\,}
Furthermore, as shown in Fig.\,\ref{fig:dx-da},
detecting a nonzero $\,\theta_{13} \gtrsim 3^\deg\,$ will strongly favor
a nonzero $\,\theta_{23}-45^\deg\,$.\,
Hence, we further encourage the improved measurements of
$\,\theta_{23}\,$ by Minos\,\cite{MINOS} and T2K\,\cite{T2K},
as well as future neutrino factory and super-beam facility\,\cite{nuFact,nuFact2}.

Note that our previous soft breaking model\,\cite{GHY} predicted a
lower bound $\,\dx \geqq |\da|\tan\ts\,$ with the slope
$\,0.64 \leqq \tan \ts \leqq 0.73\,$ at 90\%\,C.L., which is about
$\,3.4-4.2\,$ times smaller than the present model.
This means that given the same nonzero deviation of
$\,\theta_{23}-45^\deg\,$, the current model will place a much stronger lower bound
on $\,\theta_{13}\,$,\, higher than that in Ref.\,\cite{GHY} by a factor of $3.4-4.2$.
Hence, the prediction of Fig.\,\ref{fig:dx-da} is really encouraging
for the upcoming neutrino oscillation experiments, which
will probe the $\mutau$ violating observables $\,\theta_{13}\!-\!0^\deg\,$
and $\,\theta_{23}\!-\!45^\deg\,$ to much higher precision.

\vspace*{2mm}
Then, we analyze our model predictions for the low energy CP-violation
(via Jarlskog invariant $J$) and the neutrinoless double-beta decays (via
the element $|m_{ee}|$ of $M_\nu$).
From our theory construction in Sec.\,2.\,2, the original CP-phase $\,e^{i\om}\,$
in the Dirac mass-matrix of seesaw Lagrangian is the common source of both low energy Dirac
and Majorana CP-violations via the phase angles $\,\d_D^{}\,$ and $\,\delta\phi$\,.

The Dirac CP-violation is characterized by the Jarlskog invariant $J$ \cite{J} in the
light neutrino sector with nonzero CP-phase $\,\d_D^{}\,$ and can be measured by the
long baseline neutrino oscillation experiments. On the other hand, the neutrinoless
double-beta decay observable $\,|m_{ee}^{}|$\, contains
both $\d_D^{}$ and Majorana CP-phase $\,\delta\phi$\,.\,
We can express the Jarlskog invariant $\,J\,$ as follows\,\cite{J},
\beqa
\label{eq:J}
J  ~\equiv~
\f{1}{8}\sin2\theta_s \sin2\theta_a \sin2\theta_x \cos\theta_x \sin \delta_D
~=~  \frac{\dx}{4}\sin2\theta_s\sin\delta_D +\O(\dx^2,\da^2)\,,~~~~
\eeqa
where as defined earlier, $\,\dx\equiv\theta_x\,$ and $\,\da\equiv\ta -\f{\pi}{4}\,$.
The solutions (\ref{eq:sol-deltax})-(\ref{eq:sol-deltaa}) leads to the correlation
(\ref{eq:da-dx-1}). We can input the neutrino data for mixing angles,
$\,(\theta_s,\,\d_x)\equiv (\theta_{12},\,\theta_{13})\,$,\, and mass-ratio,
$\,y'\equiv\Delta m_{21}^2/\Delta m_{13}^2\,$,\, as well as scanning the model-parameter
$\,\zeta'\,$ in its perturbative range $\,|\zeta'| \leqq 0.6\,$.\,

We then study the neutrinoless double-beta decays. Our present model predicts the
inverted mass-ordering (IMO) with $\,m_3^{}=0\,$,\,
so from (\ref{eq:Mnu-Reconstruct-ee})
we can derive the mass-matrix element $|m_{ee}^{}|$
for neutrinoless double-beta decays,
\begin{eqnarray}
  M_{ee} ~&\equiv&~
  \left| m_{ee}^{} \right| ~=\,
  \left| \sum {V_{ej}^*}^2 m_j^{} \right|
  \,=~
  m_1^{}c_x^2\left|
   c^2_s   + s^2_s\sqrt{1+y'\,}\, e^{-i2\phi} \right|
\nn\\[3mm]
~& \simeq &~
  m_1^{}\[1+\f{1}{2}s_s^2y'-\delta_x^2-2s_s^2c_s^2\delta\phi^2\] ,
  \label{eq:Mee}
\end{eqnarray}
where in the last step we have expanded $\,\d_x\,$ and $\,\d\phi\,$ to the
second order since $\,y'=O(10^{-2})\,$ is relatively small as constrained by
the current data [cf.\ (\ref{eq:definition-y})].
Eq.\,(\ref{eq:Mee}) shows that the neutrinoless $\beta\beta$-decay observable
$\,M_{ee}\,$ only contains the second orders of the $\mutau$ breaking quantity
$\,\d_x\,(=\theta_{13})\,$ and the Majorana CP-phase angle $\,\d\phi\,$.\,
Hence, $\,M_{ee}\,$ is less sensitive to the $\mutau$ breaking and Majorana
CP-violation at low energies.

\begin{figure}[t]
\centering
\hspace*{-4mm}
\includegraphics[width=8.4cm,height=6.5cm,clip=true]{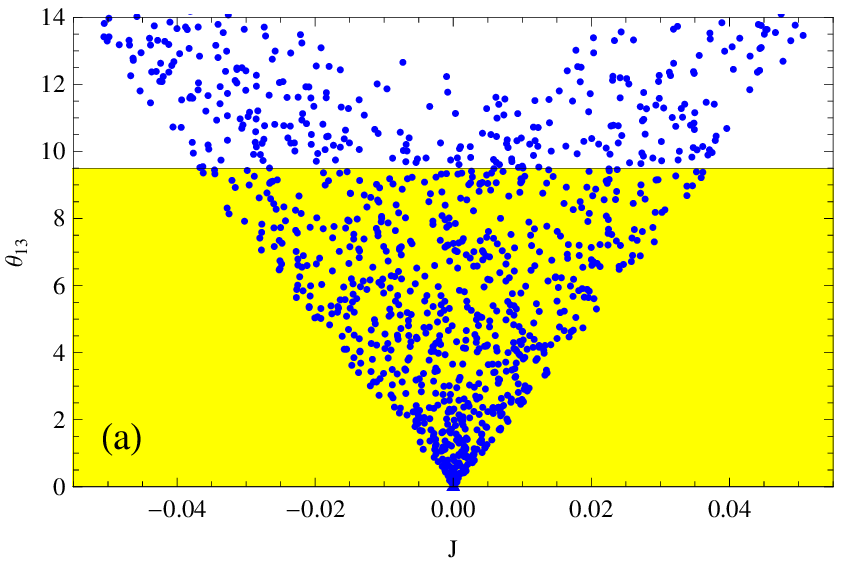}
\includegraphics[width=8.4cm,height=6.5cm,clip=true]{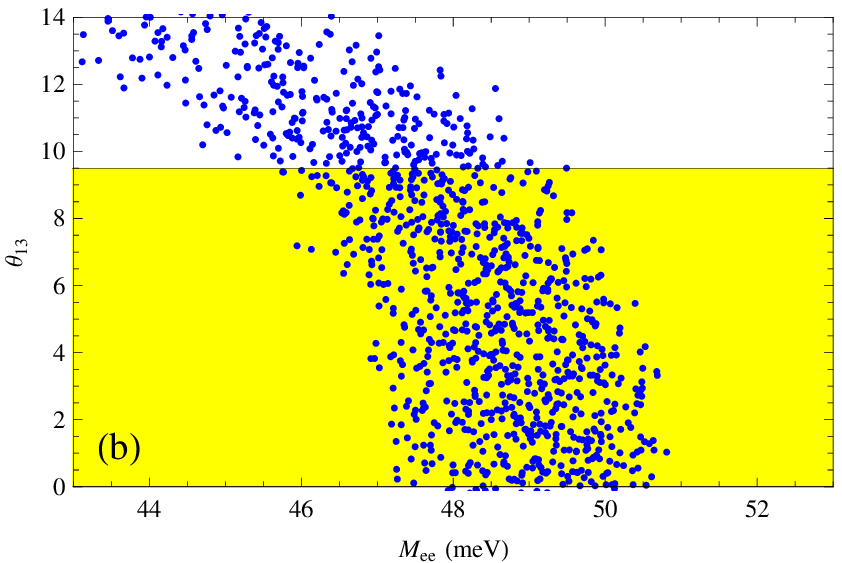}
\caption{Correlations of  $\theta_{13}$ (in degree) with the Jarlskog invariant
           $J$ [plot-(a)] and with the neutrinoless $\beta\beta$-decay observable
           $M_{ee}$ [plot-(b)].
           Each plot has computed 1500 samples.
           The shaded region (yellow) is allowed by the current data at 90\%\,C.L.
          }
  \label{fig:dx-J-Mee}
\end{figure}

We plot the correlation between $\,\theta_{13}\,$ and the Jarlskog invariant $\,J\,$
in Fig.\,\ref{fig:dx-J-Mee}(a), and the neutrinoless $\beta\beta$-decay observable
$M_{ee}$ is depicted in Fig.\,\ref{fig:dx-J-Mee}(b).
For the analysis of Fig.\,\ref{fig:dx-J-Mee}(a), we have used Eq.\,(\ref{eq:sol-deltax})
where we vary the model-parameter $\,\zeta'\in [-0.6,\,0.6]$\, in its perturbative range.
We scan all other measured parameters within their 90\%C.L. ranges.
The shaded region (yellow) in Fig.\,\ref{fig:dx-J-Mee} is allowed by the neutrino data
at 90\%\,C.L.\,  Fig.\,\ref{fig:dx-J-Mee}(a) shows that any nonzero $J$ will lead to
a lower bound on $\theta_{13}$ due to
$~\dx \geqq {4|J|}/\sin 2\ts~$ as inferred from Eq.\,(\ref{eq:J}).
Combining the current upper limit $\,\theta_{13}< 9.5^\deg\,$ (shaded region in yellow)
with our parameter space in Fig.\,\ref{fig:dx-J-Mee}(b),
we predict the allowed range,
\beqs
\label{eq:bound-J-Mee}
\beqa
\label{eq:bound-J}
&- 0.037 ~\lesssim~ J ~\lesssim~ 0.037 \,, &
\\
\label{eq:bound-Mee}
& 45.5\,\textrm{meV} ~\lesssim~ M_{ee} ~\lesssim~ 50.8\,\textrm{meV}\,, &
\eeqa
\eeqs
which can be probed by the on-going neutrinoless double beta decay
experiments\,\cite{0nu2beta}.

\vspace*{3mm}

Before concluding this subsection, we compare our prediction
(\ref{eq:da-dx-1}) with a recent independent work\,\cite{FL2010}.
In Ref.\,\cite{FL2010}, using a charged lepton perturbation,
Friedberg and Lee derived a very interesting prediction,
$\,\cos 2\theta_{23}=\tan^2\theta_{13}\,$, leading to
\beqa
\f{\pi}{4}-\theta_{23} ~\simeq~ \f{1}{2}\theta_{13}^2 ~\ll~ \theta_{13}\,,
\label{eq:FL}
\eeqa
which does not contain CP phase and predicts a nearly maximal $\theta_{23}$\,.\,
For comparison, we rewrite our predictions
(\ref{eq:IMO-da-dx-expand-ds})-(\ref{eq:NMO-da-dx-expand-ds}) in the same
notations,
\beqs
\label{eq:da-dx-expand-ds2}
\beqa
\label{eq:IMO-da-dx-expand-ds2}
\f{\pi}{4}-\theta_{23}  \,&\simeq&\,  2\cos\delta_D
\(\f{\pi}{4}-\theta_{12}\)\theta_{13} \,\ll\, \theta_{13} \,,
~~~~~~~ (\textrm{Current Prediction}),
\\[1.5mm]
\label{eq:NMO-da-dx-expand-ds2}
\f{\pi}{4}-\theta_{23} \,&\simeq&\, \cos\delta_D \,\theta_{13}
\,=\, O(\theta_{13}) \,,
~~~~~~~~~~ \hspace*{11mm}
(\textrm{Prediction~of~Ref.\,\cite{GHY}}),
\eeqa
\eeqs
where our correlations explicitly contain the CP-phase angle $\,\d_D^{}\,$.\,
Moreover, our present model predicts a deviation
$\,\f{\pi}{4}-\theta_{23}$\, to be significantly smaller than $\,\theta_{13}\,$
as in (\ref{eq:IMO-da-dx-expand-ds2}),
due to the suppression of $\,\f{\pi}{4}-\theta_{12}=0.16-0.21$ at 90\%\,C.L.
But, taking $\,\cos\d_D^{}=O(1)\,$, we see that the right-hand-side of
(\ref{eq:IMO-da-dx-expand-ds2}) is larger than that of
(\ref{eq:FL}) by a factor of
$\,4(\f{\pi}{4}-\theta_{12})/\theta_{13}=(36.0 -48.8^\deg)/\theta_{13}\,$
at 90\%\,C.L.,  which is clearly bigger than one.
On the other hand, our previous soft breaking model\,\cite{GHY} predicts the two
small $\mutau$ breaking observabes to be of the same order,
$\,\f{\pi}{4}-\theta_{23} =O(\theta_{13})$\,,\, as in (\ref{eq:NMO-da-dx-expand-ds2}).
Hence, the predictions by Friedberg-Lee\,\cite{FL2010} and by us differ in a nontrivial
and interesting way, which strongly motivate the on-going and future neutrino experiments
for tests and resolution.

\vspace*{3mm}
\subsection{Baryon Asymmetry from  $\bd{\mutau}$ Blind Seesaw and Direct Link to Low Energy}
\label{sec:solution}

 In this subsection, we study the predictions of our $\mutau$ blind seesaw model
 for cosmological baryon asymmetry (matter-antimatter asymmetry) via
 thermal leptogenesis\,\cite{lepG,lepGrev}.  We build up the direct link between
 leptogenesis CP-asymmetry and the low energy Dirac CP-phase, and further
 predict the low energy leptonic Jarlskog invariant $\,J\,$ \cite{J}.
 Imposing the WAMP data on the baryon asymmetry\,\cite{WMAP08},
 we predict a negative Jarlskog invariant, $\,J<0\,$,\,
 and derive a lower bound on the reactor mixing angle,
 $\,\theta_{13} \gtrsim 1^\deg\,$.\,
 We also analyze the {\it correlations} of the leptogenesis
 scale with the low energy observables such as the Jarlskog-invariant
 $J$ and neutrinoless $\beta\beta$-decay parameter $M_{ee}$ \cite{0nu2beta}.
 We further deduce a lower bound on the leptogenesis scale for producing the observed
 baryon asymmetry.

 Our universe is exclusively dominated by matter rather than antimatter.
 The asymmetry of baryon-anti-baryon density
 $\,n_B^{} - \ov{n}_B^{}(\simeq n_B^{}\,)$\,  relative to
 the photon density $n_\gamma^{}$ is measured to be a tiny nonzero ratio\,\cite{WMAP08},
 \begin{equation}
 \label{eq:etaB-exp}
  \eta_B^{}
  ~\equiv~
  \f{\,n_B^{} - n_{\ov B}^{}\,}{n_\gamma^{}}
  ~=~
  (6.19\pm 0.15) \times 10^{-10} \,.
 \end{equation}
The SM fails to generate the observed baryon asymmetry
because of the too small CP-violations from CKM matrix and the lack of
sufficiently strong first-order electroweak phase transition\,\cite{EW-BG},
which violate Sakharov's condition for baryongenesis\,\cite{Sakharov}.
It is important that the seesaw extension of the SM allows
the thermal leptogenesis\,\cite{lepG} with CP-violations originating from the
neutrino sector and the lepton-number asymmetry produced during
out-of-equilibrium decays of
heavy Majorana neutrino $N_j$ into the lepton-Higgs pair $\ell H$
and its CP-conjugate $\bar{\ell}H^*$.\,
Then, the lepton asymmetry can be partially converted to a baryon asymmetry
via the nonperturbative electroweak sphaleron\,\cite{sphaleron}
interactions which violate $B+L$ \cite{tHooft} but preserve $B-L$ \cite{HT,kappa-BDP},
$\,
  \eta_B^{} \,=\,
  \f{\xi}{\,f\,} N_{B-L}^f \,=\, -\f{\xi}{\,f\,} N_{L}^f  \,,
$\,
where $\xi$ is the fraction of $B\!-\!L$ asymmetry
converted to baryon asymmetry via sphaleron process\,\cite{HT}
and $\,\xi=28/79\,$ for the SM.
The dilution factor $\,f =N_\gamma^{\rm rec}/N_\gamma^* ={2387}/{86}\,$ is
computed by considering standard photon production from the onset of leptogenesis till
recombination\,\cite{kappa-BDP}.
The effect of the heavier right-handed neutrino ($N_2$) decays
will be washed out in the thermal equilibrium, only the lightest one
($N_1$) can effectively generate the net lepton asymmetry
 for $\,M_1\ll M_2\,$.\, (In the numerical analysis below, we will consider
 the parameter space with $\,M_2/M_1\geqq 5\,$,\, to ensure the full washout
 of lepton asymmetry from $N_2$-decays.)
 Thus, the net lepton asymmetry $N_L^f$ is deduced as\,\cite{kappa-BDP},
%
$\,  N_L^f \,=\, \f{3}{4}\kappa_f^{}\epsilon_1^{} \,.$\,
%
Hence, we can derive the final baryon asymmetry,
\beqa
\label{eq:etaB-f}
\eta_B^{} ~=\, -\f{3\,\xi}{\,4f\,} \kappa_f^{}\ep_1^{}
~=\, - d \,\kappa_f^{}\ep_1^{} \,,
\eeqa
where
$\,d \equiv 3\xi/(4f) \simeq 0.96\!\times\! 10^{-2}\,$,\, and the factor
$\,\kappa_f^{}\,$ measures the efficiency of out-of-equilibrium $N_1$-decays.
The $\,\kappa_f^{}\,$ is determined by solving the
Boltzmann equation numerically\,\cite{kappa-BDP,kappa-Strumia}.
In practice, useful analytical formulas for $\,\kappa_f^{}\,$ can be
inferred by fitting the numerical solution of the Boltzmann equation.
We find it convenient to use the following fitting formula of $\,\kappa_f^{}$
\cite{kappa-Strumia},
\beqa
  \kappa_f^{-1} ~\simeq~
  \(\f{\ov{m}_1^{}}{\,0.55\!\times\! 10^{-3}\,\textrm{eV}\,}\)^{1.16}
  +\f{\,3.3\!\times\! 10^{-3}\,\textrm{eV}\,}{\ov{m}_1^{}}  \,,
  \label{eq:kappa1}
\eeqa
with
$~\ov{m}_1^{}  \equiv
  {(\under{m}^\dagger_D \under{m}_D^{})_{11}^{}}/{M_1} \,$\,,
%
and $\,\under{m}_D^{}\equiv \mD U_R\,$ with $U_R$ being the rotation matrix diagonalizing
the mass-matrix $M_R$ of right-handed neutrinos.
In the present $\mutau$ blind seesaw, it is natural to set the right-handed neutrinos
in their mass-eigenbasis from the start, $\,M_R=\textrm{diag}(M_1,\,M_2)\,$,\, as we defined
in Sec.\,2.\,1. So we have $\,U_R=\II$\, with $\II$ the unit matrix, and thus
$\,\under{m}_D^{}=\mD\,$.\,
(Other fitting formulas than (\ref{eq:kappa1})
to the exact solution of $\,\kappa_f^{}\,$
in the literature\,\cite{kappa-BDP} agree with each other quite well
for the relevant range of $\ov{m}_1^{}$.)
The CP asymmetry parameter $\,\epsilon_1^{}\,$ is defined as
%
\begin{eqnarray}
\label{eq:ep1-all}
  \epsilon_1^{}
~\equiv~
  \f {~\Gamma[N_1\to\ell H]-\Gamma[N_1\to\ov{\ell}H^*]~}
     {~\Gamma[N_1\to\ell H]+\Gamma[N_1\to\ov{\ell}H^*]~}
~=~
  \f{1}{\,4\pi v^2\,} F\!\(\f{M_2}{M_1}\)
  \f{\Im\mathfrak{m}
  \left\{ [ (\under{m}^\dagger_D \under{m}_D^{})_{12}^{}]^2
  \right\}}
  {( \under{m}^\dagger_D \under{m}_D^{})_{11}^{}} \,,
\label{eq:ep1}
\end{eqnarray}
%
where $v$ denotes the vacuum expectation value of the SM Higgs boson.
As we constructed in Sec.\,2.\,2, the Dirac mass-matrix $\mD$ is complex and
provides the common origin of the $\mutau$ and CP breaking; the complexity of
$\,\mD\,$ causes the difference between the decay widths $\,\Gamma[N_1\to\ell H]\,$
and $\,\Gamma[N_1\to\ov{\ell} H^*]\,$,\, and thus a nonzero CP asymmetry
$\,\epsilon_1^{}\neq 0\,$.\,
For the SM, the function $F(x)$ in (\ref{eq:ep1}) takes the form,
%
\beqa
\label{eq:F}
  F(x) ~\equiv~
  x \left[
    1 - (1 + x^2) \ln \frac {1 + x^2}{x^2}
  + \frac 1 {1 - x^2} \right]
\,=~ - \f{3}{\,2x\,} + \O\!\( \f{1}{x^3} \),
\quad (\textrm{for}~\,  x \gg 1\,)\,.
\eeqa
%
For our numerical analysis of the thermal leptogenesis,
the mass ratio $\,M_2/M_1\gg 1\,$ and thus the above expanded formula of $F(x)$
holds with good accuracy.

\begin{figure}[t]
\centering
\includegraphics[width=10cm,height=7cm]{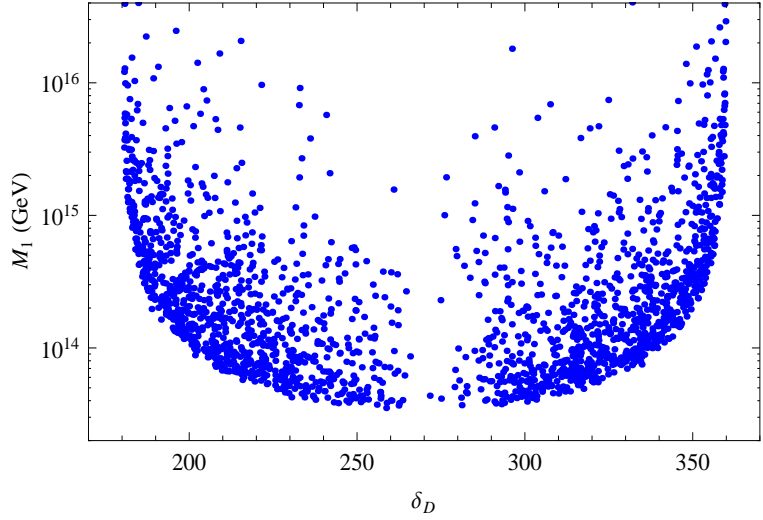}
\vspace*{-2mm}
\caption{Leptogenesis scale $M_1$ is plotted as a function of Dirac CP-phase angle
         $\,\delta_D^{}\,$,\, where the seven years of WMAP measurement (\ref{eq:etaB-exp})
         is imposed.  All experimental inputs are scanned within
         their $90\%$\,C.L.\ ranges, with 1500 samples.}
\label{fig-etaB/M1-omega}  
\end{figure}

Then, we proceed to compute the matrix elements,
\begin{subequations}
  \begin{eqnarray}
    ( \under{m}^\dag_D \under{m}_D^{} )_{11}^{}
  &~=~& \widehat m_0 M_1\(a^2+2b^2\) ~=~ \widehat m_0 M_1 \,,
  \label{eq:md11}
  \\[2mm]
    ( \under{m}^\dag_D \under{m}_D^{} )_{12}^{}
 &~=~&
  -\widehat m_0\sqrt{M_1M_2}\,bc\(\zeta' +\zeta e^{i\omega}\) .
  \end{eqnarray}
\end{subequations}
So we can deduce the effective mass-parameter $\ov{m}_1^{}$
as introduced below (\ref{eq:kappa1}),
\beqa
\label{eq:m1bar}
  \ov{m}_1^{} ~=~ \mh_{0}^{}
  ~\simeq~ \chi_1^{}\sqrt{\Delta m_{13}^2} ~,
\eeqa
and the imaginary part,
%
\begin{eqnarray}
  \Im\mathfrak{m}\!
  \left\{ [ ( \under{m}^\dag_D \under{m}_D^{} )_{12}^{} ]^2 \right\}
  &~=~& -\f{1}{2}\,\mh_{0}^{2}\,M_1\,M_2 ~y'\sin2\theta_s \sin\delta_D\,\delta_x \,,
\end{eqnarray}
%
where the RG running factor $\,\chi_1^{}=\chi(M_1,m_Z^{})\,$
is defined in Eqs.\,(\ref{eq:RG-Run-mj})-(\ref{eq:RG-Run-m123}).
Using Eq.\,(\ref{eq:m1bar}) together with the neutrino data (Table-1),
we find that the light neutrino mass-parameter
$\,\ov{m}_1^{}\,$ lies in the $3\sigma$ range,
$\,0.046 < \ov{m}_1^{}/\chi_1^{} <0.053\,$eV, where the RG factor
$\,\chi_1^{}\simeq 1.3-1.4\,$ is evaluated numerically, as explained around
the end of Sec.\,3.\,2.
So, in Eq.\,(\ref{eq:kappa1}) the second term on the right-hand-side is negligible and
$\,\kappa_f^{}\,$ is thus dominated by the first term.

\begin{figure}[t]
   \centering
  \vspace*{-5mm}
  \includegraphics[width=8.3cm,clip=true]{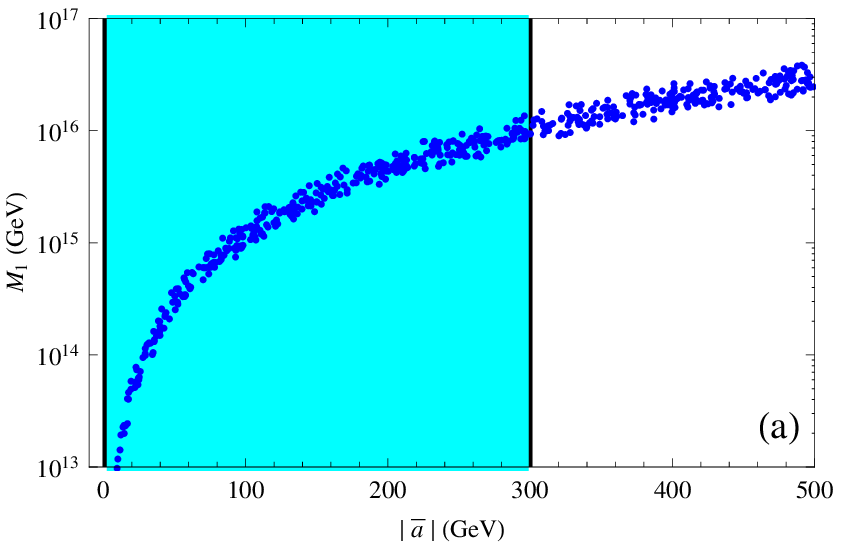}
  \includegraphics[width=8.3cm,clip=true]{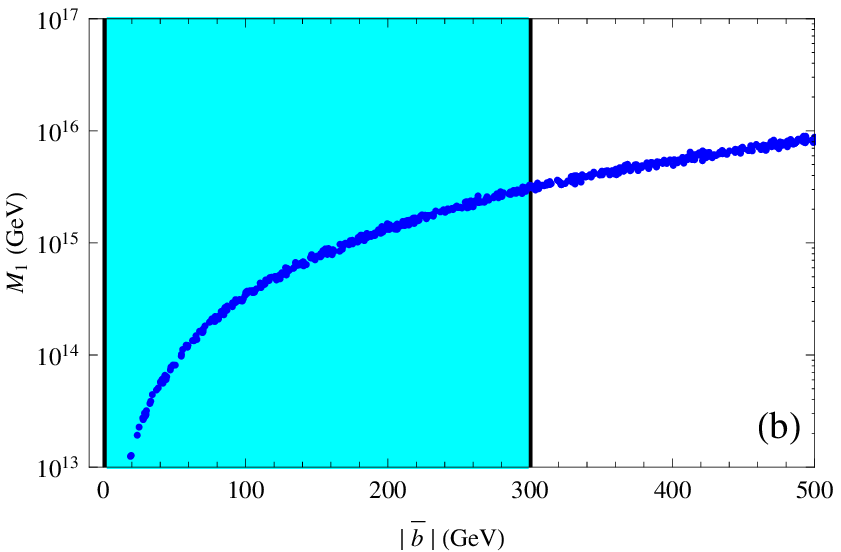}
  \includegraphics[width=8.3cm,clip=true]{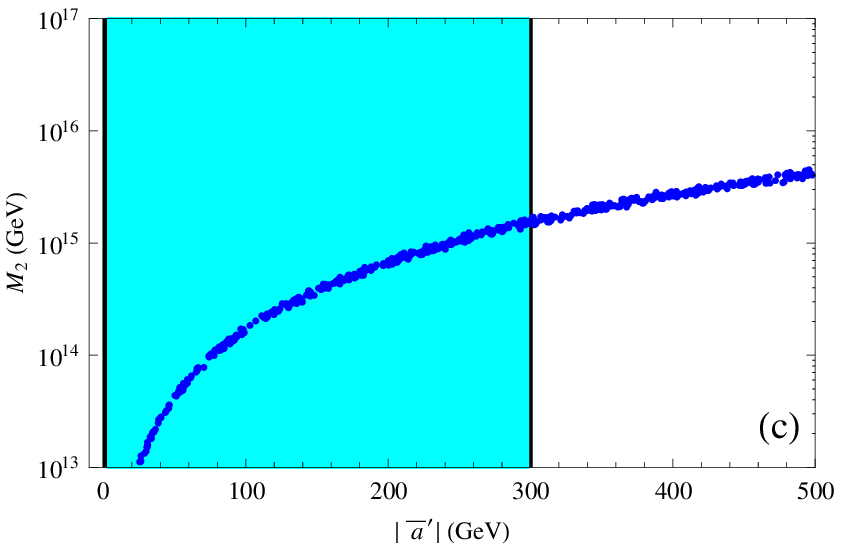}
  \includegraphics[width=8.3cm,clip=true]{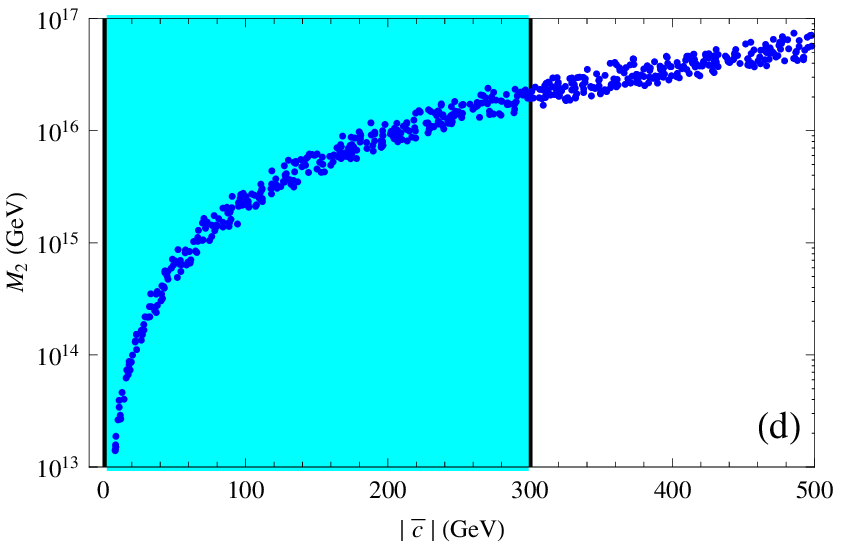}
  \caption{Seesaw scale $M_1$ and $M_2$ as functions of the elements
  $\,(\bar{a},\,\bar{b})$\, and $\,(\bar{a}',\,\bar{c})$\,
  in the Dirac mass-matrix $\,\mD\,$,\,  where the shaded regions correspond to
  the natural perturbative region
  $\,(\bar{a},\,\bar{b},\, \bar{a}',\, \bar{c})\,\in [1,\,300]$\,GeV,
  and 600 samples are generated in each plot.
  This puts an upper bound, $\,M_1\leqq 3.5\times 10^{15}\,$GeV from plot-(b),
  and $\,M_2\leqq 1.7\times 10^{15}\,$GeV from plot-(c).
  }
 \label{fig:M1-|ab|}
\end{figure}

With these and from (\ref{eq:ep1}), we derive the CP asymmetry parameter
$\,\epsilon_1^{}\,$ as follows,
%
\beqa
  \epsilon_1^{}
   &~\simeq~&\f{\,3y'\mh_{0}^{}M_1\,}{16\pi v^2}
   \sin2\theta_s\,\sin\delta_D^{}~\delta_x \,.
\label{eq:ep1-a}
\eeqa
%
Finally, inspecting Eqs.\,(\ref{eq:etaB-f}), (\ref{eq:ep1})
and (\ref{eq:F}), we can derive,
\begin{eqnarray}
  \f{\eta_B^{}}{M_1}  &~=~& - d \,\kappa_f^{}
  \f{3y'\mh_{0}^{}}{\,16\pi v^2\,}\,\sin2\theta_s\sin\delta_D^{}\,\delta_x \,.
\label{eq:etaB-M1}
\end{eqnarray}
Since the WMAP measurement (\ref{eq:etaB-exp}) finds
the baryon asymmetry $\,\eta_B^{}>0\,$,\, so we can infer the constraint,
$\,\sin\delta_D^{} <0\,$,\, which restricts the Dirac phase angle,
$\,\delta_D^{} \in (\pi,\,2\pi)\,$.

Then, from Eq.\,(\ref{eq:etaB-M1}) we compute the ratio $\,\eta_B^{}/M_1\,$
for any nonzero $\,\sin\delta_D^{}$\,,\, where we vary
all measured quantities within their $90\%$\,C.L.\ ranges.
Since $\,0<|\sin\delta_D^{}|\leqq 1$\,,\,
we can deduce a robust numerical upper bound,
\beqa
  \f{\eta_B^{}}{M_1} ~<~
  1.8 \times 10^{-23} \,\textrm{GeV}^{-1} \,.
\eeqa
Inspecting (\ref{eq:etaB-M1}) we can also reexpress the leptogenesis scale $\,M_1\,$
in terms of baryon asymmetry $\,\eta_B^{}\,$ and other physical observables,
\beqa
  M_1 ~=\,
  \frac{-16\pi v^2\,\eta_B}
        {3d \,\kappa_f^{}\mh_{0}^{}\,y'\sin2\theta_s\,\sin\delta_D\,\delta_x} \,.
  \label{eq:M1-etaB}
\eeqa
With the data of $\eta_B^{}$ from (\ref{eq:etaB-exp}),
we can plot, in Fig.\,\ref{fig-etaB/M1-omega},
the leptogenesis scale \,$M_1$\, as a function
of Dirac CP-phase $\,\delta_D$\,,\,  where
all experimentally measured quantities are scanned within their
$90\%$\,C.L.\ range (with 1500 samples).
Fig.\,\ref{fig-etaB/M1-omega} reveals a robust lower bound on $\,M_1\,$,
\beqa
\label{eq:M1-LowerBound}
  M_1 ~>~ 3.5 \times 10^{13} \,\textrm{GeV}\,.
\eeqa

Using Eqs.\,(\ref{eq:abc-m0}) and (\ref{eq:expression-abc-new}),
we connect the seesaw scale $(M_1,\,M_2)$ to the elements of
the Dirac mass-matrix \,$\mD$\,,
\begin{subequations}
\beqa
&&
M_1~=~\frac{\bar{a}^2}{\,\widehat m_0 \cos^22\theta_s\,}
   ~=~\frac{2\,\bar{b}^2}{\,\widehat m_0 \sin^22\theta_s\,} \,,
\\[3mm]
&&
M_2~=~\frac{\bar{a}^{\prime 2}}{\,\mh_0 \sin^22\theta_s\,}
   ~=~\frac{2\,\bar{c}^2}{\,\widehat m_0 \cos^22\theta_s\,} \,,
\eeqa
\end{subequations}
where the Dirac mass-parameters $\,(\bar{a},\,\bar{b},\, \bar{a}',\, \bar{c})$\,
arise from the Yukawa interactions,
$\,(\bar{a},\,\bar{b},\, \bar{a}',\, \bar{c})
 = (y_a^{},\,y_b^{},\,y_{a'}^{},\,y_c^{})v/\sqrt{2}\,$.\,
So we can plot $\,M_1\,$ as a function of the magnitude of the Dirac
mass-parameter $\,|\bar{a}|\,$ or $\,|\bar{b}|\,$
in Fig.\,\ref{fig:M1-|ab|}(a)-(b),
and $\,M_2\,$ as a function of the magnitude of the Dirac
mass-parameter $\,|\bar{a}'|\,$ or $\,|\bar{c}|\,$ in Fig.\,\ref{fig:M1-|ab|}(c)-(d),
where we have varied the measured quantities in their 90\%\,C.L. ranges.
We note that the Yukawa couplings $\,(y_a^{},\,y_b^{},\,y_{a'}^{},\,y_c^{})\,$
cannot be too small (to avoid excessive fine-tuning) or too large (to keep valid perturbation).
So, we will take the Dirac mass-parameters
$\,(\bar{a},\,\bar{b},\, \bar{a}',\, \bar{c})\,$ in the natural range
\,$[1,\,300]$\,GeV, corresponding to the Yukawa couplings $\,y_j^{}\,$ no
smaller than $\,O(10^{-2})\,$ and no larger than $\,O(y_t^{})$,  where
$\,y_t^{}=\sqrt{2}m_t^{}/v\simeq 1\,$ is the top-quark Yukawa coupling in the SM.\,
This natural perturbative range of
$\,(\bar{a},\,\bar{b},\, \bar{a}',\, \bar{c})\,$  is indicated by
the shaded area in Fig.\,\ref{fig:M1-|ab|}(a)-(d), which results in an
upper limit on the seesaw scale \,$(M_1,\,M_2)$\, due to the perturbativity requirement.
From Fig.\,\ref{fig:M1-|ab|}(b) we infer an upper bound $\,M_1\leqq 3.5\times 10^{15}\,$GeV,
while Fig.\,\ref{fig:M1-|ab|}(c) requires  $\,M_2\leqq 1.67\times 10^{15}\,$GeV.
For the above construction of natural thermal Leptogenesis we consider the parameters space
$\,M_2/M_1\geqq 5\,$, so with the upper bound of Fig.\,\ref{fig:M1-|ab|}(c) we further
deduce a stronger limit $\,M_1 \leqq 3.3\times 10^{14}\,$GeV.

\begin{figure}[t]
\vspace*{-3mm}
\centering
\includegraphics[width=14cm,height=10.5cm]{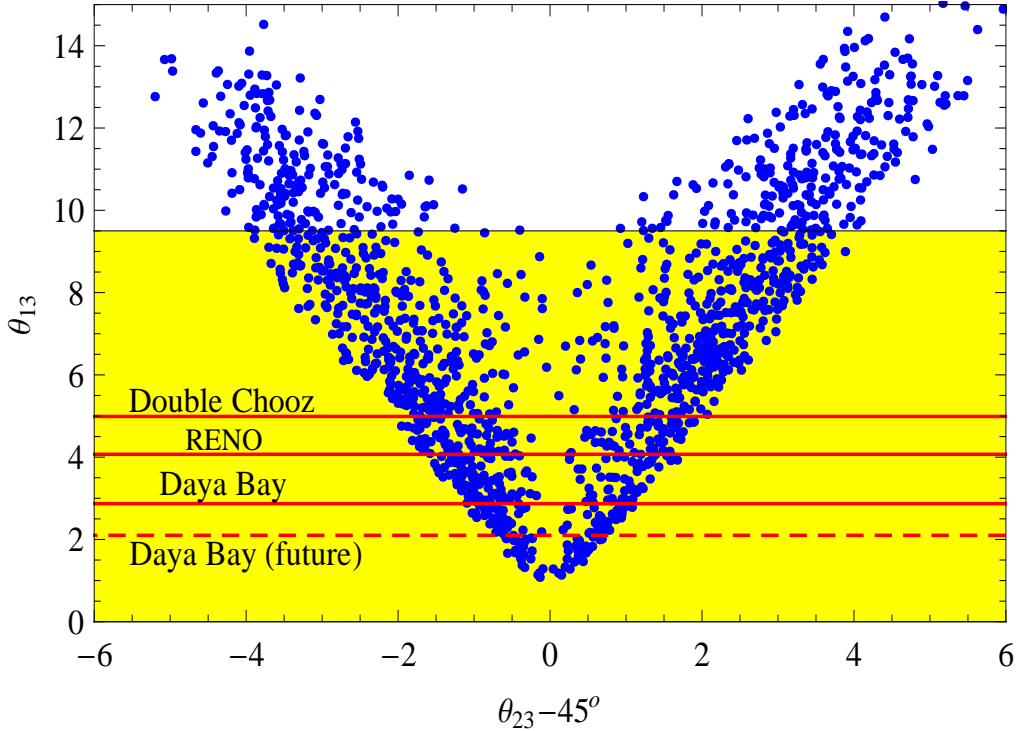}
\caption{Correlation between $\,\theta_{13}\,$ and
  $\,\theta_{23}-45^\deg$,\, where all the inputs are the same as
  Fig.\,\ref{fig:dx-da},
  except requiring successful leptogenesis in the
  present analysis, with 1500 samples.
  }
  \label{fig-deltax-deltaa-new}  
\end{figure}

\begin{figure}[t]
\vspace*{-2mm}
 \centering
  \includegraphics[width=11cm,clip=true]{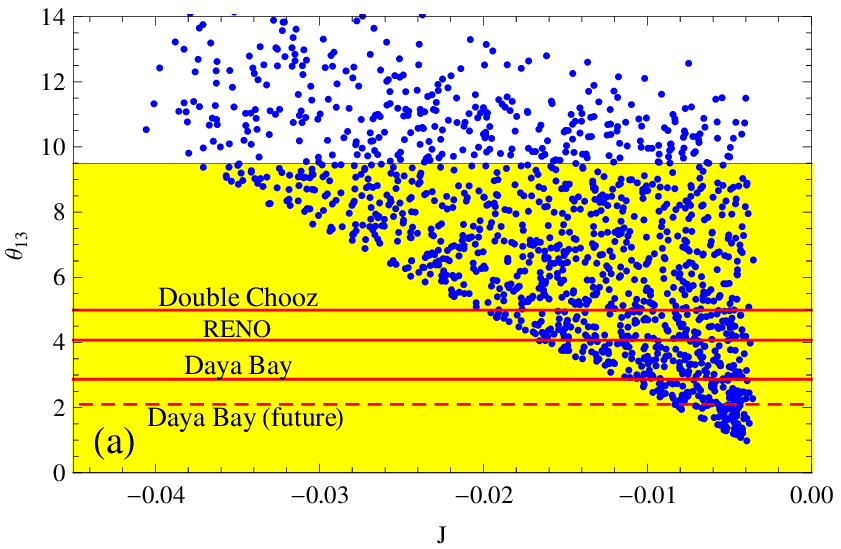}\\
  \includegraphics[width=11cm,clip=true]{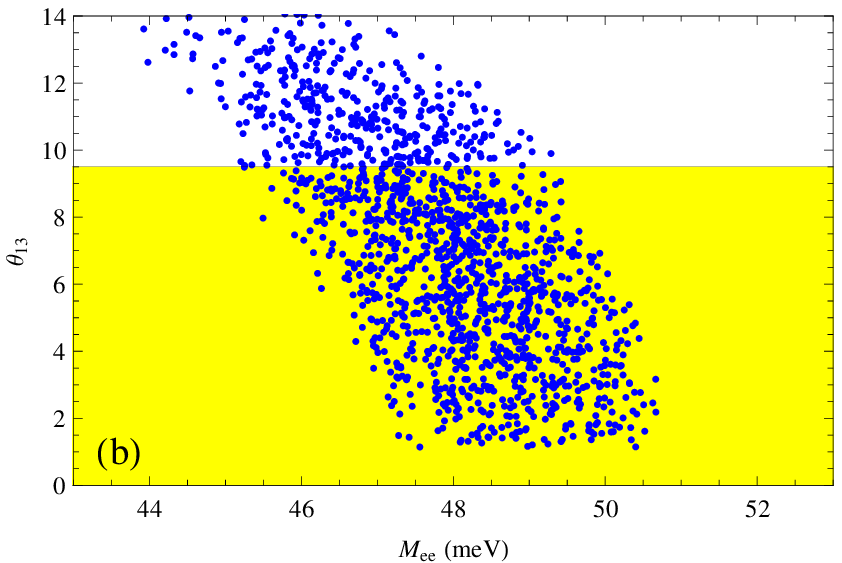}
  \caption{Correlations of $\,\theta_{13}\,$ with Jarlskog invariant
           $J$ in plot-(a) and with neutrinoless double beta decay observable
           $M_{ee}$ in plot-(b), where all inputs are the same as
           Fig.\,\ref{fig:dx-J-Mee}, except requiring the successful leptogenesis
           in the present figure, with 1500 samples for each plot.}
  \label{fig:dx-J-Mee-new}
\end{figure}

With the above constraint on the parameter space from realizing
successful thermal leptogenesis, we can rederive the correlation between
$\,\theta_{13}\,$ and $\,\theta_{23}-45^\deg\,$,\, as shown in the new
Fig.\,\ref{fig-deltax-deltaa-new}, which should be compared with
Fig.\,\ref{fig:dx-da} in Sec.\,4.\,1 (without requiring
leptogensis). We note that the realization of successful thermal
leptogenesis puts a general \,{\it lower bound}\, on the mixing angle $\,\theta_{13}$\,,\,
\beqa
\theta_{13} ~\gtrsim~ 1^\deg ~,
\eeqa
even for the region around $\,\theta_{23} = 45^\deg\,$.\,

Under successful leptogenesis, the correlations of $\,\theta_{13}$ with
the Jarlskog invariant $\,J\,$ and  the neutrinoless double beta decay observable
$\,M_{ee}$ are plotted in Fig.\,\ref{fig:dx-J-Mee-new}(a) and (b), respectively.
This should be compared to Fig.\,\ref{fig:dx-J-Mee}
where leptogenesis is not required. We see that due to the constraint from
the observed baryon asymmetry, the parameter space of $\,J> 0\,$ is forbidden in
Fig.\,\ref{fig:dx-J-Mee-new}(a).
On the other hand, the constrained range for $M_{ee}$ in
Fig.\,\ref{fig:dx-J-Mee-new}(b) is almost the same as Fig.\,\ref{fig:dx-J-Mee}(b),
since Eq.\,(\ref{eq:Mee}) shows that the observable $\,M_{ee}\,$
has rather weak dependence on small NLO parameters
$\,\d_x^{}\,(=\theta_{13})\,$ and $\,\d\phi\,$ via their squared terms.
Thus, from Fig.\,\ref{fig:dx-J-Mee-new}(a)-(b), we infer the following constraints on
$\,J\,$ and $\,M_{ee}\,$,
\beqs
\label{eq:boundLepG-J-Mee}
\beqa
\label{eq:boundLepG-J}
&- 0.037 ~\lesssim~ J ~\lesssim~ -0.0035 \,, &
\\
\label{eq:boundLepG-Mee}
& 45.5\,\textrm{meV} ~\lesssim~ M_{ee} ~\lesssim~ 50.7\,\textrm{meV}\,, &
\eeqa
\eeqs
which should be compared to Eqs.\,(\ref{eq:bound-J})-(\ref{eq:bound-Mee})
in Sec.\,4.\,1 without requiring the successful leptogensis.

\begin{figure}[t]
  \centering
\includegraphics[width=8.3cm,height=5.85cm,clip=true]{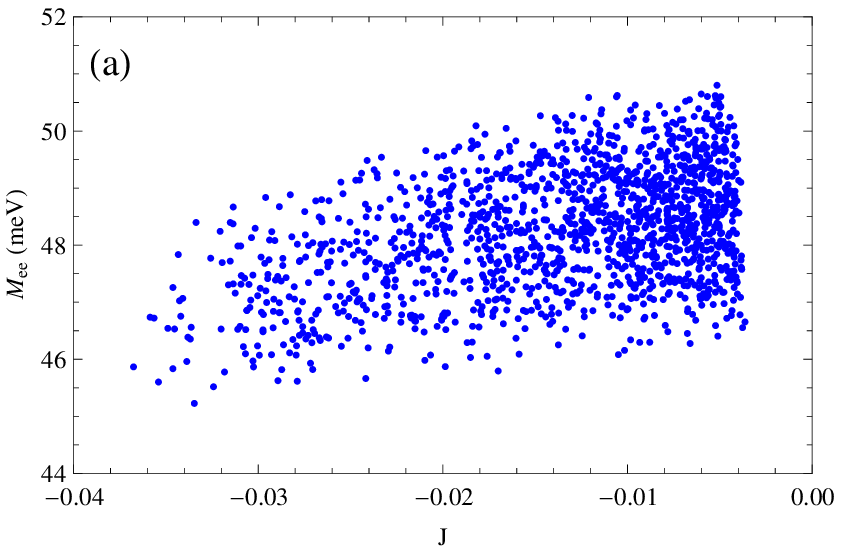}
\includegraphics[width=8.3cm,height=5.65cm,clip=true]{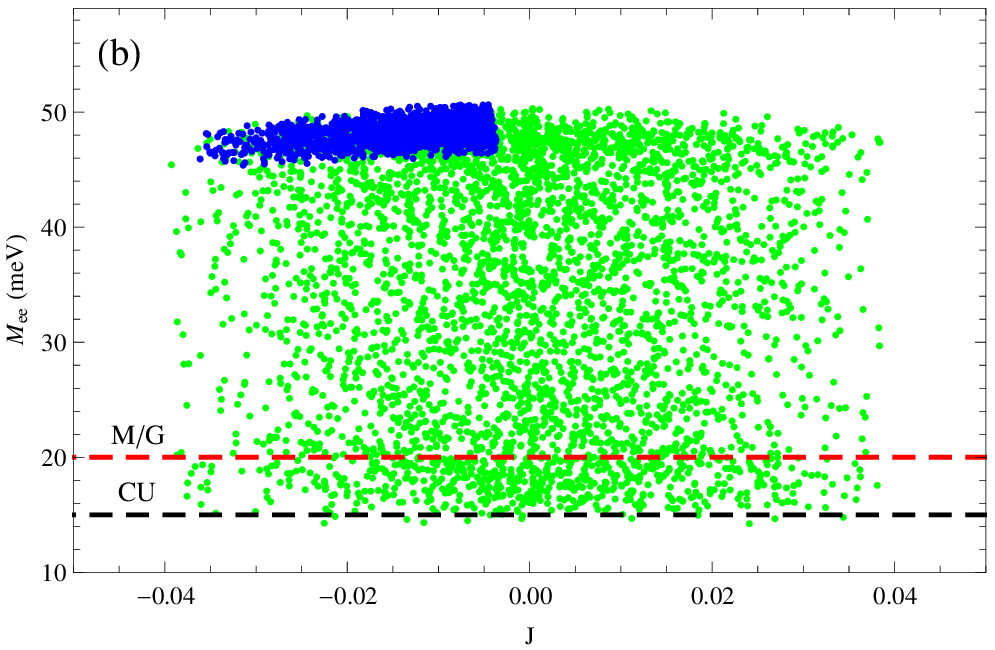}
\includegraphics[width=8.3cm,height=5.8cm,clip=true]{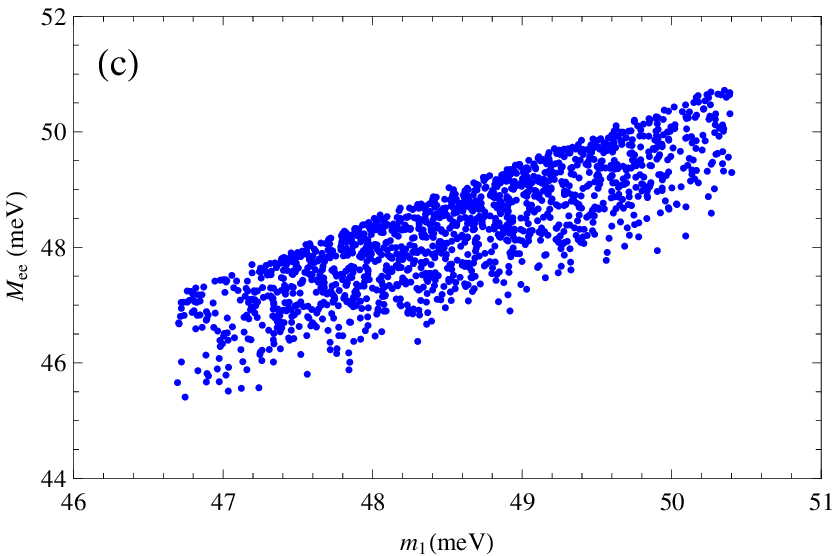}
\includegraphics[width=8.3cm,height=5.7cm,clip=true]{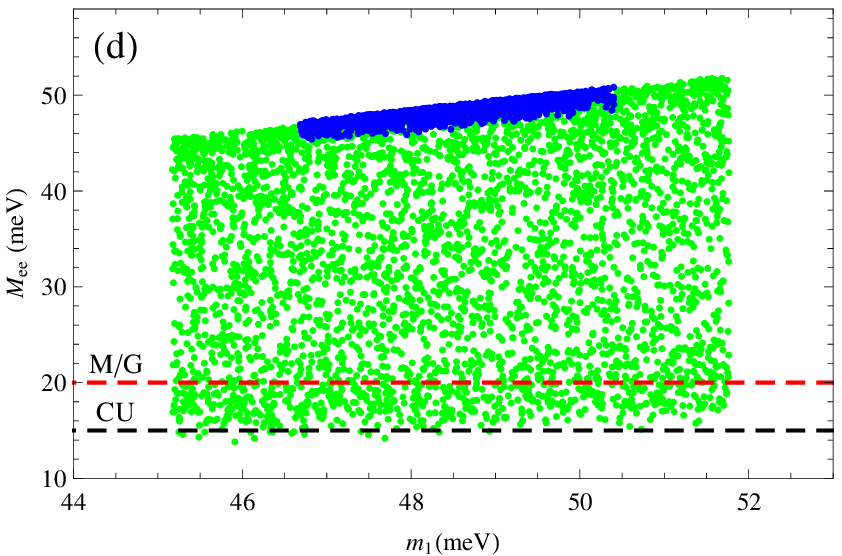}
  \caption{Upper plots (a)-(b) show the correlations between the neutrinoless
  $\beta\beta$-decay observable $M_{ee}$ and the Jarlskog invariant $J$
  with successful leptogenesis.
  Lower plots (c)-(d) depict the correlations between $M_{ee}$ and
  light neutrino mass $\,m_1^{}\,(\simeq m_2^{})$ with successful leptogenesis.
  All experimental inputs are varied within 90\%\,C.L. ranges, for 1500 samples.
  The background (green) regions in plots (b) and (d)
  represent the model-independent parameter space of the IMO scheme with
  $\,m_3\simeq 0\,$. The horizontal dashed lines in (b) and (d) depict the sensitivities
  of the future neutrinoless $\beta\beta$-decay experiments CUORE (CU)\,\cite{cuore}
  and Majorana\,\cite{major}/GERDA\,III\,\cite{gerda} (M/G),
  at $15$\,meV (black) and $20$\,meV (red), respectively.}
  \label{fig-mee-J-new}
\end{figure}

We further analyze the correlations of the neutrinoless $\beta\beta$-decay observable
$\,M_{ee}\,$ with the Jarlskog invariant $\,J\,$ and the light neutrino mass
$\,m_1^{}(\simeq m_2^{})\,$,\,
in Fig.\,\ref{fig-mee-J-new}(a-b) and Fig.\,\ref{fig-mee-J-new}(c-d),
respectively. The two left plots in Fig.\,\ref{fig-mee-J-new}(a) and (c) show the correlations
of  $\,M_{ee}\,$ with $\,J\,$  and with $m_1^{}$ after imposing the leptogenesis.
For the two right plots in Fig.\,\ref{fig-mee-J-new}(b)(d), we have replotted the
same model-predictions as in the two corresponding left plots of
Fig.\,\ref{fig-mee-J-new}(a)(c) (all in blue color).
For comparison, we have further plotted, in Fig.\,\ref{fig-mee-J-new}(b)(d) with green color,
the model-independent parameter space of $\,M_{ee}\,$ [cf.\ (\ref{eq:Mee})] versus
$\,J\,$ [cf.\ (\ref{eq:J})] or $\,m_1^{}\,(=\sqrt{\Delta m_{13}^2})\,$,
for the IMO scheme with $\,m_3\simeq 0\,$,\, where the relevant observables are varied
within their 90\%\,C.L. ranges and $\,\delta_D^{}\in \(0,\,2\pi\]$.\,
This comparison shows that our model predictions are located at the {\it upper boundaries}
of the whole parameter space,  giving rise to the largest allowed $\,M_{ee}\,$.\,
This is very distinctive and highly testable.
Furthermore, in Fig.\,\ref{fig-mee-J-new}(b)(d),
we have compared our predictions with the sensitivities of the future
neutrinoless $\beta\beta$-decay experiments CUORE (CU)\,\cite{cuore} and
Majorana\,\cite{major}/GERDA\,III\,\cite{gerda} (M/G), which are depicted by
the horizontal dashed lines at $15$\,meV (black) and $20$\,meV (red), respectively.

\begin{figure}[t]
  \centering
\includegraphics[width=8.3cm,height=5.5cm,clip=true]{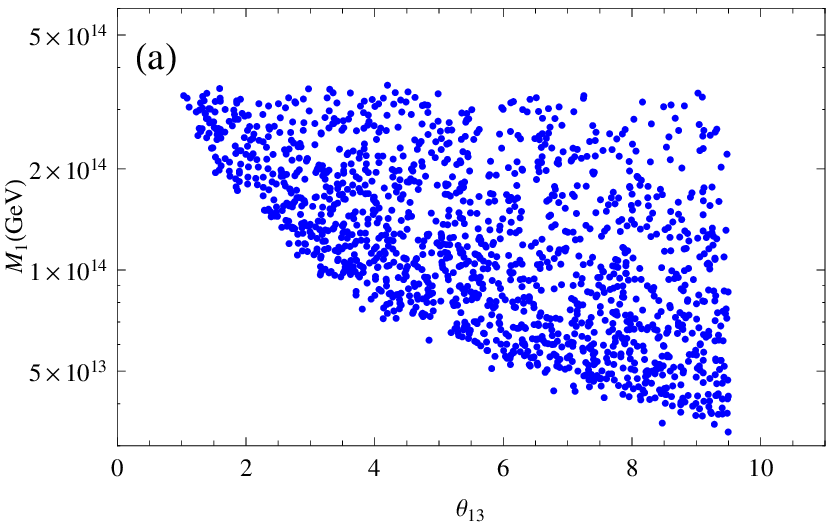}
\includegraphics[width=8.3cm,height=5.5cm,clip=true]{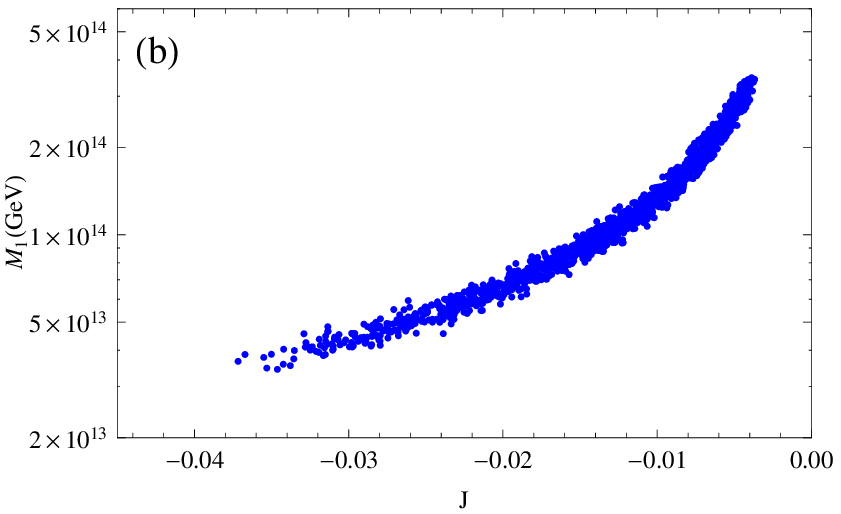}
  \caption{Correlations of leptogenesis scale $\,M_1\,$
           with the reactor mixing angle $\,\theta_{13}\,$ in plot-(a),
           and with the low energy Jarlskog invariant $\,J\,$ in plot-(b).
           Each plot contains 1500 samples. }
  \label{fig-M1-mee-j-new}
\end{figure}

The leptogenesis scale $M_1$ can be determined from the baryon asymmetry
$\,\eta_B^{}$,\, the reactor angle $\,\theta_{13}\,$,\, the Dirac phase
$\,\sin\d_D^{}\,$ and other neutrino observables as in Eq.\,(\ref{eq:M1-etaB}).
Since the low energy parameter $J$ in Eq.\,(\ref{eq:J})
is also predicted as a function of $\,\theta_{13}\,$ and $\,\sin\d_D^{}\,$,\,
so it will correlate with the leptogenesis scale $\,M_1$.\,
Hence, we can plot the correlations of the leptogenesis scale $M_1$ with the
reactor angle $\,\theta_{13}\,$ in Fig.\,\ref{fig-M1-mee-j-new}(a), and with the
Jarlskog invariant $J$ in Fig.\,\ref{fig-M1-mee-j-new}(b).
Inspecting Eqs.\,(\ref{eq:J}) and (\ref{eq:M1-etaB}),  we deduce,
$\,J\propto \d_x\sin\d_D\,$ and $\,M_1\propto (\d_x\sin\d_D)^{-1}\,$,\,
from which we arrive at, $\,M_1 \propto 1/|J|\,$.\,
This behavior is impressively reflected in Fig.\,\ref{fig-M1-mee-j-new}(b),
as expected. In addition, the relation,
$\,M_1\propto (\d_x\sin\d_D)^{-1} \geqq \theta_{13}^{-1}\,$,\, nicely explains
the lower arched edge in Fig.\,\ref{fig-M1-mee-j-new}(a).


\subsection{Extension to General Three-Neutrino Seesaw}

In this subsection, we analyze the extension to the general neutrino seesaw with
three right-handed neutrinos $\,\N'=(N_1,\,N_2,\,N_3)^T$,\, where
$\,\N'\,$ is $\mutau$ blind.
Then, in the $\mutau$ and CP symmetric limit,
the mass-matrices $\,\mD\,$ and $M_R$ are extended to $3\times3$ matrices,
 \begin{eqnarray}
 \label{eq:mD0-3}
 \mD ~=
 \(
  \ba{lll}
    \bar{a}~  & \bar{a}' & \bar{a}'' \\[2mm]
    \bar{b}~  & \bar{c}  & \bar{d}   \\[2mm]
    \bar{b}~  & \bar{c}  & \bar{d}
  \ea \)
 \equiv
 \(
  \ba{lll}
    \sigma_1^{}a~  & \sigma_2^{}a' & \sigma_3^{}a'' \\[2mm]
    \sigma_1^{}b~  & \sigma_2^{}c  & \sigma_3^{}d     \\[2mm]
    \sigma_1^{}b~  & \sigma_2^{}c  & \sigma_3^{}d
  \ea \)\!, ~~~~~~~~~
  M_R ~=~\textrm{diag}(M_1,\,M_2,\,M_3)\,,~~~~~
 \end{eqnarray}
with
 $\,\sigma_1^{}\equiv\sqrt{\widehat{m}_0^{}M_1}\,$,\,
 $\,\sigma_2^{}\equiv\sqrt{\widehat{m}_0^{}M_2}\,$,\, and
 $\,\sigma_3^{}\equiv\sqrt{\widehat{m}_0^{}M_3}\,$,\,
 where the $\mutau$ blind right-handed neutrinos $\,\N'\,$ can always
 be rotated into their mass-eigenbasis without affecting the structure
 of $\mD$.\,
Thus, we rederive the $\mutau$ and CP symmetric seesaw mass-matrix
for the light neutrinos,
\begin{eqnarray}
\label{eq:Mnu-3nuSS}
M_\nu ~=~ \widehat{m}_0^{}
\begin{pmatrix}
  a^2 \!+\! a'^2 \!+\! a''^2
& ~ab+a'c+a''d
& ~ab+a'c+a''d~ \\[2mm]
& ~b^2 \!+\! c^2 \!+\! d^2
& ~b^2 \!+\! c^2 \!+\! d^2 \\[2mm]
& & ~b^2 \!+\! c^2 \!+\! d^2
\end{pmatrix}
\equiv
\left(
\begin{array}{ccc}
 A & B_s & B_s \\
   & C_s & C_s \\
   &     & C_s
\end{array}
\right) ,
\end{eqnarray}
from which we deduce the mass-eigenvalues and mixing angles,
\begin{subequations}
 \beqa
&& \widehat m_{1,2}^{}
 \,=\, \hf\[(A+2C_s) \mp \sqrt{(A-2C_s)^2+8B_s^2}\]
\nn\\[2mm]
&& \hspace*{8mm}
\,=\, \f{\widehat m_0}{2}\[ (a^2+a'^2+a''^2+2b^2+2c^2+2d^2) \right.
\nonumber\\[2mm]
&& \hspace*{12mm}
 \mp \left.\sqrt{[(a^2+a'^2+a''^2)-2(b^2+c^2+d^2)]^2+8(ab+a'c+a''d)^2}\],
\label{eq:m12-3nu}
\\[2mm]
&&
\widehat m_3^{} ~=~ C_s - C_s ~=~ 0\,,
\label{eq:m3-3nu}
\\[2mm]
&& \tan 2\theta_{12} ~=\, \f{2\sqrt{2}B_s}{\,A-2C_s\,}
\,=\, \f{2\sqrt{2}|ab+a'c+a''d|}
        {|a^2\!+\!a'^2\!+\!a''^2\!-\!2(b^2\!+\!c^2\!+\!d^2)|}\,, ~~~~~
\label{eq:t12-3nu}
\\[2mm]
&& \theta_{23} \,=\, 45^\deg\,, ~~~~~
\theta_{13} \,=\, 0^\deg \,, ~~~~~
\label{eq:t23-t13-3nu}
\eeqa
\label{eq:LO-Mass-angle-33}
\end{subequations}
where the mass-spectrum remains the inverted mass-ordering (IMO).
The third mass-eigenvalue $\,\widehat m_3^{}\,$ vanishes because
our $\mutau$ blind seesaw (\ref{eq:mD0-3}) predicts the seesaw mass-matrix
(\ref{eq:Mnu-3nuSS}) with its 23-element equal to the
22-element and 33-element.
This is also a general feature of any $\mutau$ symmetric IMO scheme at the LO,
as to be shown in (\ref{eq:IMO-A0-C0D0-B0}) of Sec.\,5.\,1.
Furthermore, we will demonstrate shortly that the third mass-eigenvalue
$\,\widehat m_3^{}=0\,$ actually holds up to the NLO after including the $\mutau$
and CP breaking in our analysis. So this resembles very much the minimal seesaw
we studied earlier.

Similar to Eqs.\,(\ref{eq:abc-1}) and (\ref{eq:m0=m1=m2-cond}) in Sec.\,2.\,1,
we can realize the IMO at the LO of three-neutrino seesaw,
$\,\widehat m_1^{}=\widehat m_2^{}=\widehat m_0^{}\,$,\, which leads to the three
extended conditions,
\begin{subequations}
\label{eq:IMO-LO-cond-3nu}
\beqa
&&  (a^2+a'^2+a''^2)+2(b^2+c^2+d^2) ~=~ 2 \,, \\[2mm]
&&  (a^2+a'^2+a''^2)-2(b^2+c^2+d^2) ~=~ 0 \,, \\[2mm]
&&  ab+a'c+a''d ~=~ 0 \,.
\eeqa
\end{subequations}
With these we deduce from (\ref{eq:Mnu-3nuSS})
the generic LO seesaw mass-matrix for the IMO,
\beqa
 M_\nu^{(0)}
 & \,=\, & \widehat m_0^{}
 \begin{pmatrix}\,
     a^2 \!+\! a'^2 \!+\! a''^2~ & a b \!+\! a' c \!+\! a'' d~
  &  a b\!+\! a'c \!+\! a'' d~
  \\[3mm]
  &   b^2 \!+\! c^2 \!+\! d^2 & b^2 \!+\! c^2 \!+\! d^2
  \\[3mm]
  & & b^2 \!+\! c^2 \!+\! d^2 \,
  \end{pmatrix}
  =~  \widehat m_0^{}
 \begin{pmatrix}\,
  1  & 0  &  0 \,~
  \\[1.5mm]
  &   \hf  & \hf \,~
  \\[1.5mm]
  & &  \hf \,~
  \end{pmatrix}  \!,
  \label{eq:Mnu30}
\eeqa
which is the same as the LO mass-matrix (\ref{eq:Mnu0})
we derived earlier for the minimal seesaw.
Hence, despite that the LO mass-matrix $\,M_\nu^{(0)}$\, contains two new
parameters $(a'',\,d)$ at the beginning, the realization of IMO eliminates them all
and reduces $\,M_\nu^{(0)}$\, to the universal LO mass-matrix
as shown in the final form of (\ref{eq:Mnu30})
which is parameter-free except an overall mass-scale.
As a result of the IMO conditions (\ref{eq:IMO-LO-cond-3nu}), we note that the solar
angle formula (\ref{eq:t12-3nu}) gives $\,\tan 2\theta_{12}=\f{0}{0}\,$ at the LO,
which is now undetermined. So, the $\,\theta_{12}\,$ has to be derived from the
NLO contributions related to $\mutau$ breaking terms.
Before getting into detail, it is convenient to infer $\theta_{12}$
by using the l\,$'$H\^{o}pital rule, similar to what we did in Sec.\,2.\,1 for
the minimal seesaw. Thus we have,
\beqs
\beqa
 \tan 2\theta_{12} \,=\,\frac{|a|}{\sqrt{2}\,|b|\,} \,,
 \label{eq:t12-31}
\eeqa
for $\mutau$ breaking arising from the deviation in the element $\,b\,$ of $\,\mD\,$,\, or
\beqa
 \tan 2\theta_{12} \,=\,\frac{|a'|}{\sqrt{2}\,|c|\,} \,,
 \label{eq:t12-32}
\eeqa
for $\mutau$ breaking arising from the deviation in the element $\,c\,$ of $\,\mD\,$,\, or
\beqa
 \tan 2\theta_{12} \,=\,\frac{|a''|}{\sqrt{2}\,|d|\,} \,,
 \label{eq:t12-33}
\eeqa
\eeqs
for $\mutau$ breaking arising from the deviation in the element $\,d\,$ of $\,\mD\,$.

As noted in Sec.\,2.\,2, we can always rotate the first column in $\mD$ to be all real
by rephasing. For the convenience of comparison with the minimal neutrino seesaw, we
will thus formulate the common origin of $\mutau$ and CP breaking in
the element $\,c\,$ of $\,\mD$\,.\,  It is possible to construct such a breaking
in the element $\,d\,$ of $\,\mD\,$,\, but this does not affect our physical conclusions
as will be clarified below, after Eq.(\ref{eq:da-dx-3nu}).
[Since we are constructing a common origin of $\mutau$ and CP breaking from
a single source in $\mD$, we do not consider this breaking to occur
in both $c$ and $d$ elements of $\mD$ at the same time.]
So, we build the Dirac mass-matrix $\mD$ with the common $\mutau$ and CP breaking in
the following form,
\beqs
\label{eq:mD-3nu}
\begin{eqnarray}
 \label{eq:mD-3}
&&
  \mD ~=\, \(
  \ba{lll}
    \sigma_1^{}a  &~ \sigma_2^{}a'     &~ \sigma_3^{}a'' \\[2mm]
    \sigma_1^{}b  &~ \sigma_2^{}c_1^{} &~ \sigma_3^{}d   \\[2mm]
    \sigma_1^{}b  &~ \sigma_2^{}c_2^{} &~ \sigma_3^{}d
  \ea \) \!,
\\[2mm]
&& c_1^{} \,=\, c\(1-\zeta'\) ,~~~~~~
   c_2^{} \,=\, c\(1-\zeta e^{i\omega}\) .
 \label{eq:c1-c2}
\end{eqnarray}
\eeqs
Thus we can deduce the NLO part of the seesaw mass-matrix
$\,M_\nu= M_\nu^{(0)} + \delta M_\nu^{(1)}$\,
for light neutrinos,
%
\begin{eqnarray}
 \delta M_\nu^{(1)}
  & = & \widehat m_0^{}
 \begin{pmatrix}\,
    0
  & -a'c \,\zeta'
  & -a'c \,\zeta e^{i\omega}
  \\[3mm]
  & -2c^2 \zeta'
  & -c^2 (\zeta' \!+\! \zeta e^{i\omega})
  \\[3mm]
  & &-2c^2\zeta e^{i\omega}
  \end{pmatrix} \!,
  \label{eq:Mnu31}
\end{eqnarray}
%
which equals (\ref{eq:Mnu1}) as expected, since the new parameters $(a'',\,d)$ appear
in the seesaw mass-matrix $M_\nu$ only via the products $({a''}^2,\,d^2,\,a''d)$ with
no crossing terms like $\,c_{1,2}^{}a''\,$ or $\,c_{1,2}^{}d\,$.
With these, we deduce the $\mutau$ symmetric
and antisymmetric elements of  $\,\delta M_\nu^{(1)}\,$ to be the same as
Eq.\,(\ref{eq:seesaw-Delta-m}).

Using the formalism of Sec.\,3.\,1 and extending Sec.\,3.\,2, we can reconstruct
the light neutrino mass-matrix $M_\nu$ for the IMO with $\, m_3^{}\neq 0\,$,\,
via the NLO parameters,
\beqa
 (\, y',\,
     z,\,z',\,
     \d_a,\,
     \d_x,\,
     \d\ab_1^{},\,
     \d\ab_2^{},\,
     \d\ab_3^{}, \,\delta\phi,\,\delta\phi')\,,
\label{eq:NLO-parameters-3}
\eeqa
where we have defined $\,z'\equiv\frac{m_3^{}}{m_1^{}}\,$ and
$\,\phi'\equiv\phi_3-\phi_{1}=\phi'_0+\delta\phi'\,$.\,
Note that the LO phases vanish,
$\,\bar\alpha_{i0}^{}=\phi_0=\phi'_0=0\,$.\,
So the NLO elements of $M_\nu$ are reconstructed as follows,
\begin{subequations}
\begin{eqnarray}
  \d A &\,=\,&
   m_0^{}
   \[z+\frac{s_s^2}{2}y'- i2(s_s^2\,\delta\phi+\delta\bar\alpha_1)\]\!,
\label{eq:dMnu-3nu-dA}
\\[2mm]
  \d B_s
&\,=\,& \frac{m_{0}^{}}{2\sqrt{2}\,}\sin 2\ts
\[-\frac{1}{2}y'+ i2\delta\phi\] \!,
\label{eq:dMnu-3nu-dBs}
\\[2mm]
  \d C_s
&\,=\,& \f{m_0^{}}{2}
\[{z}+{z'}+\frac{c_s^2}{2}y'
  -i(2c_s^2~\delta\phi+\delta\overline\alpha_2
  +\delta\overline\alpha_3)\] \!,
\label{eq:dMnu-3nu-dCs}
\\[2mm]
\d D &\,=\,& \f{m_0^{}}{2}
\[{z}-{z'}+\frac{c_s^2}{2}y'
  -{i}(2c_s^2~\delta\phi+\delta\overline\alpha_2
  +\delta\overline\alpha_3)\] \!,
\label{eq:dMnu-3nu-dD}
\\[2mm]
\d B_a &\,=\,& -\frac{m_{0}^{}}{\sqrt{2}\,}e^{i\delta_D}\delta_x\,,
\label{eq:dMnu-3nu-dBa}
\\[2mm]
\d C_a
&\,=\,& -m_{0}
\[\delta_a+\frac{i}{2}(\delta\overline\alpha_2
  -\delta\overline\alpha_3)\] \!,
\label{eq:dMnu-3nu-dCa}
\end{eqnarray}
\label{eq:dMnu-13}
\end{subequations}
where we note that the Majorana phase $\,\d\phi'\,$ does not appear at the NLO
because it is always suppressed by another NLO parameter $\,z'=\frac{m_3^{}}{m_1^{}}\,$.\,
Moreover, since the $\mutau$ and CP breaking matrix (\ref{eq:Mnu31}) gives
Eq.\,(\ref{eq:seesaw-Delta-m}) with the equality $\,\d C_s =\d D\,$,\,
we deduce $\,z'=\f{m_3^{}}{m_1^{}}=0\,$
by comparing (\ref{eq:dMnu-3nu-dCs}) with (\ref{eq:dMnu-3nu-dD}),
and thus $\,m_3^{}=0\,$ holds up to the NLO.
Hence, we have shown that our model with the general three-neutrino seesaw under IMO
does share the essential feature of $\,m_3^{}=0\,$ with the minimal seesaw.

Then, with the NLO $\mutau$ symmetric parts from
(\ref{eq:seesaw-Delta-m}) and (\ref{eq:dMnu-13}),
we deduce the solar angle $\theta_{12}$,
\begin{equation}
  \tan2\ts ~=\, -\frac{a'}{\sqrt{2}\,c\,} \,,
  \label{eq:t12-3}
\end{equation}
which coincides with Eq.\,(\ref{eq:t12-NLO})
as we derived earlier for the minimal seesaw.

Next, connecting the $\mutau$ anti-symmetric parts in
(\ref{eq:seesaw-Delta-m}) and (\ref{eq:dMnu-13}) gives,
\beqs
\beqa
 \frac{m_0^{}}{2} a' c(\zeta'-\zeta e^{i\omega})
 &\,=\,&
 -\frac{m_0^{}}{\sqrt{2}\,}\,e^{i\delta_D}\delta_x \,,
 \label{eq:dBa-3}
 \\[2mm]
  - m_0^{} c^2 (\zeta'-\zeta e^{i\omega})
 &\,=\,& - m_0^{}\[\delta_a+\frac{i}{2}(\delta\bar\alpha_2
                      -\delta\bar\alpha_3)\] \!,
\label{eq:dCa-3}
\eeqa
\label{eq:dBa-dCa-3}
\eeqs
from which we arrive at
\begin{subequations}
\begin{eqnarray}
\cos\delta_D~\delta_x
&\,=\,&
\frac{a'c}{\sqrt{2}\,}\(\zeta' -\zeta \cos\omega\) ,
\label{eq:solve-II-a-a3}
\\[2mm]
\sin\delta_D~\delta_x
&\,=\,&
-\frac{a'c}{\sqrt{2}\,}\(\zeta \sin\omega\) ,
\label{eq:solve-II-a-b3}
\\[2mm]
\delta_a
&\,=\,&
c^2\(\zeta' -\zeta \cos\omega\) ,
\label{eq:solve-II-a-c3}
\\[2mm]
\delta\bar\alpha_2 -\delta\bar\alpha_3
&\,=\,&
-2c^2 \(\zeta \sin\omega\) .
\label{eq:solve-II-a3}
\end{eqnarray}
\end{subequations}
Here for the left-hand-sides of (\ref{eq:dBa-3})-(\ref{eq:dCa-3}) we have used the
Eq.\,(\ref{eq:RG-Run-m123}) to evolve the overall mass-parameter $\,\mh_0^{}\,$
from seesaw scale down to the corresponding $\,m_0^{}\,$ at low energy.

Finally, using Eqs.\,(\ref{eq:t12-3}), (\ref{eq:solve-II-a-a3}) and (\ref{eq:solve-II-a-c3}),
we derive the key correlation between two low energy $\mutau$ breaking observables
$\,\da\,$ and $\,\dx$\,,
\begin{eqnarray}
\label{eq:da-dx-3nu}
{\da} ~=\, -\cot 2\theta_s \cos\delta_D \,\dx \,,
\end{eqnarray}
which coincides with (\ref{eq:da-dx-1}) as we derived earlier for the minimal seesaw.

We note that it is also possible to construct the common origin of $\mutau$ and CP breaking
in the element $\,d\,$ of $\,\mD\,$,\, instead of the element $\,c\,$.\,
Then we can rewrite the Dirac mass-matrix (\ref{eq:mD-3nu}) as
\beqs
\label{eq:mD-3nu-d}
\begin{eqnarray}
 \label{eq:mD-3-d}
&&
  \mD ~=\, \(
  \ba{lll}
    \sigma_1^{}a  &~ \sigma_2^{}a'  &~ \sigma_3^{}a'' \\[2mm]
    \sigma_1^{}b  &~ \sigma_2^{}c   &~ \sigma_3^{}d_1^{}   \\[2mm]
    \sigma_1^{}b  &~ \sigma_2^{}c   &~ \sigma_3^{}d_2^{}
  \ea \) \!,
\\[2mm]
&& d_1^{} \,=\, d\(1-\zeta'\) ,~~~~~~
   d_2^{} \,=\, d\(1-\zeta e^{i\omega}\) .
 \label{eq:c1-c2}
\end{eqnarray}
\eeqs
This results in the following NLO seesaw mass-matrix,
\beqa
 \delta M_\nu^{(1)}
  & = & \widehat m_0^{}
 \begin{pmatrix}\,
 0
  & -a''d \,\zeta'
  & -a''d \,\zeta e^{i\omega}
  \\[3mm]
  & -2d^2 \zeta'
  & -d^2 (\zeta' \!+\! \zeta e^{i\omega})
  \\[3mm]
  & &-2d^2\zeta e^{i\omega}
  \end{pmatrix} \!,
  \label{eq:Mnu31-d}
\eeqa
from which we derive the solar angle,
\beqa
\tan2\ts ~=\, -\frac{a''}{\sqrt{2}\,d\,} \,,
\label{eq:t12-3-d}
\eeqa
and the reconstruction conditions,
\begin{subequations}
\begin{eqnarray}
\cos\delta_D~\delta_x
&\,=\,&
\frac{a''d}{\sqrt{2}\,}\(\zeta' -\zeta \cos\omega\) ,
\label{eq:solve-II-a-a3-d}
\\[2mm]
\sin\delta_D~\delta_x
&\,=\,&
-\frac{a''d}{\sqrt{2}\,}\(\zeta \sin\omega\) ,
\label{eq:solve-II-a-b3-d}
\\[2mm]
\delta_a
&\,=\,&
d^2\(\zeta' -\zeta \cos\omega\) ,
\label{eq:solve-II-a-c3-d}
\\[2mm]
\delta\bar\alpha_2 -\delta\bar\alpha_3
&\,=\,&
-2d^2 \(\zeta \sin\omega\) .
\label{eq:solve-II-a3-d}
\end{eqnarray}
\end{subequations}
So, from Eqs.\,(\ref{eq:t12-3-d}), (\ref{eq:solve-II-a-a3-d}) and (\ref{eq:solve-II-a-c3-d}),
we can readily derive the correlation between two $\mutau$ breaking observables,
\begin{eqnarray}
\label{eq:da-dx-3nu-d}
{\da} ~=\, -\cot 2\theta_s \cos\delta_D \,\dx \,,
\end{eqnarray}
which coincides with (\ref{eq:da-dx-3nu}).

In summary, the general three-neutrino seesaw (with right-handed neutrinos being
$\mutau$ blind) still predicts the inverted mass-ordering (IMO) for light neutrinos
[cf.\ Eqs.\,(\ref{eq:m12-3nu})-(\ref{eq:m3-3nu})].
Despite that the LO conditions (\ref{eq:IMO-LO-cond-3nu}) for the IMO
contains two new parameters $(a'',\,d)$, the LO seesaw mass-matrix (\ref{eq:Mnu30})
is shown to take the same form as in the minimal seesaw.
Furthermore, the NLO $\mutau$ and CP breaking part of
our seesaw mass-matrix (\ref{eq:Mnu31}) or (\ref{eq:Mnu31-d})
exhibits the same structure as in the minimal seesaw. This makes
our final physical prediction of the correlation (\ref{eq:da-dx-3nu}) or
(\ref{eq:da-dx-3nu-d}) coincides with (\ref{eq:da-dx-1}).

\vspace*{4mm}
\section{\large Hidden Symmetry and Dictation of Solar Mixing Angle}
\label{sec:Z2s}

So far, by analyzing the $\mutau$ symmetry and its breaking,
we have studied the atmospheric mixing angle $\theta_{23}$ and the reactor
mixing angle $\theta_{13}$ in great detail.
As shown in Table-1, the solar mixing angle $\theta_{12}$
is best measured\,\cite{SNO,KamLAND} among the three mixing angles.
In this section we will clarify the connection between $\mutau$ breaking and the
determination of the solar mixing angle $\,\theta_{12}$\, for both inverted mass-ordering (IMO)
(cf.\ Sec.\,2) and normal mass-ordering (NMO)\,\cite{GHY}.
Then, we analyze the general model-independent $\ZZ_2\otimes\ZZ_2$ symmetry structure in the
light neutrino sector, and map it into the seesaw sector, where one of the $\ZZ_2$ symmetries
corresponds to the $\mutau$ symmetry $\ZZ_2^{\mmutau}$ and another the hidden symmetry $\ZZ_2^s$
(which we revealed in \cite{GHY} for the NMO of light neutrinos and is supposed to dictate
 $\,\theta_{12}\,$).
We will further derive the general consequences of this $\,\ZZ_2^s$\, and its possible violation
in the presence of $\mutau$ breaking for cases either with or without neutrino seesaw,
regarding the $\,\theta_{12}\,$ determination.

\vspace*{3mm}
\subsection{$\bd{\mutau}$ Breaking versus $\bd{\theta_{12}}$ Determination: Inverted Mass-Ordering}
\label{sec:5.1}

In Ref.\,\cite{GHY} we proved that the solar mixing angle
$\,\theta_{12}\,(\equiv\theta_s)$\, is not affected by the soft $\mutau$ breaking from
the neutrino seesaw, and we revealed a hidden symmetry $\ZZ_2^s$ for both
the seesaw Lagrangian and the light neutrino mass-matrix which dictates $\,\theta_s\,$,\,
where the normal mass-ordering (NMO) is realized.
In this subsection, we generally analyze mass-eigenvalues and mixing angles for
the $\mutau$ symmetric mass-matrix of light neutrinos under the inverted mass-ordering (IMO).
Then we explain why the $\mutau$ breaking is invoked for the $\,\theta_s\,$ determination
and why the hidden symmetry $\ZZ_2^s$ will be violated.
The $\mutau$ blind seesaw constructed in Sec.\,2 belongs to an
explicit realization of the IMO scheme.

Let us start with the general $\mutau$ symmetric mass-matrix for light neutrinos,
\beqs
\beqa
M_\nu^{(s)} ~=~
\left(
\begin{array}{ccc}
 A & B_s & B_s \\
   & C_s & D \\
   &     & C_s
\end{array}
\right) ,
\eeqa
\eeqs
which can be diagonalized as follows\,\cite{GHY,Grimus2001},
\beqs
\beqa
\label{eq:5.1-m12}
m_{1,2}^{} &~=~& \f{1}{2}\[
\[A+(C_s\!+\!D)\] \mp \sqrt{\[A-(C_s\!+\!D)\]^2+8B_s^2\,}
\] ,
\\[1.5mm]
\label{eq:5.1-m3}
m_3^{}  &\!\!=\!\!&  C_s - D \,,
\\[2mm]
\label{eq:5.1-ts-ta-tx}
\tan 2\ts  &\!\!=\!\!& \f{2\sqrt{2}\,B_s}{\,A-(C_s\!+\!D)\,}\,, ~~~~~
\ta \,=\, 45^\deg\,, ~~~~~
\tx \,=\, 0^\deg \,.
\eeqa
\eeqs
Substituting (\ref{eq:5.1-ts-ta-tx}) into (\ref{eq:5.1-m12}), we arrive at
\beqa
\label{eq:5.1-m12-2}
m_{1,2}^{} &~=~& \hf \left\{
[A+(C_s\!+\!D)]\mp |A-(C_s\!+\!D)|\sec 2\ts \right\}.
\eeqa
For the IMO scheme, we have the mass-spectrum $\,\mb\gtrsim \ma \gg \mc\,$,\,
where a small $\,m_3^{}\neq 0\,$ is also generally allowed for the analysis below.
So we can derive, for the general IMO scheme,
\beqa
\label{eq:IMO-ratio}
\left|\f{A-(C_s\!+\!D)}{A+(C_s\!+\!D)}\right|
~=~ \f{\,\mb -\ma\,}{\mb +\ma}\cos 2\ts
~\simeq~ \f{\Delta m_{21}^2}{\,4\Delta m_{13}^2\,}\cos 2\ts
~=~ (2.1-3.8)\times 10^{-3}
, ~~~~~~~~
\eeqa
where in the last step we have used the neutrino data (Table-1) to
estimate the allowed range of this ratio at 90\%C.L.
Literally,  Eq.\,(\ref{eq:IMO-ratio}) shows a fine-tuned cancellation
between the mass-matrix elements $A$ and $(C_s+D)$ down to the level of $\,10^{-3}$.\,
As will be clear in Sec.\,5.\,2. 2 by using the general reconstruction formalism
for the IMO scheme, we find that the LO form of the $\mutau$ symmetric mass-matrix
$M_\nu^{(0)}$ predicts the exact relations [cf. Eq.\,(\ref{eq:IMO-Mnu-0th})],
\beqa
\label{eq:IMO-A0-C0D0-B0}
&& A^{(0)}-(C_s^{(0)}+D^{(0)})
\,=\,\[1 -\(\hf+\hf\)\]m_0^{}\,=\,0\,, ~~~~~~~~
\\[2mm]
&& B_s^{(0)} \,=\, 0\,,~~~~
C_s^{(0)}-D^{(0)} ~=~ 0 \,,
\non
\eeqa
which ensures $\,\ma =\mb\,$ and $\,\mc =0\,$ at the LO.
So, the small ratio (\ref{eq:IMO-ratio}) naturally arises from the NLO elements
$\,\[\d A-(\d C_s+\d D)\]\neq 0\,$,\,
and thus there is no real fine-tuning in (\ref{eq:IMO-ratio}).
This also means that at the LO the solar angle $\ts$ is undetermined from
the formula (\ref{eq:5.1-ts-ta-tx}),
$\,\tan 2\ts =\f{0}{0}\,$, and the real determination of $\ts$ is given by the
NLO elements of $M_\nu^{(s)}$,
\beqa
\label{eq:IMO-tan2ts-NLO}
\tan 2\ts ~=~
\f{2\sqrt{2}\,\d B_s}{\,\d A-(\d C_s \!+\!\d D)\,} \,,
\eeqa
as we will explicitly verify in the next subsection for the general IMO scheme
[cf.\ Eqs.\,(\ref{eq:IMO-tan2ts-0})-(\ref{eq:IMO-Ms-tan2t12})].

For the $\mutau$ blind seesaw defined in Sec.\,2.\,1,
we find that the light neutrino mass-spectrum must be
inverted ordering, as given in Eqs.\,(\ref{eq:m12})-(\ref{eq:m3}).
So, following the consistency with neutrino data (\ref{eq:IMO-ratio})
and matching the reconstruction formalism (\ref{eq:IMO-A0-C0D0-B0})
for the IMO scheme, we can explicitly
realize the degeneracy $\,\ma =\mb\,$ at the LO
by imposing the condition (\ref{eq:abc-1}) on the elements of $\,\mD$\,.\,
(Here $\,\mc =0\,$ is an outcome of the minimal seesaw.)
Thus, as expected, we find a problem for the $\,\theta_s\,$ determination
in the $\mutau$ symmetric limit,
\beqa
\label{eq:LO-t12-new}
\tan 2\theta_s ~=~ \f{2\sqrt{2}|ab+a'c|}{\,|a^2+a'^2-2(b^2+c^2)|\,} ~=~ \f{0}{0}\,,
\eeqa
which is just an explicit realization of our above general IMO analysis
[cf.\ (\ref{eq:IMO-A0-C0D0-B0})].
Hence, it is clear that $\,\theta_s\,$ must be inferred from the NLO formula
(\ref{eq:IMO-tan2ts-NLO}), where the NLO elements will be predicted by a
given model, {\it e.g.,} by the first four expressions in Eq.\,(\ref{eq:seesaw-Delta-m})
in the $\mutau$ blind seesaw with all NLO corrections
arising from the $\mutau$ breaking\,\cite{footnote-5}.
Thus the explicit expression of $\,\ts\,$ from such underlying models will
depend on how the $\mutau$ breaking is constructed.
This is contrary to the neutrino seesaw with normal mass-ordering (NMO) of light neutrinos
as studied in Ref.\,\cite{GHY}, where we find that the formula of $\,\tan 2\ts\,$
[cf.\ (\ref{eq:5.1-ts-ta-tx}) above] is well defined in the $\mutau$ symmetric limit.

As we noted in Sec.\,2.\,1, the structure $\,\f{0}{0}\,$ in Eq.\,(\ref{eq:LO-t12-new})
allows us to use the l\,$'$H\^{o}pital rule on (\ref{eq:LO-t12-new}) by taking the
first derivatives on both its numerator and denominator. We need to decide
for which parameter in (\ref{eq:LO-t12-new}) the derivatives should be taken.
There are only two possible choices, either $c$ or $b$, since the $\mutau$ breaking
under the $\mutau$ blind seesaw could appear in either $c$ or $b$ element of $\mD$,
as we explicitly constructed in Eqs.\,(\ref{eq:mD-B-1}) and (\ref{eq:mD-B-3}).
Thus, applying the l\,$'$H\^{o}pital rule to (\ref{eq:LO-t12-new}) we have
\beqa
\label{eq:t12-2eqs}
\tan 2\theta_s ~=\,
\left\{
\ba{ll}
\dis\frac{|a'|}{\sqrt{2}\,|c|}\,,~~~ & (\mutau~\textrm{breaking~in}~c)\,,
\\[4mm]
\dis\frac{|a|}{\sqrt{2}\,|b|}\,,~~~ & (\mutau~\textrm{breaking~in}~b)\,,
\ea
\right.
\eeqa
which, as expected, gives finite expressions for $\,\theta_s\,$,\, depending only
on the LO parameters of the Dirac mass-matrix $\,\mD$.\,
This also agrees to Eqs.\,(\ref{eq:t12})-(\ref{eq:t12-b}) in Sec.\,2.\,1.
But Eq.\,(\ref{eq:t12-2eqs}) shows that $\theta_s$ does depend on how the $\mutau$
breaking is built in the seesaw Lagrangian, and the two different constructions of
$\mutau$ breaking for $\mD$ lead to two different $\theta_s$ formulas above.
This is an essential difference from the soft $\mutau$ breaking
model in Ref.\,\cite{GHY}, where $\theta_s$ is dictated by the hidden symmetry
$\,\ZZ_2^s\,$ under which the soft $\mutau$ breaking term in $\,M_R\,$ is an exact singlet.
In the next subsections we will analyze the general model-independent $\ZZ_2\ot\ZZ_2$ symmetry
in the light neutrino sector, and then map it into the seesaw sector. This allows us to
explore, {\it at a deeper level, the $\ZZ_2^s$ symmetry and its possible partial violation
under the $\mutau$ breaking in a unified way,
concerning $\,\theta_s\,$ determination.}

\vspace*{3mm}
\subsection{\,${\ZZ_2^s}$ Symmetry under General $\bd{\mutau}$ Breaking
            and General Determination of $\bd{\theta_{12}}$}
\label{sec:5.2}

This subsection consists of two parts.
In Sec.\,5.\,2.\,1, we analyze the general model-independent $\ZZ_2\otimes\ZZ_2$ symmetry structure
of the light neutrino sector, in both the mass and flavor eigenbases. We will show that,
in the flavor eigenbasis of light neutrinos, one of the
$\ZZ_2$'s is the $\,\ZZmt\,$ symmetry which predicts the mixing angles
$\,(\tbc,\,\tac)=(45^\deg,\,0^\deg)\,$,\,
and another is the $\,\ZZs\,$ symmetry which generally dictates
the solar angle $\,\tab\,$ by its group parameter
(allowing deviations from the conventional tri-bimaximal mixing ansatz).
With general $\mutau$ breaking parameters, we will derive a nontrivial correlation
between the two $\mutau$ breaking observables
which is {\it necessary for holding the $\ZZs$ symmetry.}
In Sec.\,5.\,2.\,2, we will further analyze the general $\mutau$ breaking in the
light neutrino mass-matrix $M_\nu$ and derive a nontrivial consistency
condition to hold the $\ZZs$ symmetry. From this condition and using the general
reconstruction formalism of Sec.\,3.\,1, we will deduce
the {\it same correlation} between the $\mutau$ breaking observables, for both
the normal mass-ordering and inverted mass-ordering of light neutrinos
({\it without} approximating the lightest neutrino mass to zero)\,\cite{footnote-6}.

\vspace*{2mm}
\subsubsection{\,$\ZZ_2^s$ Symmetry for General Determination of
                 Solar Angle $\bd{\theta_{12}}$}
\label{sec:5.2.1}

Let us inspect the flavor symmetries in the lepton and neutrino sectors.
In general, the lepton and neutrino sectors are expected to obey different flavor symmetries.
After spontaneous symmetry breaking, the residual symmetry groups for the lepton and neutrino
mass-matrices may be denoted as $\,\GG_\ell\,$ and $\,\GG_\nu\,$, respectively.
Consider the symmetry transformations
$\,F_j\in \GG_\ell\,$ and $\,G_j\in \GG_\nu\,$ for left-handed leptons and neutrinos.
Thus the mass-matrices of leptons ($M_\ell$) and light neutrinos ($M_\nu$)
will satisfy the invariance equations\,\cite{S4},
\beqa
F_j^\dag M_\ell^{}M_\ell^\dag F_j ~=~ M_\ell^{}M_\ell^\dag\,,
&~~~&
G_j^T M_\nu G_j ~=~ M_\nu \,.
\label{eq:G-M}
\eeqa
The above mass-matrices can be diagonalized by unitary rotations for left-handed leptons and
neutrinos,
\beqa
 U_\ell^\dag M_\ell^{}M_\ell^\dag U_\ell \,=\, D_\ell^{}
 \,\equiv\, \textrm{diag}\(m_e^2,\,m_\mu^2,\,m_\tau^2\),
~~~~
 U_\nu^T M_\nu U_\nu \,=\, D_\nu \,\equiv\,
 \textrm{diag}\(\ma,\,\mb,\,\mc\) .
\label{eq:V-M}
\eeqa
Then, combining the invariance equations (\ref{eq:G-M}) and diagonalization equations
(\ref{eq:V-M}) result in
\beqa
U_\ell^\dag F_j^\dag M_\ell^{}M_\ell^\dag F_jU_\ell^{}
\,=\, d_\ell^\dag D_\ell^{} d_\ell^{} \,=\, D_\ell^{} \,,
~~~~
U_\nu^T G_j^T M_\nu G_jU_\nu^{} \,=\, d_\nu^T D_\nu^{} d_\nu^{}
\,=\, D_\nu^{}\,,
\label{eq:G-V}
\eeqa
where $\,d_\ell^{}\,$ and $\,d_\nu^{}\,$ are diagonal phase-matrices obeying
$\,d_\ell^\dag d_\ell^{}={\cal I}_3\,$  and
$\,d_\nu^2={\cal I}_3^{}\,$ (with ${\cal I}_3^{}$ the $3\times 3$ unit matrix), which
require
$\,d_\ell^{}=\textrm{diag}(e^{i\gaa},\,e^{i\gab},\,e^{i\gac})\,$ and
$\,d_\nu^{} =\textrm{diag}(\pm 1,\,\pm 1,\,\pm 1)\,$.\,
So, up to an overall phase factor, the
$\,\{d_\ell^{(j)}\}\,$ forms the generic Abelian group
$\,U(1)\otimes U(1)=\GG_\ell\,$ for leptons,
and $\,\{d_\nu^{(j)}\}\,$  has only two independent $\,\dnu\,$,
\beqa
 d_\nu^{(1)}\,=\, \textrm{diag}(1, 1, -1)\,,~~~~~~
 d_\nu^{(2)}\,=\, \textrm{diag}(-1, 1, 1)\,,
\label{eq:dnu-1-2}
\eeqa
forming the generic discrete group $\,\ZZ_2\otimes \ZZ_2=\GG_\nu\,$ for neutrinos.
From Eq.\,(\ref{eq:G-V}) the following consistency solutions are deduced,
\beqa
F_j \,=\, U_\ell^{}d_\ell^{(j)}U_\ell^\dag \,,
~~~~~
G_j \,=\, U_\nu^{} d_\nu^{(j)}U_\nu^\dag \,.
\label{eq:Z2-ma}
\eeqa
This proves that $\{F_j\}$ and $\{\dlj\}$
are just connected by the similarity transformations, and are thus
two {\it equivalent representations} of the same
group $\,\GG_\ell\,$;\, similarly, $\{G_j\}$ and $\{\dnuj\}$ are two
{\it equivalent representations} of the same group $\,\GG_\nu\,$.\,
We may call the representation $\{\dlj\}$ and $\{\dnuj\}$ the ``kernel representations",
with which the equivalent ``flavor representations" $\{F_j\}$ and $\{G_j\}$ can be
generated as in (\ref{eq:Z2-ma}) via the disgonalization matrices
$U_\ell$ and $U_\nu$, respectively.
Hence, we are free to choose an equivalent lepton symmetry group
representation $\,\{F_j\} = \{d_\ell^{(j)}\}\,$ with $\,U_\ell^{}={\cal I}_3^{}\,$,\,
and accordingly, rewrite the representation of neutrinos symmetry group,
\beqa
 G_j ~=~ V\, d_\nu^{(j)}\,V^\dag \,,
\label{eq:Z2-mb}
\eeqa
with $\,V=U_\ell^\dag U_\nu^{}=U_\nu\,$
equal to the physical PMNS mixing matrix
as defined in Eq.\,(\ref{eq:V-def}) of Sec.\,3\cite{footnote-7}.
Let us rewrite the PMNS matrix (\ref{eq:V-def}),
$\,V = U''U U' = V'U' \,$,\,
with $\,V' \equiv U''U\,$ as introduced in Eq.\,(\ref{eq:Mnu-V'-Dt}).
So we see that the Majorana phase-matrix $\,U'\,$ cancels in $G_j$,
\beqa
 G_j ~=~ V' d_\nu^{(j)}\,V^{\prime\dag} \,.
\label{eq:Gj-dnuj-V'}
\eeqa

According to the most general reconstruction formulation in Sec.\,3.\,1,
we can expand the matrix $V'$ to NLO in terms of the small parameters,
$\,
 (\delta_a,\,\delta_x,\,\delta\alpha_i^{})
\,$,\,
where $(\delta_a,\,\delta_x)$ characterizes the low energy $\mutau$ breaking and
the CP-angle $\delta\alpha_i^{}$ arises from the phase matrix $U''$ (which is not
directly observable and only needed for the consistency of diagonalizing the mass
matrix $M_\nu$). There is no need to expand the Dirac CP-phase
$\,e^{i\d_D^{}}\,$ itself since it is always associated with the small $\mutau$
breaking parameter $\,\d_x$\,.\,
So, under this expansion we derive
\beqa
\label{eq:V'=Vs+dV'}
 V' ~=~ V_s + \delta V' \,,
\eeqa
with
\beqs
\beqa
V_s &~=~&
\begin{pmatrix}
c_s & -s_s & 0
\\[2mm]
\frac{s_s}{\sqrt{2}\,} & \frac{c_s}{\sqrt{2}\,} & -\frac{1}{\sqrt{2}\,}
\\[2mm]
\frac{s_s}{\sqrt{2}\,} & \frac{c_s}{\sqrt{2}\,} & \frac{1}{\sqrt{2}\,}
\end{pmatrix} \!,
\\[3mm]
\delta V' &\!\!\!=\!\!\!&
\begin{pmatrix}
ic_s\delta\alpha_1&is_s\delta\alpha_1&-\delta_xe^{-i\delta_D}\\[2mm]
-\frac{s_s\delta_a+c_s\delta_xe^{i\delta_D}+is_s\delta\alpha_2}{\sqrt{2}}
&\frac{-c_s\delta_a+s_s\delta_xe^{i\delta_D}-ic_s\delta\alpha_2}{\sqrt{2}}&
-\frac{\delta_a-i\delta\alpha_2}{\sqrt{2}}\\[2mm]
\frac{s_s\delta_a+c_s\delta_xe^{i\delta_D}-is_s\delta\alpha_3}{\sqrt{2}}
&\frac{c_s\delta_a-s_s\delta_xe^{i\delta_D}-ic_s\delta\alpha_3}{\sqrt{2}}&
-\frac{\delta_a+i\delta\alpha_3}{\sqrt{2}}
\end{pmatrix} \!.
\eeqa
\eeqs
Let us first consider the $\mutau$ symmetric limit with $\,V'=V_s\,$.\,
So substituting $\,V_s\,$ into Eq.\,(\ref{eq:Gj-dnuj-V'}) we deduce,
\beqa
G_{\mu\tau} ~\equiv~ G_1^{} ~=~ V_s^{} d_\nu^{(1)} V^{\dag}_s
~=\,
\begin{pmatrix}
1 & 0 & 0
\\
0 & 0 & 1
\\
0 & 1 & 0
\end{pmatrix}
\!,
\eeqa
which, as expected,
just gives the $\ZZ_2^{\mmutau}$ symmetry-transformation matrix $G_\nu$ for light neutrinos
as we explicitly constructed in (\ref{eq:T3T2}) earlier
for the seesaw Lagrangian (\ref{eq:L-seesaw}).

Next, we derive the symmetry-transformation matrix $G_s^0$ corresponding to $\,d_\nu^{(2)}\,$
of (\ref{eq:dnu-1-2}) in the $\mutau$ symmetric limit with ($\,\d V'=0\,$),
\beqs
\label{eq:Gs0-all}
\beqa
\label{eq:Gs0}
G_s^0 &~=~& V_s^{} \,d_\nu^{(2)}\, V_s^\dag
= \begin{pmatrix}
s_s^2\!-\!c_s^2 & -\sqrt{2}\,s_sc_s & -\sqrt{2}\,s_sc_s
\\[1.5mm]
& c_s^2 & -s_s^2
\\[1.5mm]
& & c_s^2
\!\!
  \end{pmatrix}
\\[2mm]
&\!=\!& \f{1}{1+k^2}
\begin{pmatrix}
 k^2\!-\!1 & -\sqrt{2}k & -\sqrt{2}k
\\[1.5mm]
& 1 & -k^2
\\[1.5mm]
& & 1
\end{pmatrix} \!,
\label{eq:Gs0-k}
\eeqa
\eeqs
which is symmetric since $V_s^{}$ and $d_\nu^{(j)}$ are real.
(For the same reason $G_{\mmutau}$ is also symmetric.)
In the last step, for convenience we have defined,
\beqa
(s_s^{},\, c_s^{}) ~=~ \f{(k,\,1)}{\sqrt{1+k^2}\,}\,,
\eeqa
with $\,k\,$ (or equivalently, $\tan\ts$) serving as the group parameter of $\,\ZZ_2^s$\,,
\beqa
\label{eq:t12-k}
\tan\theta_s \,=\, k \,,
\eeqa
where we can always choose the convention of $\,\theta_s\in\[0,\,\f{\pi}{2}\]\,$ such that,
$\,\tan\theta_s = k \geqq 0\,$.\,
Noting $\,(d_\nu^{(j)})^2={\cal I}_3\,$ and using the relation
$\,G_s^0 = V_s^{}d_\nu^{(2)}V_s^\dag\,$,\,
we can readily verify $\,(G_s^0)^2 = {\cal I}_3\,$  and thus indeed
$\,G_s^0\in \ZZ_2^s\,$.\,
Hence, the solar angle $\,\theta_s\,$ is dictated by the group parameter $\,k\,$ of the
3-dimensional representation of the hidden symmetry $\,\ZZ_2^s\,$ \cite{footnote-8}.
We stress that the $G_s^0$ in (\ref{eq:Gs0-k}), as the 3d representation of $\ZZ_2^s$ ,
is uniquely fixed by the $\mutau$ symmetric matrix $\,V_s\,$; we call
$\,\ZZ_2^s\,$ a {\it hidden symmetry} since it generally exists for any
$\mutau$ symmetric neutrino mass-matrix $M_\nu^{(s)}$
[cf.\ Eq.\,(\ref{eq:Ms-dMa}) below], i.e.,
{\it any $\mutau$ symmetric neutrino sector must
automatically contain the hidden $\,\ZZ_2^s\,$ symmetry
which dictates the solar angle $\theta_s$ as in (\ref{eq:t12-k}).}

As pointed out in Ref.\,\cite{GHY}, a particular choice of
$\,k=\f{1}{\sqrt{2}\,}\,$
gives the conventional tri-bimaximal ansatz\,\cite{TBM}
$\,\tan\theta_s=\f{1}{\sqrt{2}\,}\,$ ($\theta_s\simeq 35.3^\deg$),
but {\it other choices of the group parameter $\,k\,$
allow deviations from the conventional tri-bimaximal mixing,} e.g.,
we can make a very simple choice of $\,k=\f{2}{3}\,$,\,
leading to $\,\tan\theta_s=\f{2}{3}\,$ ($\theta_s\simeq 33.7^\deg$),
which agrees to the neutrino data equally well (cf.\ Table-1 in Sec.\,2)
or even better (cf.\ Table-2 in ``Note Added in Proof").
The $\ZZ_2^s$ itself, as {\it the minimal hidden symmetry for $\,\ts$\,},\,
is not restrictive enough
to fix its group parameter $\,k$\,.\, But, extending the $\,\ZZ_2^{\mmutau}\ot\ZZ_2^s\,$
symmetry into a larger simple group can fix a particular $\,k\,$ value and thus the
solar angle $\,\ts$\,.\,
As we demonstrated in Sec.\,6.3 of Ref.\,\cite{GHY}, a simple example is to enlarge
$\,\ZZ_2^{\mmutau}\ot\ZZ_2^s\,$ to the permutation group $S_4$ \cite{S4},
under which we can infer $\,k=\f{1}{\sqrt{2}\,}\,$,\, corresponding to the tri-bimaximal mixing
$\,\theta_s=\arctan\f{1}{\sqrt{2}\,}\,$.

Then, we examine how such a $\ZZ_2^s$ symmetry could possibly survive
after including general $\mutau$ breaking terms in $\,V'=V_s^{}+\d V'\,$.\,
Expanding the small $\mutau$ breaking parameters up to NLO,
we can derive the symmetry-transformation matrix $G_s$ corresponding to
$\,d_\nu^{(2)}\,$ of (\ref{eq:dnu-1-2}),
\beqa
\label{eq:Gs=Gs0+dGs}
 G_s &~\equiv~& G_2 ~=~ V'd_\nu^{(2)} V'^\dag
    ~=~  V_s^{}d_\nu^{(2)}V_s^\dag
       + (V_s^{}d_\nu^{(2)}\delta V^{\prime\dag}
       +\delta V' d_\nu^{(2)} V_s^\dag )
 \nonumber\\[0mm]
   &\!\equiv\!&  G_s^0 + \delta G_s \,,
\eeqa
where $\,\d G_s = \textrm{Re}[\d G_s] + i\textrm{Im}[\d G_s]\,$
with
\beqs
\label{eq:RE-IM-dGs}
\beqa
\label{eq:RE-dGs}
\textrm{Re}[\delta G_s]
& ~=~ &
\left(\!\!\!\!
\begin{array}{ccc}
0
& -\frac{s_{2s}^{} \delta _a+2  c_s^2 \cos\delta_D\delta _x}{\sqrt{2}}
& \frac{s_{2s}^{} \delta _a+2 c_s^2 \cos\delta_D \delta _x}{\sqrt{2}}
\\[2mm]
-\frac{s_{2s}^{}\delta _a+2c_s^2  \cos\delta_D \delta _x}{\sqrt{2}}
& -2 s_s^2 \delta_a \!-\! s_{2s}^{} \cos\delta_D \delta _x
& 0
\\[2mm]
 \frac{s_{2s}^{} \delta_a+2c_s^2  \cos\delta_D \delta _x}{\sqrt{2}}
 & 0 & 2 s_s^2 \delta_a \!+\! s_{2s}^{}\cos\delta_D  \delta _x
\end{array}
\!\!\!\right) \!,
\\[4mm]
\label{eq:IM-dGs}
\textrm{Im}[\delta G_s]
& \!\!\!\!=\!\!\!\!\! &
\left(\!\!\!
\begin{array}{ccc}
0 & \frac{s_{2s}^{}(\delta \alpha_1-\delta \alpha_2)-2 c_s^2 \sin\delta_D \delta _x}{\sqrt{2}}
  & \frac{s_{2s}^{}(\delta \alpha_1-\delta \alpha_3)+2 c_s^2 \sin\delta_D \delta _x}{\sqrt{2}}
\\[2mm]
 -\frac{s_{2s}^{}(\delta \alpha_1-\delta \alpha_2)-2 c_s^2 \sin\delta_D \delta _x}{\sqrt{2}}
 & 0 & s_s^2(\delta \alpha_2\!-\!\delta \alpha_3) \!+\! s_{2s}^{}\sin\delta_D \delta _x
\\[2mm]
 -\frac{s_{2s}^{}(\delta \alpha_1-\delta \alpha_3)+2 c_s^2 \sin\delta_D \delta _x}{\sqrt{2}}
 & -s_s^2(\delta\alpha_2\!-\!\delta \alpha_3)\!-\! s_{2s}^{}\sin\delta_D \delta _x
 & 0
\end{array}
\!\!\!\!\right) \!,
\nonumber\\[-1mm]
\eeqa
\eeqs
where $\,s_{2s}^{}\equiv \sin 2\theta_s\,$.\,
Because the symmetry transformation $\,G_s\in \ZZ_2^s\,$,\,
we have the condition $\,G_s^2 = {\cal I}_3\,$.\,
Then, expanding this up to the NLO, we have verified the consistency condition, 
\beqa
\label{eq:cond-dGs}
\{G_s^0,\,\delta G_s\} ~=~ 0 \,.
\eeqa
Requiring that the $\,\ZZ_2^s$\, symmetry persists under $\mutau$ breaking, i.e.,
the form of $\,G_s\,$ remains unaffected by the $\mutau$ violation, we have
the condition,  
\beqa
\label{eq:Gs=Gs0}
 G_s^{} \,=\, G_s^0 \,, ~~~~\text{or,}~~~~ \delta G_s^{} \,=\, 0 \,. 
\eeqa
Thus, with (\ref{eq:RE-dGs})-(\ref{eq:IM-dGs}), 
we can derive the following solutions, 
\beqa
\label{eq:Re-dGs-sol}
\f{\da}{\dx} ~=\, -\cot\theta_s  \cos\delta_D \,,
\eeqa
from the real part condition $\,\textrm{Re}[\delta G_s^{}] = 0\,$, and
\beqs
\label{eq:Im-dGs-sol}
\beqa
\label{eq:Im-dGs-sol-1}
2\delta\alpha_1^{} &=& \delta\alpha_2^{} + \delta\alpha_3^{} \,,
\\[1.5mm]
\label{eq:Im-dGs-sol-2}
\delta\alpha_2^{}-\delta\alpha_3^{} &=&\! -2\cot\ts\sin\d_D\,\dx
~=~ 2\tan\d_D^{}\,\da \,,
\eeqa
\eeqs
from the imaginary part condition 
$\,\textrm{Im}[\delta G_s^{}] = 0\,$,\,  where in the last step of
(\ref{eq:Im-dGs-sol-2}) we have made use of (\ref{eq:Re-dGs-sol}) for simplification.
Note that the correlation (\ref{eq:Re-dGs-sol})
precisely agrees to what derived from our soft breaking model in
Eq.\,(4.12a) of Ref.\,\cite{GHY}; but now it is re-derived by {\it requiring that the
$\ZZ_2^s$ symmetry persists in the presence of general low energy $\mutau$ breaking.}
In addition, the above Eq.\,(\ref{eq:Im-dGs-sol-2}) also coincides with Eq.\,(4.12b)
of Ref.\,\cite{GHY}.
As we will demonstrate in the next subsection, the $\ZZ_2^s$ symmetry 
is independent of the soft $\mutau$
breaking in the seesaw model of Ref.\,\cite{GHY}.
We note that in the current construction of common $\mutau$ and CP breaking with seesaw
mechanism (Sec.\,2.\,2), such a $\ZZ_2^s$ symmetry is not fully respected, hence the correlation
(\ref{eq:Re-dGs-sol}) no longer holds and we have predicted a {\it modified correlation
(\ref{eq:da-dx-1}), which can be tested against (\ref{eq:Re-dGs-sol}) by the on-going and
upcoming neutrino oscillation experiments.}

To summarize, as we have demonstrated above from general low energy reconstruction formulation,
the transformations $\,G_{\mu\tau}=G_1^{}\,$ and $\,G_s =G_2^{}\,$ in the $\mutau$ symmetric limit
correspond to the discrete groups $\,\ZZ_2^{\mmutau}\otimes\ZZ_2^s\,$,\, which are {\it equivalent
to} and {\it originate from}
the generic symmetry $\,\ZZ_2\otimes\ZZ_2\,$
in the neutrino mass-eigenbasis because they are connected
by the similarity transformations via (\ref{eq:Z2-ma}).
The $\mutau$ symmetry $\,\ZZ_2^{\mmutau}\,$ has been known before, and the hidden symmetry
$\,\ZZ_2^s\,$ (as {\it the minimal group dictating the solar angle $\theta_s$}) was revealed
by Ref.\,\cite{GHY} in the context of neutrino seesaw.
In this work, we further find that requiring the symmetry $\,\ZZ_2^s\,$ to persist in the presence of
most general $\mutau$ breaking terms will predict a new correlation (\ref{eq:Re-dGs-sol})
between the small $\mutau$ breaking parameters
$\,(\d_a,\,\d_x)\,$.\,
As we will prove below, the $\,\ZZ_2^s\,$ symmetry is respected by a class
of soft $\mutau$ breaking seesaw models in Ref.\,\cite{GHY}, but is {\it partially
violated} in the present $\mutau$ breaking seesaw model (Sec.\,2.\,2).

\vspace*{2mm}
\subsubsection{\,$\ZZ_2^s$ Symmetry and Neutrino Mass-Matrix with General $\bd{\mutau}$ Breaking}
\label{sec:5.2.2}

In this subsection,
we directly analyze the generally reconstructed light neutrino mass-matrix $M_\nu$ under
the hidden symmetry $\,\ZZ_2^s\,$ and the determination of solar angle $\,\theta_s\,$.\,
The mass-matrix (\ref{eq:Mnu33}) can be {\it uniquely} decomposed into the
$\mutau$ symmetric and anti-symmetric parts,
\beqa
\label{eq:Mnu=Ms+dMa}
M_\nu^{} &=&
M_\nu^{(s)}+\delta M_\nu^{(a)} \,,
\eeqa
with
\beqs
\beqa
\label{eq:Ms-dMa}
&&
M_\nu^{(s)}~ =~
\begin{pmatrix}
A  & B_s &  B_s
  \\[0mm]
   &   C_s  & D
  \\[0mm]
   & & C_s
\end{pmatrix} \!,
~~~~~~~
\delta M_\nu^{(a)} ~=~
\begin{pmatrix}
     0  & \delta B_a  &  - \delta B_a
  \\[0mm]
        & \delta C_a  & 0
  \\[0mm]
        & & - \delta C_a
\end{pmatrix} \!,
\\[1.5mm]
&&
B_{s}  ~\equiv~ \hf \({B_1 + B_2}\) \,,
~~~~~
C_{s}  ~\equiv~ \hf\({C_1 + C_2}\) \,,
\\[1.5mm]
&&
\d B_{a}  ~\equiv~ \hf\({B_1 -  B_2}\) \,,
~~~~~
\d C_{a}  ~\equiv~ \hf\({C_1 -  C_2}\) \,,
\eeqa
\eeqs
where we generally allow $\,\ma\mb\mc\neq 0\,$.\,
Then, from (\ref{eq:G-M}),
the invariance equation of $M_\nu$ under $G_s^{}$ corresponds to
\beqa
G_s^\dag (M_\nu^{(s)}+\delta M_\nu^{(a)}) G_s
~=~ M_\nu^{(s)}+\delta M_\nu^{(a)} \,,
\eeqa
which uniquely gives,
\beqs
\beqa
\label{eq:Gs-Ms}
G_s^\dag M_\nu^{(s)}G_s &~=~& M_\nu^{(s)} \,,
\\
\label{eq:Gs-Ma}
G_s^\dag \delta M_\nu^{(a)}G_s &\!=\!& \delta M_\nu^{(a)} \,.
\eeqa
\eeqs
Note that two possibilities may exist:
{\bf (i)}.\ The $\ZZ_2^s$ symmetry is a {\it full symmetry} of
the light neutrino mass-matrix $M_\nu$
if {\it both (\ref{eq:Gs-Ms}) and (\ref{eq:Gs-Ma}) hold.}
{\bf (ii)}.\ The $\ZZ_2^s$ symmetry is a {\it partial symmetry} of $M_\nu$
if the $\mutau$ anti-symmetric part $\,M_\nu^{(a)}\,$ breaks (\ref{eq:Gs-Ma}).

We can prove that the $\ZZs$ is always a symmetry of the
$\mutau$ symmetric part $\,M_\nu^{(s)}\,$ and generally holds (\ref{eq:Gs-Ms}).
Substituting (\ref{eq:V'=Vs+dV'}) into (\ref{eq:Mnu-V'-Dt}) and noting that
the decomposition (\ref{eq:Mnu=Ms+dMa}) is unique,
we can reconstruct the $\mutau$ symmetric and anti-symmetric parts of $M_\nu$,
respectively,
\beqs
\beqa
M_\nu^{(s)} &~=~& V_s^* \Dt_\nu V_s^\dag \,,
\\[1mm]
\d M_\nu^{(a)} &\!=\!&
       V_s^* \Dt_\nu \d V^{\pp\dag} +
\d V^{\pp *} \Dt_\nu V_s^\dag +
\d V^{\pp *} \Dt_\nu \d V^{\pp\dag}
\non\\
&\!=\!&
       V_s^* \Dt_\nu \d V^{\pp\dag} +
\d V^{\pp *} \Dt_\nu V_s^\dag + O(\d_j^2) \,,
\eeqa
\eeqs
where $\d_j$ denotes all possible NLO parameters under consideration (such as
$\d_x$, $\d_a$ and $y'$, etc).
This shows that the $\mutau$ symmetric part $M_\nu^{(s)}$ is diagonalized
by $V_s$.\, Hence, the corresponding $\ZZs$ transformation matrix is just
$\,G_s^0\,$,\, as given by  (\ref{eq:Gs0-all}).
The $\,G_s^0\,$ must be the symmetry of $M_\nu^{(s)}$ and thus
always holds the invariance equation (\ref{eq:Gs-Ms}). 
This proves that the
{\it solar mixing angle $\,\ts\,$ (as contained in the rotation matrix $V_s$ and symmetry
transformation matrix $G_s^0$) is generally dictated by the $\,\ZZs\,$ symmetry,
independent of any specific model.}

On the other hand, the validity of (\ref{eq:Gs-Ma}) is highly nontrivial
because {\it the requirement of $\,G_s^{}=G_s^0\,$ [cf. (\ref{eq:Gs=Gs0})] 
does not generally hold under $\mutau$ breaking,} and it has to be checked case by case.
As we will prove in Sec.\,5.\,3, the $\mutau$ anti-symmetric part $M_\nu^{(a)}$
will break $\,\ZZs\,$ in the current $\mutau$ blind seesaw (Sec.\,2),
while it preserves $\,\ZZs\,$ in the soft $\mutau$ breaking seesaw of Ref.\,\cite{GHY}.

Using the expression of $G_s^{}$ [Eqs.\,(\ref{eq:Gs0}) and (\ref{eq:Gs=Gs0})],
we can derive the solution from (\ref{eq:Gs-Ms}) for the $\mutau$ symmetric part,
\beqa
\label{eq:t12-sol-s}
\tan 2\theta_s ~=~ \frac{2\sqrt{2}~B_s}{\,A-(C_s\!+\!D)\,} \,,
\eeqa
and another solution from (\ref{eq:Gs-Ma}) for the $\mutau$ anti-symmetric part,
\beqa
\label{eq:t12-sol-a0}
\tan\theta_s ~=\,
-\sqrt{2}\frac{~\delta B_a}{\delta C_a} \,,
\eeqa
which further leads to,
\beqa
\label{eq:t12-sol-a}
\tan 2\theta_s ~=\,
-\frac{\,2\sqrt{2}\,\delta B_a\delta C_a\,}{\,\delta C_a^2-2\delta B_a^2\,} \,.
\eeqa
Hence, {\it if}\, the $\,\ZZ_2^s\,$ would be a full symmetry of $\,M_\nu\,$ (including
its $\mutau$ breaking part),
the two solutions (\ref{eq:t12-sol-s}) and (\ref{eq:t12-sol-a})
for the solar angle $\,\theta_s\,$ must be identical,
leading to a {\it nontrivial consistency condition,}
\beqa
\label{eq:t12-consistencyCD}
\tan 2\theta_s ~=~
\frac{2\sqrt{2}~B_s}{A-(C_s+D)}
~\,\doteqdot~
-\frac{\,2\sqrt{2}\,\delta B_a\delta C_a\,}{\,\delta C_a^2-2\delta B_a^2\,}\,.
\eeqa
An explicit counter example to this condition will be given in Sec.\,5.\,3.\,2.

In the following, we apply the most general reconstruction formalism (Sec.\,3.\,1)
to compute the $\mutau$ symmetric and anti-symmetric parts of light
neutrino mass-matrix $\,M_\nu =M_\nu^{(s)} + M_\nu^{(a)}\,$.
With these, we will explicitly verify Eq.\,(\ref{eq:t12-sol-s}) by using the
elements of $\mutau$ symmetric $\,M_\nu^{(s)}\,$,\,
and we further derive physical consequences of the consistency condition
(\ref{eq:t12-consistencyCD}) by using the elements of
$\,\mutau\,$ anti-symmetric $M_\nu^{(a)}$.

\vspace*{2mm}
\noindent
\underline{\tt $\blacklozenge$\,Reconstruction Analysis for General Normal Mass-Ordering Scheme~}
\\[2mm]
Eq.\,(\ref{eq:Mnu-Reconstruct}) reconstructs
all the elements of $M_\nu$ in terms of three mass-eigenvalues,
three mixing angles and relevant CP-phases.
The normal mass-ordering (NMO) has the spectrum
$\,m_1^{} <m_2^{} \ll m_3^{}\,$,\,
so we can define the small ratios,
\beqa
y_1^{} \,\equiv\, \f{\ma}{\mc}\,, ~~~~~
y_2^{} \,\equiv\, \f{\mb}{\mc}\,, ~~~~~
y_3^{} \,\equiv\, \f{\,\mc \!-\! m_{30}^{}\,}{m_{3}^{}}\,.
\eeqa
Thus we have the independent NLO parameters for the NMO analysis,
$\, (y_1,~y_2,~z,~\d_a,~\d_x,~\d\a_i^{},~\d\phi_i^{})\,$.\,
Expanding them perturbatively, we derive the LO form of
the $\mutau$ symmetric mass-matrix $M_\nu$,
\begin{eqnarray}
\label{eq:NMO-Mnu-0th}
M_\nu^{(0)}
~=~ m_{30}^{}
\left(
\begin{array}{rrr}
 0 & 0 & 0
\\
   &  \mbox{$\hf$} & -\mbox{$\hf$}
\\[1mm]
   &    &  \mbox{$\hf$}
\end{array}
\right) \!,~~~~~~
  \end{eqnarray}
with
$\,
\alpha_{10}^{}=\alpha_{20}^{}=\alpha_{30}^{}\equiv\alpha_{0}^{}\,,~~
\alpha_{30}^{}+\phi_{30}^{}=n\pi
$\,,\,
and the NLO elements in $\,\d M_\nu$\,,
\beqs
\label{eq:NMO-Ms}
\begin{eqnarray}
\label{eq:NMO-Ms-A}
&& \delta A ~=~
e^{-i2\alpha_{0}^{}}
\(e^{-i2\phi_{10}^{}}c_s^2y_1^{} + e^{-i2\phi_{20}^{}}s_s^2y_2^{}\) m_{30}^{} \,,
\\[0mm]
\label{eq:NMO-Ms-Bs}
&& \delta B_s ~=~
\mbox{$\f{1}{2\sqrt{2}\,}$}
{e^{-i2\alpha_{0}^{}}
\left(e^{-i2\phi_{10}^{}} y_1^{} - e^{-i2\phi_{20}^{}}y_2^{}\right)
\sin 2\theta_{s}}\,m_{30}^{}\,,
\\[0mm]
\label{eq:NMO-Ms-CsD}
&& \delta C_s+\delta D ~=~
e^{-i2\alpha_{0}^{}}
\left(e^{-i2\phi_{10}^{}}s_s^2y_1^{} + e^{-i2\phi_{20}^{}}c_s^2y_2^{}\right)
m_{30}^{} \,,
\\[0mm]
\label{eq:NMO-Ma-Ba}
&& \delta B_a ~=~
\mbox{$\f{1}{\sqrt{2}\,}$}
e^{-i\d_D^{}} \d_x\, m_{30}^{} \,,
\\[0mm]
\label{eq:NMO-Ma-Ca}
&& \delta C_a ~=~
\mbox{$\f{1}{2}$}
\[2 \delta _a - i \(\d\a_2^{} - \d\a_3^{}\)\] m_{30}^{} \,.
  \end{eqnarray}
\eeqs
From (\ref{eq:NMO-Mnu-0th}), we have
$\,A^{(0)}_s = B_s^{(0)} =C_s^{(0)}\!+ D^{(0)}_{}=0\,$.\,
Thus, using the $\mutau$ symmetric NLO elements (\ref{eq:NMO-Ms-A})-(\ref{eq:NMO-Ms-CsD}),
we can compute the ratio,
\begin{eqnarray}
\label{eq:NMO-tan2ts-0}
\frac{2\sqrt{2}\,B_s}{\,A_s-(C_s\!+\!D)\,}
~=~ \f{(e^{-i2\phi_{10}^{}} y_1^{} - e^{-i2\phi_{20}^{}}y_2^{})\sin 2\theta_{s}}
      {\,(e^{-i2\phi_{10}^{}} y_1^{} - e^{-i2\phi_{20}^{}}y_2^{})(c_s^2- s_s^2)\,}
~=~ \tan2\theta_{s} \,,
\end{eqnarray}
which explicitly verifies our Eq.\,(\ref{eq:t12-sol-s}) [as generally derived from the
invariance equation (\ref{eq:Gs-Ms}) under $\ZZs$] for the current NMO scheme.
This is an explicit proof up to NLO that for a general NMO scheme the $\mutau$ symmetric mass-matrix
$\,M_\nu^{(s)} = M_\nu^{(0)} + \d M_\nu^{(s)}\,$
does hold the $\,\ZZ_2^s\,$ symmetry.

Then, using the $\mutau$ anti-symmetric elements (\ref{eq:NMO-Ma-Ba})-(\ref{eq:NMO-Ma-Ca}),
we derive the ratio,
\begin{eqnarray}
\label{eq:NMO-Ba/Ca}
-\sqrt{2}\frac{\delta B_a}{\delta C_a}
~=\, -\frac{e^{-i\delta_D}\delta_x}
           {\,\d_a - \mbox{$\f{i}{2}$}(\d\alpha_2^{} \!-\! \d\alpha_3^{})\,}
~=~ \tan\ts \,,
\end{eqnarray}
where in the last step we have used Eq.\,(\ref{eq:t12-sol-a0}) under {\it the assumption}
that $\,\ZZs\,$ symmetry also holds for the $\mutau$ anti-symmetric mass-matrix
$\,M_\nu^{(a)}\,$,\, i.e., the validity of the invariance equation (\ref{eq:Gs-Ma}).
Analyzing the real and imaginary parts of (\ref{eq:NMO-Ba/Ca}),
we deduce two relations,
\beqs
\beqa
\label{eq:NMO-da-dx}
\da  &=& - \dx\,\cot\ts^{}\cos\d_D \,,
\\
\label{eq:NMO-da2-da3}
\d\a_2^{}-\d\a_3^{} &=& 2\tan\d_D\,\da \,.
\eeqa
\eeqs
These are in perfect agreement with (\ref{eq:Re-dGs-sol}) and (\ref{eq:Im-dGs-sol-2}),
which are generally derived under a single assumption that
the $\,\ZZs\,$ symmetry persists in the presence of $\mutau$ breaking.
But, as will be shown in Sec.\,5.\,3.\,2, {\it this assumption does not generally hold,
and the current $\mutau$ blind seesaw (Sec.\,2.\,2) provides a nontrivial counter example.}

\vspace*{2mm}
\noindent
\underline{\tt $\blacklozenge$\,Reconstruction Analysis for General Inverted Mass-Ordering Scheme~}
\\[2mm]
For the inverted mass-ordering (IMO), the light neutrinos have the spectrum
$\,\mb\gtrsim \ma \gg \mc\,$,\, so we can define the small ratios,
\beqa
z_1^{} ~\equiv~\f{\ma - m_0^{}}{m_1^{}}\,, ~~~~~
z_2^{} ~\equiv~\f{\mb - m_0^{}}{m_1^{}}\,, ~~~~~
z_3^{} ~\equiv~\f{\mc}{m_1^{}}\,,
\eeqa
where we have $\,z_1^{}=z\,$ and $\,z_2^{}\simeq z+\hf y'\,$ in connection to
the NLO parameters $(y',\,z)$ introduced in
Eqs.\,(\ref{eq:definition-y})-(\ref{eq:z-y}) of Sec.\,3.\,2.
Then we have the independent NLO parameters for the IMO analysis,
$\,
(z_1^{},~z_2^{},~z_3^{},~\da,~\dx,~\d\a_i^{},~\d\phi_i^{}).
$\,
Expanding them perturbatively, we derive the LO form of the symmetric mass-matrix $M_\nu$,
\begin{eqnarray}
\label{eq:IMO-Mnu-0th}
M_\nu^{(0)} ~=~ m_{0}^{}
\left(
\begin{array}{ccc}
 1 & 0 & 0 \\
   & \mbox{$\hf$} & \mbox{$\hf$} \\[1mm]
   &   & \mbox{$\hf$}
\end{array}
\right) ,
  \end{eqnarray}
with
$\,
\alpha_{10}^{}=\alpha_{20}^{}=\alpha_{30}^{}=\alpha_{0}^{}\,,\,
~\,\phi_{10}^{}=\phi_{20}^{}=-\alpha_{0}^{}\,$,\,
and the NLO elements of $M_\nu$,
\beqs
\begin{eqnarray}
\label{eq:IMO-Ms-A}
&&
\delta A ~=~ m_0^{}
\[c_s^2 z_1^{} + s_s^2 z_2^{}
  -i2(c_s^2 \d\phi_1^{} + s_s^2\d\phi_2^{} +\d\alpha_1^{})\] ,
\\[0mm]
\label{eq:IMO-Ms-Bs}
&&
\delta B_s ~=~ \mbox{$\f{1}{2\sqrt{2}\,}$} m_0^{}\sin 2\theta_{s}
\[z_1^{}-z_2^{} - i2\(\d\phi_1^{} - \d\phi_2^{}\)\] ,
\\[0mm]
\label{eq:IMO-Ms-CsD}
&&
\delta C_s+\delta D ~=~ m_0^{}
\[s_s^2 z_1^{} + c_s^2 z_2^{} -i\(2s_s^2\d\phi_1^{} +2c_s^2\d\phi_2^{}
  + \d\alpha_2^{} +\d\alpha_3^{}\)\] ,
\\[0mm]
\label{eq:IMO-Ma-Ba}
&&
\delta B_a ~=\, -\mbox{$\f{1}{\sqrt{2}\,}$} m_0^{}e^{i\d_D}\,\dx \,,
\\[0mm]
\label{eq:IMO-Ma-Ca}
&&
\delta C_a ~=\, -m_0^{}
\[\da + \mbox{$\f{i}{2}$} \(\d\alpha_2^{}-\d\alpha_3^{}\)\] .
\end{eqnarray}
\eeqs
From (\ref{eq:IMO-Mnu-0th}), we have
$\,B_s^{(0)}=0\,$ and
$\,A^{(0)}_s-(C_s^{(0)}\!+\! D^{(0)}_{})=0\,$.\,
So using the $\mutau$ symmetric NLO elements (\ref{eq:IMO-Ms-A})-(\ref{eq:IMO-Ms-CsD}),
we can compute the ratio,
\begin{eqnarray}
\label{eq:IMO-tan2ts-0}
\frac{2\sqrt{2}~B_s}{A_s-(C_s+D)} ~=~
\frac{\sin 2\ts \[z_1^{}-z_2^{} - i2\(\d\phi_1^{} - \d\phi_2^{}\)\]}
     {\,\cos 2\ts \[ z_1^{}-z_2^{} -i2\(\d\phi_1^{} - \d\phi_2^{}\)\]
     -i\( 2\d\alpha_1^{} -\d\alpha_2^{} -\d\alpha_3^{}\)\,} \,,
\end{eqnarray}
from which we deduce the consistent solution,
\beqs
\begin{eqnarray}
\label{eq:IMO-Ms-tan2t12}
&&
\frac{2\sqrt{2}\,B_s}{\,A_s-(C_s\!+\!D)\,} ~=~
\tan 2\theta_s \,,
\\[2mm]
\label{eq:IMO-Ms-alpha123}
&&
2\d\alpha_1^{} ~=~ \d\alpha_2^{} +\d\alpha_3^{} \,,
\end{eqnarray}
\eeqs
which explicitly verifies our Eq.\,(\ref{eq:t12-sol-s}) [as generally derived from the
invariance equation (\ref{eq:Gs-Ms}) under $\ZZs$] for the current IMO scheme.
Also the above solution (\ref{eq:IMO-Ms-alpha123}) exactly coincide with the general
Eq.\,(\ref{eq:Im-dGs-sol}).
The above is an explicit proof up to NLO that for a general IMO scheme
the $\mutau$ symmetric mass-matrix
$\,M_\nu^{(s)} = M_\nu^{(0)} + \d M_\nu^{(s)}\,$
does hold the $\,\ZZ_2^s\,$ symmetry.

Then, with the $\mutau$ anti-symmetric elements (\ref{eq:IMO-Ma-Ba})-(\ref{eq:IMO-Ma-Ca}),
we further evaluate the ratio,
\begin{eqnarray}
\label{eq:IMO-Ba/Ca}
-\sqrt{2}\frac{\delta B_a}{\delta C_a} ~=\,
-\frac{e^{i\d_D}\,\dx}{\da +\mbox{$\f{i}{2}$}\(\d\alpha_2^{}-\d\alpha_3^{}\)}
~=~ \tan\ts \,,
\end{eqnarray}
where in the last step we have applied (\ref{eq:t12-sol-a0}) under {\it the assumption}
that the $\mutau$ anti-symmetric mass-matrix $\,M_\nu^{(a)}\,$ also respects the
$\,\ZZs\,$ symmetry, i.e., the invariance equation (\ref{eq:Gs-Ma}) holds.
Inspecting the real and imaginary parts of (\ref{eq:IMO-Ba/Ca}),
we deduce the following,
\beqs
\beqa
\label{eq:IMO-da-dx}
\da  &~=~& - \dx\,\cot\ts^{}\cos\d_D \,,
\\
\label{eq:IMO-da2-da3}
\d\a_2^{}-\d\a_3^{} &=& 2\tan\d_D\,\da \,,
\eeqa
\eeqs
which coincide with Eqs.\,(\ref{eq:NMO-da-dx})-(\ref{eq:NMO-da2-da3}) as we derived
for the NMO scheme. We see that both (\ref{eq:IMO-da-dx})-(\ref{eq:IMO-da2-da3})
and (\ref{eq:NMO-da-dx})-(\ref{eq:NMO-da2-da3})
precisely agree to (\ref{eq:Re-dGs-sol}) and (\ref{eq:Im-dGs-sol-2})
which are generally derived under {\it a single assumption that
$\,\ZZs\,$ is a symmetry of the full mass-matrix
$\,M_\nu =M_\nu^{(s)}+\d M_\nu^{(a)}\,$ including its $\mutau$ breaking part
$\,\d M_\nu^{(a)}\,$.}\,
But, as we will prove in Sec.\,5.\,3.\,2, the above assumption
is not generally true and for the $\mutau$ blind seesaw with IMO (Sec.\,2.\,2)
the $\,\ZZs\,$ symmetry is violated by $\,\d M_\nu^{(a)}\,$.

\vspace*{2.5mm}

So far we have explicitly proven the relations
(\ref{eq:Re-dGs-sol}) and (\ref{eq:Im-dGs-sol}) for general NMO and IMO schemes
via the general {\it model-independent} reconstruction formalism (Sec.\,3.\,1), where the only
assumption is that the $\,\ZZs\,$ symmetry fully persists in the presence of $\mutau$ breaking.
In the next subsection, we will map the $\,\ZZ_2^{\mmutau}\otimes\ZZ_2^s\,$
symmetry into the neutrino seesaw Lagrangian, and demonstrate that the hidden $\,\ZZ_2^s\,$
symmetry is a {\it full symmetry} of our soft $\mutau$ breaking model in Ref.\,\cite{GHY}
where the physical prediction (\ref{eq:Re-dGs-sol}) holds;
while for the current $\mutau$ blind seesaw model the $\,\ZZ_2^s\,$ is only a
{\it partial symmetry} (respected by the $\mutau$ symmetric part $\,M_\nu^{(s)}\,$),
and is violated by the $\mutau$ anti-symmetric part $\,\d M_\nu^{(a)}$,\,
leading to our prediction of the modified new correlation (\ref{eq:da-dx-1}) in Sec.\,4.\,1,
in contrast to (\ref{eq:IMO-da-dx}) or (\ref{eq:Re-dGs-sol}).

\vspace*{3mm}
\subsection{Mapping $\,{\ZZ_2\otimes\ZZ_2}\,$ Hidden Symmetry into Neutrino Seesaw}
\label{sec:5.3}

Consider the general seesaw Lagrangian in the form of (\ref{eq:L-seesaw})
with two or three right-handed neutrinos. After spontaneous electroweak symmetry breaking,
consider the invariance of (\ref{eq:L-seesaw})
under the residual symmetry 
transformations,
\beqa
\label{eq:Z2-s}
  \nuL \,\to\,  G_j^{} \,\nuL \,,
&&~~~
  \mathcal{N} \,\to\, G_j^R \mathcal{N} \,,
\eeqa
where $G_j$ is 3-dimensional unitary matrix, and $G_j^R$
is $2\times 2$ or $3\times 3$ matrix (depending on two or three right-handed neutrinos
invoked in the neutrino seesaw).
Accordingly, we have the following invariance equations
for the Dirac and Majorana neutrino mass-matrices,
\beqs
\label{eq:G-GR-mDMR}
\beqa
\label{eq:G-mD}
  G_j^T \mD G_j^R &~=~& \mD \,,
\\
\label{eq:GR-MR}
  {G_j^R}^T M_R^{} G_j^R &\!=\!& M_R^{} \,,
\eeqa
\eeqs
from which we deduce the invariance equation for the seesaw mass-matrix of light neutrinos,
\beqa
\label{eq:G-Mnu}
G_j^T M_\nu G_j^{} ~=~ M_\nu \,,
\eeqa
where $\,M_\nu = \mD M_R^{-1} m_D^T\,$.\,
Let us diagonalize the Majorana mass-matrices $M_\nu$ and $M_R^{}$ as follows,
\beqa
\label{eq:Mnu-MR-diag}
U_\nu^T M_\nu U_\nu^{} \,=\, D_\nu\,,
&~~~~&
U_R^T M_R^{} U_R^{} \,=\, D_R^{} \,,
\eeqa
in which $\,D_\nu = \textrm{diag}(\ma,\,\mb,\,\mc)\,$ and
$\,D_R^{} =\,\textrm{diag}(M_1,\cdots\!,M_n)$\, with $n=2$ for the minimal seesaw
or $n=3$ for three-neutrino-seesaw.
Thus, from (\ref{eq:G-GR-mDMR})-(\ref{eq:Mnu-MR-diag}),
we can express $\,G_j^{}\,$ and $\,G_j^R\,$ as,
\beqa
\label{eq:Gj-GjR}
G_j^{} \,=\, U_\nu^{} d_\nu^{(j)}U_\nu^\dag \,,
&~~~~&
G_j^R \,=\, U_R^{} d_R^{(j)} U_R^\dag \,,
\eeqa
where kernel representation $\,\{d_\nu^{(j)}\}\,$ is given in (\ref{eq:dnu-1-2}),
and corresponds to the product group $\,\ZZmt\ot\ZZs\,$
via the {\it equivalent flavor representation} $\,\{G_j^{}\}\,$
for the light neutrino sector.
For $\,d_R^{(j)}\,$ in (\ref{eq:Gj-GjR}), we give its nontrivial forms,
\beqs
\beqa
 d_R^{(1)}\,=\, \textrm{diag}(-1,\,1)\,,
 \hspace*{4.45cm}
 &~~~~& (\textrm{for minimal seesaw}),
\label{eq:dR2-1}
\\
 d_R^{(1)}\,=\, \textrm{diag}(1, 1, -1)\,,~~~\,
 d_R^{(2)}\,=\, \textrm{diag}(-1, 1, 1)\,,
 &~~~~& (\textrm{for 3-neutrino-seesaw}),~~~~~~~~~~~~
\label{eq:dR3-1-2}
\eeqa
\eeqs
where $\,d_R^{(1)}\,$ forms a $\,\ZZ_2'\,$ symmetry for right-handed neutrinos in
the minimal seesaw, and $\,\{d_R^{(1)},\,d_R^{(2)}\}$ form a product group
$\,\ZZmtp\ot\ZZsp\,$ for right-handed neutrinos in the three-neutrino-seesaw.
The trivial case with $\,d_R^{(j)}\,$ equal to unity matrix
is not listed here which corresponds to the singlet representation $\,G_j^R=\II\,$.\,
Since the low energy oscillation data do not directly enforce a $\,\ZZmtp\,$
symmetry for heavy right-handed neutrinos, we find two possibilities when mapping
the $\,\ZZmt\,$ to the seesaw sector:
{\bf (i)}.\ the right-handed neutrinos have correspondence with the light neutrinos
in each fermion family and transform simultaneously with the light neutrinos
under the $\,\ZZmt\,$ to ensure the invariance equation (\ref{eq:G-mD}); this
means $\,\ZZmtp = \ZZmt\,$.\,
{\bf (ii)}.\ the right-handed neutrinos are singlet of the usual $\,\ZZmt\,$ symmetry
(called ``$\mutau$ blind"), so the extra symmetry $\,\ZZmtp$ in the ${\cal N}$
sector is fully independent of the $\,\ZZmt\,$ for light neutrinos; this means
that under $\,\ZZmt\,$ the invariance equation (\ref{eq:G-mD}) has
$\,G_1\in \ZZmt\,$ for light neutrinos and $\,G^R=\II\,$ for right-handed neutrinos.
As generally shown in Sec.\,5.\,2, the $\,\ZZs\,$ symmetry dictates the solar angle
$\ts$ for light neutrinos. The extra group $\,\ZZsp\,$ in the right-handed neutrino
sector also has two possibilities: one is $\,\ZZsp\,=\ZZs\,$,\, and another is for
the right-handed neutrinos being singlet of the $\,\ZZs\,$ symmetry with $\,G_s^R=\II\,$.

\vspace*{2mm}
\subsubsection{Neutrino Seesaw with Common Soft $\bd{\mutau}$ and CP Breaking}
\label{sec:5.3.1}

In Ref.\,\cite{GHY},
we studied the common soft $\mutau$ and CP breaking in
the minimal neutrino seesaw,  where the right-handed neutrinos
$\,{\cal N}=(N_\mu,\,N_\tau)^T\,$ obeying the same $\,\ZZmt (=\ZZmtp)\,$  at the LO,
and small soft $\mutau$ breaking is uniquely constructed in $M_R^{}$ at the NLO.
In the $\mutau$ symmetric limit, we inferred that the diagonalization matrix
$U_R^{}$ is a $2\times 2$ orthogonal rotation with its rotation angle
$\theta_R\equiv\theta_{23}^R=\f{\pi}{4}$ \cite{GHY}, as expected.
Thus, inputting (\ref{eq:dR2-1}) for $\,d_R^{(1)}\,$,\,
we deduce from (\ref{eq:Gj-GjR}),
\beqa
G_{\mmutau}^R \,=\,
\begin{pmatrix}
0 & 1
\\[1mm]
1 & 0
\end{pmatrix} \!,
\eeqa
which is just the $\ZZ_2^{\mmutau}$ transformation matrix for right-handed neutrinos.
With the two right-handed neutrinos $\,{\cal N}=(N_\mu,\,N_\tau)^T\,$ shown above,
there is no rotation angle $\,\theta_{12}^R\,$ and also no corresponding $\,\ZZsp\,$ symmetry.
So the right-handed neutrinos can only belong to the singlet representation $\,G_s^R=\II_2^{}\,$
under $\,\ZZs\,$ symmetry, with $\,d_R^{(2)}=\II_2\,$.
In our soft $\mutau$ breaking model\,\cite{GHY}, the Dirac mass-matrix,
\beqa
 \mD ~=\,
  \begin{pmatrix}
  a & a \\
  b & c \\
  c & b
  \end{pmatrix} \!,
\eeqa
exhibits the exact $\,\ZZ_2^{\mmutau}\,$ symmetry,
so it should obey the hidden $\,\ZZ_2^s\,$ as well,
\beqa
\label{eq:Gs0-mD}
G_s^T \mD G_s^R \,=\, \mD \,,
\eeqa
where $\,G_s^{}=G_s^0\,$ is given by (\ref{eq:Gs0-all}) and $\,G_s^R=\II_2^{}\,$.\,
This further leads to the invariance equation for the seesaw mass-matrix of light neutrinos,
\beqa
\label{eq:Gs0-seesawMnu}
G_s^T M_\nu G_s ~=~ M_\nu \,,
\eeqa
where $\,M_\nu = \mD M_R^{-1} m_D^T\,$,\, and the invariance equation for $M_R$ is trivial here
since $\,G_s^R=\II_2^{}\,$.\,
[Given the form of $\,G_s^{}=G_s^0\,$ as constructed in (\ref{eq:Gs0-all}),
we can also explicitly verify the equations (\ref{eq:Gs0-mD}) and (\ref{eq:Gs0-seesawMnu}).]
Hence, the group parameter $k$ of $\,\ZZ_2^s\,$ and the corresponding solar angle
$\theta_s$ via Eq.\,(\ref{eq:t12-k}) are fully fixed by the elements of
the $\mutau$ symmetric $\,\mD\,$,\,
and is independent of the soft $\mutau$ breaking in $M_R^{}$ (which is the $\ZZ_2^s$
singlet). This is a general proof based on group theory, without relying on making any expansion of
the $\mutau$ breaking terms in $\,M_R^{}$\,.\,
As can be explicitly solved from Eq.\,(\ref{eq:Gs0-mD}) above,
we have\,\cite{GHY},
\beqa
\label{eq:t12-sol-soft}
\tan\theta_s ~=~ |k| ~=~ \f{\sqrt{2}|a|}{\,|b+c|\,} \,.
\eeqa

As another nontrivial check, we inspect the consistency condition
(\ref{eq:t12-consistencyCD}). With the form of $M_\nu$ in Ref.\,\cite{GHY},
we explicitly verify that (\ref{eq:t12-consistencyCD}) indeed holds,
\beqa
\label{eq:t12-check}
\tan2\theta_s ~=~ \frac{2\sqrt{2}~B_s}{\,A-(C_s\!+\!D)\,}
~=\, -\frac{\,2\sqrt{2}\,\delta B_a\delta C_a\,}{\delta C_a^2-2\delta B_a^2}
~=~ \frac{2\sqrt{2}\,a(b+c)}{\,2a^2-(b+c)^2\,} \,,
\eeqa
where both the $\mutau$ symmetric mass-matrix $M_\nu^{(s)}$ and the
anti-symmetric part $\,\d M_\nu^{(a)}\,$ determine the same solar angle $\,\theta_s\,$.\,
The last equality in (\ref{eq:t12-check}) can be derived also from the solution
(\ref{eq:t12-sol-soft}) above, they are all consistent. Hence, the $\,\ZZ_2^s\,$ is a
full symmetry of the seesaw sector and the light neutrino mass-matrix $M_\nu$
in this soft $\mutau$ breaking model.

We note that this $\,\ZZ_2^s\,$ symmetry has a nice geometric interpretation.
The two vectors, $\,u_1^{}=(a,\,b,\,c)^T\,$ and $\,u_2^{}=(a,\,c,\,b)^T\,$,\,
in the Dirac mass-matrix $\,\mD = (u_1^{},\,u_2^{})\,$,\,
determine a plane $S$, obeying the plane-equation,
\beqa
\label{eq:S-equation}
x - \f{k}{\sqrt{2}\,}(y+z) ~=~ 0\,,
\eeqa
where the parameter $\,k\,$ is given in (\ref{eq:t12-k}).
As shown in Ref.\,\cite{GHY}, the 3-dimensional representation $G_s$
is just the reflection transformation respect to the plane $S$.
For the case of three-neutrino-seesaw, the $\mutau$ symmetric Dirac mass
is extended to a $3\!\times\! 3$ matrix,
\beqa
\label{eq:mD'-3x3}
m_D' ~=
\(\ba{lll}
a' & a & a
\\[1mm]
b' & b & c
\\[1mm]
b' & c & b
\ea\)
=~ (u_0^{},\,u_1^{},\,u_2^{}) \,.
\eeqa
Thus, to hold $\,m_D'\,$ invariant under the $\,\ZZ_2^s\,$ symmetry,
we just need to require its first column $\,u_0^{}=(a',\,b',\,b')^T\,$
to lie in the $S$ plane, i.e.,
\beqa
\label{eq:a'b'-cond}
\f{a'}{\,\sqrt{2}\,b'\,} ~=~ \f{\sqrt{2}\,a}{\,b+c\,} ~=~ k \,,
\eeqa
where $\,k=\tan\ts\,$ as in (\ref{eq:t12-k}).
This means that the Dirac mass matrix (\ref{eq:mD'-3x3}) only contains one more
independent parameter than that of the minimal seesaw; furthermore, $\,m_D'\,$
is rank-2 and thus $\,\det M_\nu = (\det m_D')^2(\det M_R)^{-1}=0\,$
always holds, as in the minimal seesaw.

\vspace*{2mm}
\subsubsection{\,$\bd{\mutau}$ Blind Seesaw with Common $\bd{\mutau}$ and CP Breaking}
\label{sec:5.3.2}

As constructed in Sec.\,2, the $\mutau$ blind seesaw defines the right-handed neutrinos
${\cal N}$ as singlet of $\ZZmt$ symmetry. This means that we must have the
$\ZZmt$ transformation matrix $\,G_{\mmutau}^R=\II_2\,$ and
$\,d_R^{\mmutau}=d_R^{(2)}=\II_2\,$.
Consider the general Dirac and Majorana mass-matrices in the minimal seesaw,
\beqa
 \mtt_D^{} ~=\,
  \begin{pmatrix}
  \att & \att' \\[1.5mm]
  \bta & \cta \\[1.5mm]
  \btb & \ctb
  \end{pmatrix} \!,~~~~~~~
  \Mtt_R ~=\,
  \begin{pmatrix}
  M_{11} & M_{12} \\[1.5mm]
  M_{12} & M_{22}
  \end{pmatrix} \!.
\eeqa
The Majorana mass-matrix $\Mtt_R$ can be diagonalized by the unitary rotation $\,U_R\,$,
\beqa
U_R^T \Mtt_R U_R^{} ~=~ M_R ~\equiv~ \textrm{diag}(M_1,\,M_2)\,,
\eeqa
Then we can derive the seesaw mass-matrix for light neutrinos,
\begin{eqnarray}
  M_\nu^{} &~\simeq~&
 \mtt_D^{} \Mtt_R^{-1} \mtt_D^T
 ~=~ m_DM_R^{-1} m_D^T \,,
\end{eqnarray}
where $\,\mD = \mtt_D^{} U_R\,$ takes the form as in (\ref{eq:mD-2}).
For the $\mutau$ blind seesaw with ${\cal N}$ being $\ZZmt$ singlet,
we can always start with the mass-eigenbasis of ${\cal N}$
with $\,M_R^{}=\textrm{diag}(M_1,\,M_2)\,$,\, which means that the rotation
$U_R^{}$ becomes automatically diagonal and real, $\,U_R=\II_2^{}\,$.\,
Then, the extra symmetry $\,\ZZ_2^{\prime}\,$ of $\,M_R\,$
must be independent of the $\ZZmt$ of light neutrinos, i.e.,
$\,\ZZ_2^{\prime}\neq \ZZmt\,$.\,
So the natural choice is $\,\ZZ_2^{\prime}=\ZZ_2^s\,$.\,
The $\ZZ_2^{\prime}$ can have a nontrivial
$\,d_R^s = d_R^{(1)}=\textrm{diag}(-1,\,1)\,$ as in (\ref{eq:dR2-1}).
Thus,  the corresponding symmetry transformation for $\,\Mtt_R^{}\,$ is
\beqa
G_s^{ R} ~=~ U_R^{}\, d_R^s U_R^\dag
~=~ d_R^{(1)} ~=~ \textrm{diag}(-1,\,1) \,.
\eeqa
There is also a singlet representation of $\ZZ_2^{\prime}$, corresponding to
$\,d_R^{} =\II_2^{}$\,.

Then, let us inspect the possible $\,\ZZ_2^s\,$ symmetry for the Dirac mass-matrix
by including the $\mutau$ breaking effects
[cf. (\ref{eq:mD-2}) and (\ref{eq:mD-B-1}) in Sec.\,2.\,2].
This means to hold the invariance equations in (\ref{eq:G-GR-mDMR}),
\beqa
\label{eq:mD-Gs-inv}
G_s^T \mtt_D^{} G_s^R ~=~ \mtt_D^{}\,, ~~~~
{G_s^R}^T \Mtt_R G_s^R ~=~ \Mtt_R \,,
\eeqa
which will become, in the mass-eigenbasis of right-handed neutrinos,
\beqa
\label{eq:mD-Gs-inv}
G_s^T  \mD d_R^s ~=~ \mD,~~~~ d_R^{s\,T} M_R d_R^s ~=~ M_R \,.
\eeqa
Since $M_R$ and $d_R$ are both diagonal, the invariance equation for $M_R$
always holds.  So we can rewrite the above invariance equation for $\mD$ as,
\beqa
G_s^T \mbD d_R^s ~=~ \mbD \,,
\eeqa
where $\,\mbD \equiv \mD (\mh_0^{}M_R)^{-\f{1}{2}}\,$.
[The $\mutau$ symmetric form of $\,\mbD\,$ was given in Eq.\,(\ref{eq:mD-bar}).]
Using the notation $\mbD$, we can reexpress the seesaw mass-matrix,
$\,M_\nu = \mh_0^{}\(\mbD\ov{m}_D^T\)\,$.\,
So we can further deduce the invariance equations under $G_s^{}$ and $d_R^s$,
respectively,
\beqa
\label{eq:mbDD-Gs}
G_s^T \overline m_D\overline m_D^T G_s^{} ~=~ \overline m_D\overline m_D^T  \,,
~~~~~~~~
d_R^{s\,T}  \overline m_D^T \mbD d_R^s ~=~ \overline m_D^T \overline m_D \,.
\eeqa

Next, we inspect the two equations in (\ref{eq:mbDD-Gs})
to check the validity of the $\ZZ_2^s$ symmetry
after embedding the $\mutau$ breaking into $\,\mD\,$
[such as those constructed in (\ref{eq:mD-B-1}) for instance].
From (\ref{eq:mbDD-Gs}), we will explicitly prove that the $\,G_s\,$ is a symmetry
only for the $\mutau$ symmetric part of $\,M_\nu  \propto \(\mbD\ov{m}_D^T\)\,$;
while $\,d_R^s\,$ is violated by the $\mutau$ breaking terms in $\,\ov{m}_D^T \mbD\,$.
Hence, the $\ZZ_2^s$ symmetry is only a {\it partial symmetry} of the light neutrinos,
valid for the $\mutau$ symmetric part $\,M_\nu^{(s)}\,$.

We can write down the mass-matrix $\,\mbD\,$ with the most general $\mutau$ breaking,
\beqa
\overline m_D
&~=~&
\(\!\!
  \ba{ll}
   a      &~~  a' \\[1.5mm]
   b_1^{} &~~  c_1^{} \\[1.5mm]
   b_2^{} &~~  c_2^{}
  \ea \!\!\)
\,=\,
  \(\!\!
  \ba{ll}
   a &~~  a' \\[1.5mm]
   b &~~  c  \\[1.5mm]
   b &~~  c
  \ea \!\!\)+
  \(\!\!
  \ba{cc}
  0 &~~  0 \\[1.5mm]
  -\frac{\d b_1^{}+\d b_2^{}}{2} & -\frac{\d c_1^{}+\d c_2^{}}{2} \\[1.5mm]
  -\frac{\d b_1^{}+\d b_2^{}}{2} & -\frac{\d c_1^{}+\d c_2^{}}{2}
  \ea \!\!\)  +
  \(\!\!
  \ba{cc}
  0 &~~  0 \\[1.5mm]
  -\f{\d b_1^{}-\d b_2^{}}{2}  & -\f{\d c_1^{}-\d c_2^{}}{2} \\[1.5mm]
  +\f{\d b_1^{}-\d b_2^{}}{2}  & +\f{\d c_1^{}-\d c_2^{}}{2}
  \ea
  \!\!\)
  \nonumber\\[2.5mm]
&\!\!\!=\!\!\!&
\overline m_D^{(0)}+\delta \overline  m_D^{(s)}+\delta \overline  m_D^{(a)}
 ~=~ \overline  m_D^{(s)}+\delta \overline  m_D^{(a)}
\eeqa
where
$\,b_1^{}\equiv b-\delta b_1^{}\,$,\,
$\,b_2^{}\equiv b-\delta b_2^{}\,$,\,
$\,c_1^{}\equiv c-\delta c_1^{}\,$,\, and
$\,c_2^{}\equiv c-\delta c_2^{}\,$.\,

For the symmetric mass-matrix product,
$\,\mbD\ov{m}_D^T = M_\nu^{}/\mh_0^{}\equiv \ov{M}_\nu^{}\,$,\,
we compute, up to the NLO,
\begin{eqnarray}
\mbD\ov{m}_D^T  &~=~&
\begin{pmatrix}
     1  &0  & 0
  \\[1.5mm]
  &   \frac{1}{2}   & \frac{1}{2}
  \\[1.5mm]
  & &\frac{1}{2}
  \end{pmatrix}
  -(\delta b_1^{} \!+\delta b_2^{})\!
  \begin{pmatrix}
     0  & \frac{a }{2}  & \frac{a }{2}
  \\[1.5mm]
  &   b & b
  \\[1.5mm]
  & &b
  \end{pmatrix}
  -(\delta c_1^{} \!+\delta c_2^{})
  \begin{pmatrix}
     0  & \frac{a' }{2}  & \frac{a' }{2}
  \\[2mm]
  &   c & c
  \\[2mm]
  & &c
  \end{pmatrix}\nonumber \\[2mm]
  && -(\delta b_1^{} \!-\delta b_2^{})
  \(\!\! \ba{rrr}
     0  & \frac{a}{2}  & -\frac{a}{2}
  \\[2mm]
  &  b~ & 0~
  \\[2mm]
  & & -b
  \ea \!\!\)
  - (\delta c_1^{} \!-\delta c_2^{})
  \(\!\! \ba{rrr}
     0  & \frac{a'}{2}  & -\frac{a'}{2}
  \\[2mm]
  &  c~ & 0~
  \\[2mm]
  & & -c
  \ea \!\!\) ~~~~~~~~~~
\non\\[3mm]
&\!\!\!\equiv\!\!\!&
\ov{M}_\nu^{(0)} +\d\ov{M}_\nu^{(s)} + \d\ov{M}_\nu^{(a)}
~=~ \ov{M}_\nu^{(s)} + \d\ov{M}_\nu^{(a)} \,.
\label{eq:mbDmbDT}
\end{eqnarray}
where the $\,\ov{M}_\nu^{(s)}\,$ denotes the sum of the first three matrices and
$\,\d\ov{M}_\nu^{(a)}\,$ equals the sum of the last two matrices.
For deriving the LO matrix $\,\ov{M}_\nu^{(0)}\,$ in (\ref{eq:mbDmbDT}) we have used
the relations (\ref{eq:abc-2}) for the IMO scheme.
There exist two basic realizations for the common breaking of $\mutau$ and CP symmetries in
$\,\mD\,$ or $\,\mbD\,$:\,
one is for $\,\d b_1^{}=\d b_2^{}=0\,$ and
$\,(\d c_1^{},\, \d c_2^{})=c(\zeta',\,\zeta e^{i\omega})\,$,\,
which corresponds to $\,\mD\,$ in (\ref{eq:mD-B-1});
and another is for $\,\d c_1^{}=\d c_2^{}=0$\, and
$\,(\d b_1^{},\, \d b_2^{})=b(\zeta',\,\zeta e^{i\omega})\,$,\,
which corresponds to $\,\mD\,$ in (\ref{eq:mD-B-3}).
As we pointed out earlier, the invariance of the product (\ref{eq:mbDmbDT})
under $\,G_s\in \ZZ_2^s\,$ [cf.\ (\ref{eq:mbDD-Gs})] would be justified so long as
our general consistency condition (\ref{eq:t12-consistencyCD}) could hold.
So, with (\ref{eq:mbDmbDT}) we can explicitly compute
$\,\tan 2\theta_s\,$ from the two expressions in (\ref{eq:t12-consistencyCD})
including the $\mutau$ symmetric and anti-symmetric mass-matrix elements, respectively.
We thus arrive at
\beqs
\label{eq:t12-check-c}
\beqa
\label{eq:t12-ss}
\tan2\theta_s^{(s)} &~=~&
\frac{2\sqrt{2}\,B_s}{\,A-(C_s\!+\!D)\,} ~=\, -\frac{a'}{\sqrt{2}\,c\,} \,,
\\[3mm]
\label{eq:t12-aa}
\tan2\theta_s^{(a)} &=&\!
-\frac{2\sqrt{2}\,\delta B_a\delta C_a}{\delta C_a^2-2\delta B_a^2}
~=~ \frac{2\sqrt{2}\,a'c}{\,a'^2\!-\! 2c^2\,}
~=~\tan4\theta_s^{(s)} \,,
\eeqa
\eeqs
for $\delta b_1=\delta b_2=0$, and
\beqs
\label{eq:t12-check-b}
\beqa
\label{eq:t12-ss-b}
\tan2\theta_s^{(s)} &~=~&
\frac{2\sqrt{2}\,B_s}{\,A-(C_s\!+\!D)\,} ~=\, -\frac{a}{\sqrt{2}\,b\,} \,,
\\[3mm]
\label{eq:t12-aa-b}
\tan2\theta_s^{(a)} &=&\!
-\frac{2\sqrt{2}\,\delta B_a\delta C_a}{\delta C_a^2-2\delta B_a^2}
~=~ \frac{2\sqrt{2}\,a b}{\,a^2\!-\! 2b^2\,}
~=~\tan4\theta_s^{(s)} \,,
\eeqa
\eeqs
for $\delta c_1=\delta c_2=0$.
The above explicitly demonstrates the inequality $\,\theta_s^{(a)}\neq \theta_s^{(s)}\,$,\,
and thus proves the violation of the consistency condition (\ref{eq:t12-consistencyCD}).
This is because the $\mutau$ anti-symmetric mass-matrix
$\,\d M_\nu^{(a)}=\mh_0^{}\d\ov{M}_\nu^{(a)}\,$
in (\ref{eq:mbDmbDT}) breaks the $\ZZ_2^s$ symmetry.
Hence, $\ZZ_2^s$ is not a full symmetry of the mass-matrix $\,M_\nu\,$.\,
Nevertheless, we find that the $\mutau$ symmetric part
$\,M_\nu^{(s)}=\mh_0^{}\ov{M}_\nu^{(s)}\,$ in (\ref{eq:mbDmbDT}) does respect the
$\ZZ_2^s$ symmetry, and its invariance equation (\ref{eq:Gs-Ms}) leads to the correct solution
(\ref{eq:t12-sol-s}) and thus (\ref{eq:t12-ss}) for the solar angle $\,\ts$\,.
Substituting (\ref{eq:t12-k}) into (\ref{eq:t12-ss}) or (\ref{eq:t12-ss-b}),
we derive the equation,
$\,
\dis\,k^2 +\f{2}{r_0^{}}k - 1 = 0\,,
\,$
with $\,r_0^{}\equiv\dis\f{a'}{\sqrt{2}c\,}\,$ corresponding to (\ref{eq:t12-ss}),
or $\,r_0^{}\equiv\dis\f{a}{\sqrt{2}b\,}\,$ corresponding to (\ref{eq:t12-ss-b}).
So we can fix the $\ZZ_2^s$ group parameter $\,k\,$
in terms of the ratio of seesaw mass-parameters in $\,\mD\,$,
\beqa
\label{eq:k-mutaublind}
k ~=\, -1 \pm \sqrt{1+r_0^2} \,.
\eeqa

Finally, we compute
the other symmetric product $\,\ov{m}_D^T\mbD\,$,\, up to the NLO,
\beqa
\ov{m}_D^T \mbD
  &~=~&
\begin{pmatrix}
    1  & 0 \\[2mm]
    0  & 1
\end{pmatrix}
  -(\d b_1^{}\!+\d b_2^{})
\begin{pmatrix}
   2b  &~  c  \\[2mm]
    c  &~  0
\end{pmatrix}
-(\d c_1^{}+\d c_2^{})
\begin{pmatrix}
   0   &~  b  \\[2mm]
  b    &~  2c
\end{pmatrix}  \!.
\label{eq:mdtmd}
\eeqa
The last two matrices of (\ref{eq:mdtmd}) arise from the $\mutau$ breaking,
which make $\,\ov{m}_D^T\mbD\,$ {\it non-diagonal} at the NLO, and thus explicitly
violate the second invariance equation of (\ref{eq:mbDD-Gs}).
This violation of $\,\ZZ_2^s\,$
does not directly lead to observable effect at low energies since the seesaw mass-matrix
$\,M_\nu\,$ for light neutrinos is given by the first product $\,\mbD\ov{m}_D^T \,$
in Eq.\,(\ref{eq:mbDmbDT}).
Also, we could choose to assign the right-handed neutrinos to be {\it singlet} under
the $\,\ZZ_2^s\,$ from the light neutrinos, i.e., $d_R^{}=\II_2^{}$,\,
then the invariance equation for $\,\mbD\ov{m}_D^T \,$ becomes trivial.
But the first invariance equation in (\ref{eq:mbDD-Gs}) under $\,G_s\in \ZZ_2^s\,$
is still broken by the $\mutau$ anti-symmetric mass-matrix
$\,\d M_\nu^{(a)}=\mh_0^{}\d\ov{M}_\nu^{(a)}\,$ in (\ref{eq:mbDmbDT})
for light neutrinos, as shown by Eq.\,(\ref{eq:t12-check-c}) or
(\ref{eq:t12-check-b}) above.

\vspace*{3mm}

From the analyses above, we conclude that
the hidden symmetry $\ZZ_2^s$ is a {\it partial symmetry} of the present model,
respected by the $\mutau$ symmetric part $\,M_\nu^{(s)}\,$
of the light neutrino mass-matrix,
and thus determines the solar angle $\ts$ as in
Eqs.\,(\ref{eq:t12-ss}) and (\ref{eq:k-mutaublind}). This also agrees to the result
(\ref{eq:t12}) [Sec.\,2.\,1] or (\ref{eq:t12-NLO}) [Sec.\,4.\,1] which we derived earlier.
As a final remark, we stress that the violation of the hidden $\,\ZZ_2^s\,$ symmetry by
the $\mutau$ anti-symmetric mass-matrix $\,\d M_\nu^{(a)}=\mh_0^{}\d\ov{M}_\nu^{(a)}\,$
in (\ref{eq:mbDmbDT}) has an important physical impact:
it predicts a {\it modified new correlation (\ref{eq:da-dx-1}),}
and can be {\it experimentally distinguished from Eq.\,(\ref{eq:Re-dGs-sol})}
as predicted before by our soft $\mutau$ breaking of neutrino seesaw\,\cite{GHY}.

\vspace*{4mm}
\section{\large Conclusion}
\label{sec:conclusion}

In this work, we have studied the common origin of $\mutau$ breaking and
CP violations in the neutrino seesaw with right-handed Majorana neutrinos being
$\mutau$ blind. The oscillation data strongly support $\mutau$ symmetry as a good
approximate symmetry in the light neutrino sector, leading to the zeroth order pattern,
$\,(\theta_{23},\,\theta_{13})=(45^\deg,\,0^\deg)\,$.\,
Hence the $\mutau$ breakings, together with the associated CP violations,
are generically small. For the $\mutau$ blind seesaw,
we have convincingly formulated their common origin into Dirac mass
matrix $\mD$ (Sec.\,2.\,2),
leading to the unique inverted mass-ordering (IMO) of light neutrinos
and distinct neutrino phenomenology. This is parallel to our previous work\,\cite{GHY}
where the common origin of $\mutau$ and CP breaking arises from
the Majorana mass matrix of the singlet right-handed neutrinos and
uniquely leads to the normal mass-ordering (NMO) of light neutrinos.

In Sec.\,3, we gave the model-independent reconstruction of low energy
$\mutau$ and CP breakings with inverted neutrino mass-spectrum.
With this we derived various predictions of the $\mutau$ blind neutrino seesaw
in Sec.\,4. In particular, we deduced a {\it modified new correlation} (\ref{eq:da-dx-1})
between the two small $\mutau$ breaking observables $\,\theta_{23}-45^\deg\,$
and $\,\theta_{13}-0^\deg\,$,\, as depicted in Fig.\,2 and is very different from that in
Ref.\,\,\cite{GHY}.
Eq.\,(\ref{eq:da-dx-1}) is shown to also hold for the general three-neutrino seesaw
in Sec.\,4.\,3. This correlation can be experimentally tested against
Eq.\,(\ref{eq:NMO-da-dx-1}) as deduced from our soft $\mutau$ breaking seesaw mechanism\,\cite{GHY}.
As shown in Fig.\,2 and Fig.\,6,
our predicted range of $\,\theta_{13}\,$ can saturate its present experimental upper bound.
Imposing the current upper limit on $\theta_{13}$, we derived a restrictive range
of the deviation, $\,-4^\deg\leqq \theta_{23}-45^\deg\leqq 4^\deg\,$ at 90\%C.L.,
in Eq.\,(\ref{eq:t23-45-limit}).
In Sec.\,4.\,2, we have further generated the observed
matter-antimatter asymmetry (the baryon asymmetry)
from thermal leptogenesis in the $\mutau$ blind seesaw. Under the successful leptogenesis,
we derived the constrained correlation between
$\,\theta_{23}-45^\deg\,$ and $\,\theta_{13}-0^\deg\,$,\,
as presented in Fig.\,6. This figure predicts a lower bound on the key mixing angle,
$\,\theta_{13}\gtrsim 1^\deg\,$,\, which will be explored soon by the on-going reactor neutrino
experiments at Daya Bay\,\cite{DayaBay}, Double-Chooz\,\cite{2CHOOZ} and RENO\,\cite{RENO}.
Fig.\,7(a) further constrains the Jarlskog invariant $J$ into the negative
range, $\,-0.037\lesssim J\lesssim -0.0035\,$,\, while Fig.\,7(b) predicts
the range of neutrinoless $\beta\beta$-decay observable,
$\,45.5\,\textrm{meV} \lesssim M_{ee} \lesssim 50.7\,\textrm{meV}$,\,
which can be probed by the on-going neutrinoless $\beta\beta$-decay experiments\,\cite{0nu2beta}.
A lower bound on the leptogenesis scale $M_1$ is inferred from Fig.\,4,
$\, M_1 > 3.5 \times 10^{13}$\,GeV,  and is given in Eq.\,(\ref{eq:M1-LowerBound}).
The correlations of the leptogenesis scale $M_1$ with the reactor angle $\,\theta_{13}\,$
and the Jarlskog invariant $J$ are analyzed in Fig.\,9(a)-(b).

Finally, we have studied the determination of solar mixing angle $\,\theta_{12}\,$ and its
connection to a hidden flavor symmetry $\,\ZZ_2^s\,$ and its possible breaking in Sec.\,5.
The general model-independent $\,\ZZ_2\otimes\ZZ_2\,$ symmetry structure
of light neutrino sector was analyzed in Sec.\,5.2.1.
We first reconstructed the 3-dimensional representation $\,G_s^0\,$
for $\,\ZZ_2^s\,$ group in
the $\mutau$ symmetric limit as in Eq.\,(\ref{eq:Gs0-all}). We proved that
{\it hidden symmetry $\,\ZZ_2^s\,$ holds for any $\mutau$ symmetric mass-matrix $\,M_\nu\,$
of light neutrinos} and determines the solar angle $\,\theta_{12}\,$ via its group parameter,
$\,k=\tan\theta_{12}\,$,\, as in Eq.\,(\ref{eq:t12-k}).
Then, requiring that $\,\ZZ_2^s\,$ persists
{\it in the presence of general $\mutau$ breaking,} 
i.e., $\,G_s =G_s^0\,$ as in (\ref{eq:Gs=Gs0}),  we deduce a 
unique correlation equation (\ref{eq:Re-dGs-sol}) which strikingly coincides with
Eq.\,(\ref{eq:NMO-da-dx-1}), as predicted by our soft $\mutau$ breaking seesaw\,\cite{GHY}.
In Sec.\,5.2.2, we further analyzed the validity of $\,\ZZ_2^s\,$ symmetry from general
model-independent reconstructions of light neutrino mass-matrix $\,M_\nu\,$.\,
We derived the general consistency condition (\ref{eq:t12-consistencyCD})
for the validity of $\,\ZZ_2^s\,$ symmetry in the presence of all possible $\mutau$ breakings.
Under this condition, we derived the nontrivial correlation (\ref{eq:NMO-da-dx})
or (\ref{eq:IMO-da-dx}) between the two $\mutau$ breaking observables
$\,\theta_{23}-45^\deg\,$ and $\,\theta_{13}-0^\deg\,$,\,
which agrees to Eq.\,(\ref{eq:Re-dGs-sol}) as derived earlier from pure group theory approach.
We stress that the agreement between (\ref{eq:Re-dGs-sol}) [or (\ref{eq:NMO-da-dx})]
and the prediction (\ref{eq:NMO-da-dx-1}) from our soft $\mutau$ breaking seesaw
is not a coincidence. As we explained in Sec.\,5.3.1,  the true reason lies in the fact that
the soft $\mutau$ breaking is uniquely embedded in the right-handed Majorana mass-matrix $M_R$
which is a singlet of the $\,\ZZ_2^s\,$ group and thus does not violate $\,\ZZ_2^s\,$.\,
On the other hand, for the $\mutau$ blind seesaw, the $\mutau$ breaking is solely confined
in the Dirac mass-matrix $\mD$ which would have
nontrivial transformation (\ref{eq:mD-Gs-inv}) or (\ref{eq:mbDD-Gs})
{\it if} $\,\ZZ_2^s\,$ could actually hold.
As we have verified in Sec.\,5.3.2, the invariance equation (\ref{eq:mbDD-Gs})
hold only for the $\mutau$ symmetric part of the light neutrino mass-matrix $\,M_\nu\,$,\,
and is {\it partially violated by its $\mutau$ anti-symmetric part} [cf.\ Eq.\,(\ref{eq:mbDmbDT})].
In consequence, we found:
(i) the solar mixing angle $\,\theta_{12}\,$ is dictated by the group parameter
$\,k\,$ of the hidden symmetry $\,\ZZ_2^s\,$ acting on the $\mutau$ symmetric mass-matrix
$\,M_\nu^{(s)}\,$ [cf. Eqs.\,(\ref{eq:t12-ss}) and (\ref{eq:k-mutaublind})];
(ii) the consistency condition (\ref{eq:t12-consistencyCD}) no longer holds, and we predicted
a {\it modified new correlation (\ref{eq:da-dx-1}),} which can be experimentally distinguished
from Eq.\,(\ref{eq:NMO-da-dx-1}) as predicted by our soft $\mutau$ breaking seesaw\,\cite{GHY}.
In contrast to our previous prediction (\ref{eq:NMO-da-dx-1}),  Fig.\,6 points to
an important feature of the new correlation (\ref{eq:da-dx-1}) by showing a more rapid increase
of $\,\theta_{13}\,$ as a function of
$\,\theta_{23}-45^\deg\,$ [cf.\ also (\ref{eq:t13-lowerB-IMO})];\,
{\it this allows $\,\theta_{13}\,$ to saturate the current experimental upper limit,}
and confines the deviation $\,\theta_{23}-45^\deg\,$ into a more
restrictive range, $\,-4^\deg\leqq \theta_{23}-45^\deg \leqq 4^\deg\,$ at 90\%C.L.,
as in Eq.\,(\ref{eq:t23-45-limit}).
These distinctive predictions of the present $\mutau$ blind seesaw can be systematically tested
against those of our previous soft $\mutau$ breaking seesaw\,\cite{GHY},
by the on-going and upcoming neutrino experiments.

\vspace*{7mm}
\noindent
\addcontentsline{toc}{section}{Note Added in Proof\,}
\underline{\bf Note Added in Proof\,:}
\\[2mm]
After the submission of this paper to arXiv:1104.2654 on April\,14, 2011,
two long-baseline accelerator experiments newly announced evidences for $\,\theta_{13}$\,
via the $\,\nu_\mu^{}\to\nu_e^{}\,$ appearance channel,
one by the T2K Collaboration\,\cite{Abe:2011sj} on June\,14, 2011 and another
by the Minos Collaboration\,\cite{MINOS2011-t13} on June\,24, 2011.
Minos reported 62 $e$-like events above an estimated background of 49 events,
and favors a nonzero $\,\theta_{13}$\, at $1.5\sigma$ level. The resultant
confidence interval yields,
$\,0 \leqq \sin^2 2\theta_{13} < 0.12\,(0.19)\,$ at 90\%\,C.L. for NMO\,(IMO)
with $\d_D^{}=0$; and the best-fit value is
$\,\sin^2 2\theta_{13}=0.04\,(0.08)\,$ for NMO\,(IMO).
On the other hand, the T2K experiment observed 6 $e$-like events with an estimated
background of 1.5 events, indicating a nonzero $\,\theta_{13}\,$
at $2.5\sigma$ level. This gives the 90\%\,C.L. limits,
$\,0.03\,(0.04) <\sin^2 2\theta_{13} < 0.28\,(0.34)\,$  for NMO\,(IMO)
with $\d_D^{}=0$; and the best-fit value is
$\,\sin^2 2\theta_{13}=0.11\,(0.14)\,$ for NMO\,(IMO).
These new data indicate a relatively large $\,\theta_{13}\,$ mixing angle,
\beqs
\beqa
\textrm{T2K:} &~~~&
5.0^\deg ~<~ \theta_{13}\,(9.7^\deg) ~<~ 16.0^\deg \,,
~~~~~~~ \textrm{(for NMO)},
\\
&~~~&
5.8^\deg ~<~ \theta_{13}\,(11.0^\deg) ~<~ 17.8^\deg \,,
~~~~~\, \textrm{(for IMO)};
\\[1mm]
\textrm{Minos:} &~~~&
0^\deg ~\leqq~ \theta_{13}\,(5.8^\deg) ~<~ 10.1^\deg \,,
~~~~~~~ \hspace*{3.1mm}
\textrm{(for NMO)},
\\
&~~~&
0^\deg ~\leqq~ \theta_{13}\,(8.2^\deg) ~<~ 12.9^\deg \,,
~~~~~~~ \hspace*{3.1mm}
\textrm{(for IMO)};
\eeqa
\eeqs
at 90\%\,C.L., where the central values are shown in the parentheses.
We would like to point out that {\it the new data from T2K and Minos further
support our theory predictions} which give the unique inverted mass-ordering (IMO)
and favors a naturally larger $\,\theta_{13}\,$
even for a rather small deviation of $\,\theta_{23}-45^\deg\,$,\, as shown in
Eq.\,(\ref{eq:t13-lowerB-IMO}) and our Fig.\,\ref{fig:dx-da} (Sec.\,4.1) 
or Fig.\,\ref{fig-deltax-deltaa-new} (Sec.\,4.2).

Shortly afterwards, a new global analysis of oscillation data
has been performed\,\cite{Fogli:2011qn} to include the latest T2K and Minos data.
With this we can update our Table-1 accordingly, and translate the
improvements\,\cite{Fogli:2011qn} into the new Table\,2.

\begin{table*}[h]
\begin{center}
\label{tab:1} {\small
\begin{tabular}{c||ccccc}
\hline\hline
& & & \\
{\tt Parameters} & {\tt Best Fit}
 & 90\%\,{\tt C.L.}  & 99\%\,{\tt C.L.}
 & ${1\sigma}$ {\tt Limits} & ${3\sigma}$ {\tt Limits}
\\
& & & & & \\[-2mm]
\hline
& & & & & \\[-2mm]
 $ \Delta m^2_{21}(10^{-5}{\rm eV}^2)$ &
$7.58 $  &$7.15-7.94$&$7.07-8.09$& $7.32-7.80$&$6.99-8.18$
\\[1mm]
\hline& & & & & \\[-2mm]
$ \Delta m^2_{13}(10^{-3}{\rm eV}^2)$ &
$2.35 $  &$2.20-2.55$&$2.10-2.63$& $2.26-2.47$&$2.06-2.67$
\\[1mm]
\hline
& & & & & \\[-2mm]
$\theta_{12}$ & $33.6^\circ$  &$32.0^\circ-35.4^\circ$ &
$31.0^\circ-36.4^\circ$&  $32.6^\circ-34.7^\circ$ & $30.6^\circ-36.8^\circ$
\\[1mm]
\hline
& & & & & \\[-2mm]
$\theta_{23}$ & $40.4^\circ$ &
$37.5^\circ-47.9^\circ$ & $36.3^\circ-51.3^\circ$ & $38.6^\circ-45.0^\circ$ &
$35.7^\circ-53.1^\circ$ \\[1mm]
\hline
& & & & & \\[-2mm]
$\theta_{13}$ & $8.3^\circ$  &
$5.09^\circ-10.4^\circ$ & $3.5^\circ  -  11.6^\circ$ &
$6.5^\circ-9.6^\circ$ & $1.8^\circ-12.1^\circ$ \\
\hline\hline
\end{tabular}
}
\caption{The updated global analysis\,\cite{Fogli:2011qn} by including the latest data from
Minos\,\cite{MINOS2011-t13} and T2K\,\cite{Abe:2011sj} long-baseline accelerator experiments.
(Using the new reactor fluxes will slightly shift the mixing angles $\theta_{12}$ and $\theta_{13}$
a bit as shown in \cite{Fogli:2011qn}.)}
\end{center}
\vspace*{-6mm}
\end{table*}

\begin{figure}[t]
\centering
\includegraphics[width=13cm,height=10.5cm,clip=true]{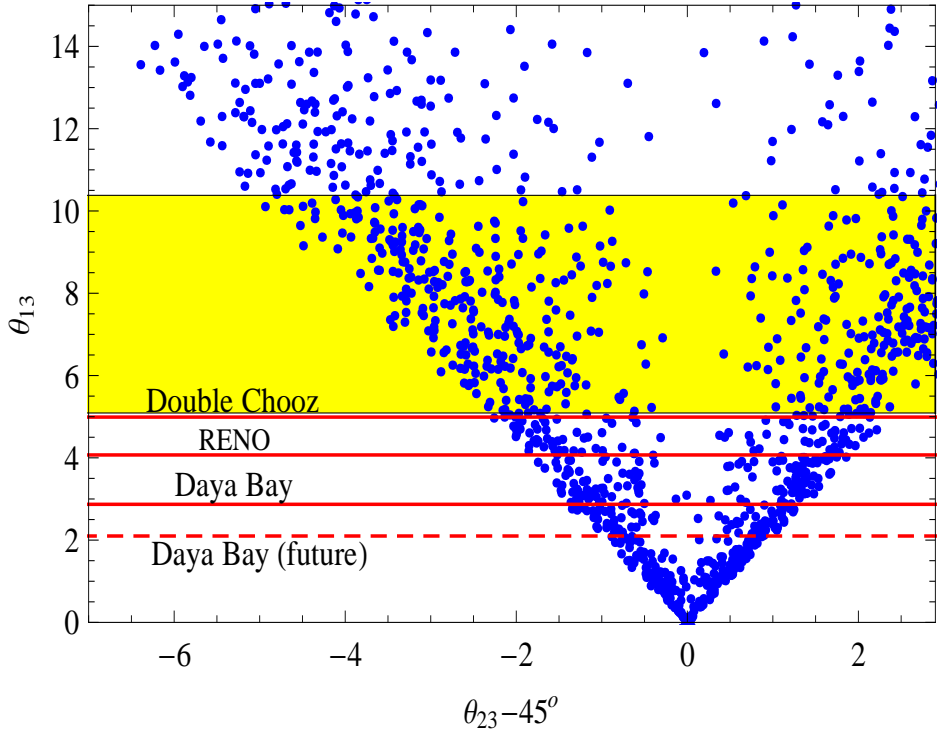}
\vspace*{-2.5mm}
  \caption{Update of Fig.\,2 (Sec.\,4.1) by using
  the improved global fit in Table-2, with 2000 samples.
  The shaded region (yellow) shows the updated constraint on $\theta_{13}^{}$
  at 90\%\,C.L.
  }
  \label{fig10:d13-d23-new}
\end{figure}

\begin{figure}[H]
\vspace*{-4mm}
\centering
\includegraphics[width=13cm,height=10.5cm,clip=true]{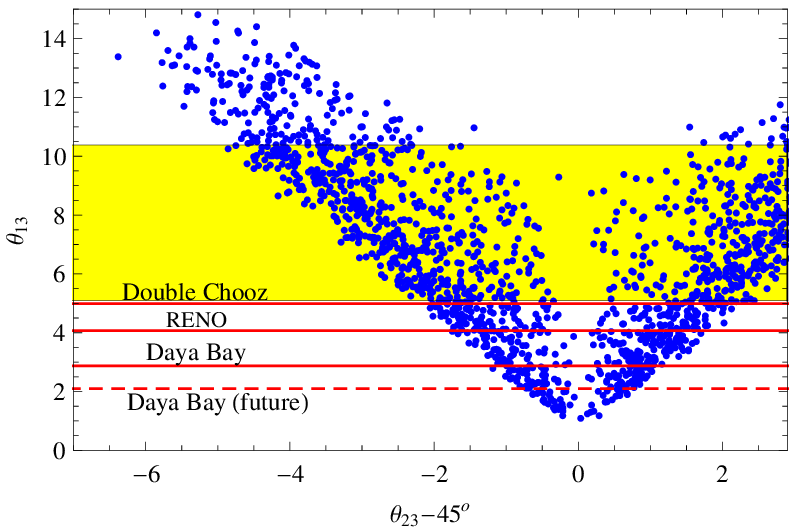}
\vspace*{-2.5mm}
\caption{Update of Fig.\,6 (Sec.\,4.2) by using
 the improved global fit in Table-2, with 2000 samples.
 The shaded region (yellow) shows the updated constraint on $\theta_{13}^{}$
 at 90\%\,C.L.
  }
  \label{fig11:d13-d23-new}
\end{figure}

\begin{figure}[h]
\centering
\includegraphics[width=10cm,height=7cm]{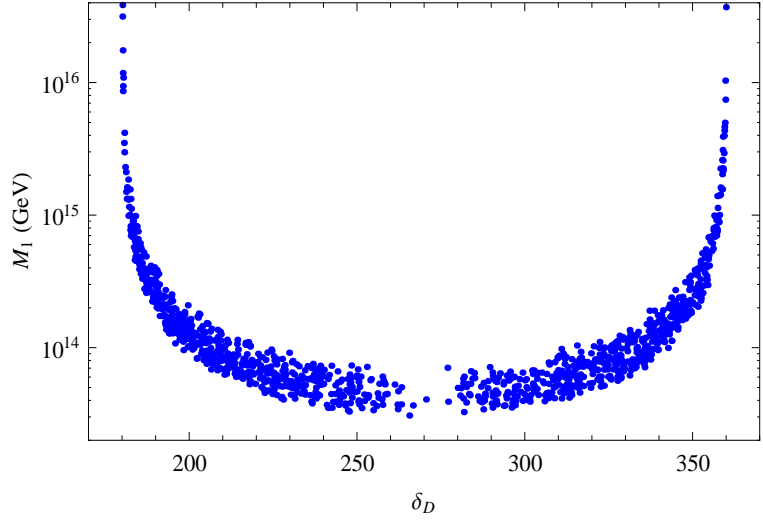}
\caption{Update of Fig.\,4 (Sec.\,4.2)
 by using the improved global fit in Table-2, with 1200 samples.}
\label{fig12:M1-dD-new}
\end{figure}

With Table-2, we have systematically updated our numerical analyses in Sec.\,4.
We find that the predictions of Fig.\,\ref{fig:dx-da}, 
Fig.\,\ref{fig-deltax-deltaa-new} and Fig.\,\ref{fig-etaB/M1-omega} exhibit more constrained
parameter space in an interesting way, while the other figures remain largely the same as before.
For comparison, we use the updates in Table-2 and replot 
Figs.\,\ref{fig:dx-da},\,\ref{fig-deltax-deltaa-new},\,\ref{fig-etaB/M1-omega} as    
new Fig.\,\ref{fig10:d13-d23-new}, Fig.\,\ref{fig11:d13-d23-new} 
and Fig.\,\ref{fig12:M1-dD-new}, respectively.
In Figs.\,\ref{fig10:d13-d23-new} and \ref{fig11:d13-d23-new}, 
we see that the updated 90\%\,C.L.\ constraint on $\theta_{13}^{}$
(yellow area) just picks up the central region of our predicted theory parameter-space.
Comparing the two plots in Figs.\,\ref{fig10:d13-d23-new}-\ref{fig11:d13-d23-new}, 
we see that imposing successful leptogensis in Fig.\,\ref{fig11:d13-d23-new} 
makes the parameter space more centered along the two wings, and the region
around $\,\theta_{23}\sim 45^\deg\,$ is clearly disfavored.
Since the new global fit of Table-2 gives the 90\%\,C.L. limits,
$\,-7.5^\deg<\theta_{23}-45^\deg <2.9^\deg\,$, with a central value
$\,\theta_{23}-45^\deg =-4.6^\deg\,$,\, it is clear that the left-wing of the
theory parameter-space is more favored over the right-wing.
Furthermore, imposing the $\,\theta_{23}\,$ and $\,\theta_{13}\,$ limits from Table-2
on our parameter-space in Fig.\,11, we deduce the allowed range at 90\%\,C.L.,
$\,-4.8^\deg < \theta_{23}-45^\deg < 2.9^\deg\,$,\, which is shifted towards
the negative side by about $1^\deg$ as compared to Eq.\,(\ref{eq:t23-45-limit}).

Then, Fig.\,\ref{fig12:M1-dD-new} 
shows that the predicted parameter region in the $M_1-\d_D^{}$ plane is much
more centered along the two edges in Fig.\,4, and a high leptogenesis scale
$\,M_1 > 10^{15}\,$GeV is strongly excluded except for the tiny regions of the
CP-angle $\,\d_D^{}\,$ very close to $180^\deg$ and $360^\deg$.

\begin{figure}[h]
\vspace*{3mm}
\centering 
\includegraphics[width=8.3cm,height=7cm]{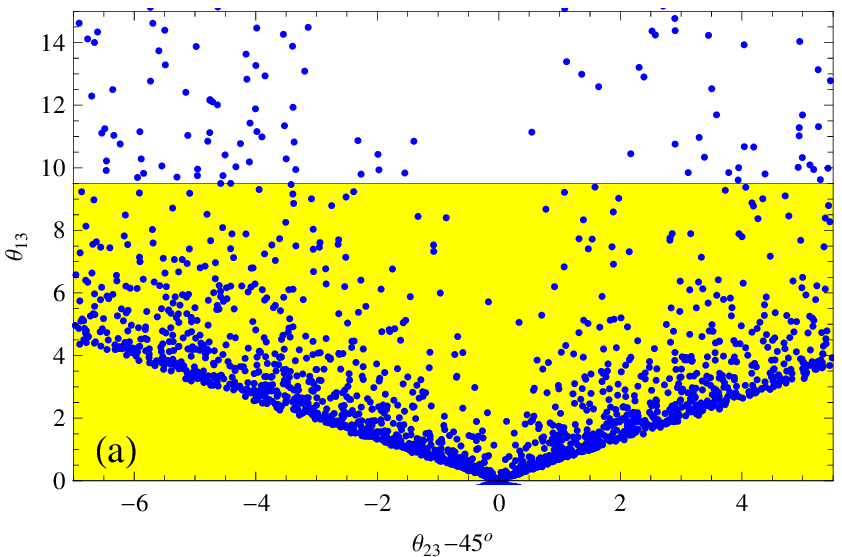}
\includegraphics[width=8.3cm,height=7cm]{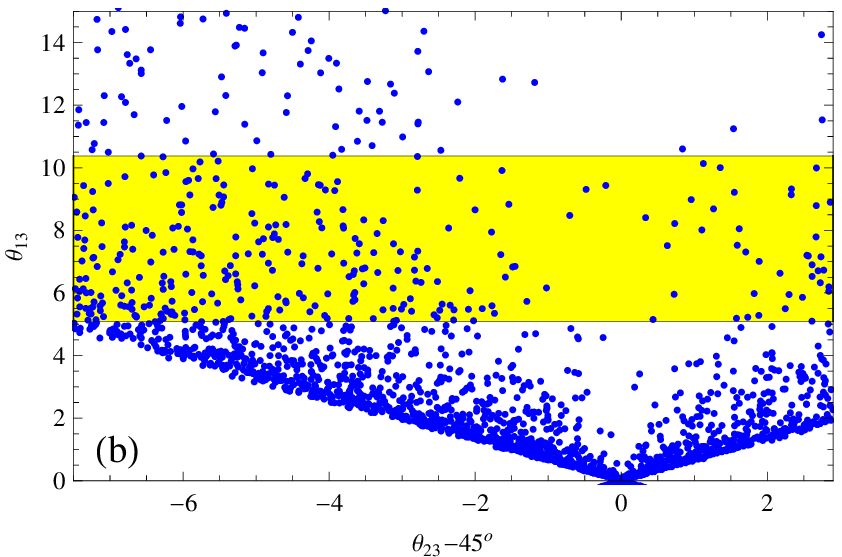}
\caption{Correlation of $\theta_{13}$ and $\theta_{23}-45^\deg$ as predicted
 by our Eq.\,(\ref{eq:Re-dGs-sol}) [Sec.\,5.2] without seesaw and under the assumption of
 exact $\,\ZZ_2^s\,$ symmetry.
 Plot-(a) shows the correlation by using the global fit in Table-1, while plot-(b) depicts
 the correlation under the improved global fit in Table-2, with 2000 samples in each plot.
 The shaded regions (yellow) give the allowed 90\%\,C.L.\ ranges by the corresponding global fit.}
\label{fig13:t13-t23-p1}
\end{figure}

Finally, as a comparison,
we further analyze the prediction from our Eq.\,(\ref{eq:Re-dGs-sol}) [Sec.\,5.2]
which we derived for light neutrinos alone (without invoking seesaw)
and under the assumption of an exact $\,\ZZ_2^s\,$ symmetry.
As we already proved in Sec.\,5.3,
this $\,\ZZ_2^s\,$ only holds in a class of models including our soft $\mutau$
and CP breaking model in Ref.\,\cite{GHY}, but can be violated in other class of models
including the current $\mutau$ blind seesaw model.
Hence, {\it the $\,\ZZ_2^s\,$ symmetry cannot generally hold in a model-independent way.}
With Eq.\,(\ref{eq:Re-dGs-sol}), we plot the correlation between the $\mutau$ breaking parameters
$\theta_{13}$ and $\theta_{23}-45^\deg$ in Fig.\,\ref{fig13:t13-t23-p1}(a)-(b).
In plot-(a) we show the correlation by using the global fit in Table-1, while in plot-(b) we depict
the correlation under the improved global fit in Table-2. Each plots contains 2000 samples.
The shaded regions (yellow) display the allowed 90\%\,C.L.\ parameter space
by the corresponding global fit. Note that Eq.\,(\ref{eq:Re-dGs-sol}) holds for both
normal mass ordering and inverted mass ordering of light neutrinos.
Fig.\,13 shows that our predicted parameter space can easily saturate the current upper limit on
$\theta_{13}$, and thus accommodates a relatively large $\theta_{13}$ as indicated by the new
data from T2K\,\cite{Abe:2011sj} and Minos\,\cite{MINOS2011-t13}.
The prediction of Fig.\,\ref{fig13:t13-t23-p1} differs from the above Fig.\,10 significantly,
because the coefficient in Eq.\,(\ref{eq:IMO-da-dx-expand-ds})
[corresponding to Fig.\,10] has a nontrivial suppression
factor relative to that of Eq.\,(\ref{eq:NMO-da-dx-expand-ds}) or Eq.\,(\ref{eq:Re-dGs-sol})
[corresponding to Fig.\,\ref{fig13:t13-t23-p1}].
Furthermore, we note that
the above Fig.\,\ref{fig13:t13-t23-p1} 
should also be compared to our previous Fig.\,2 in Ref.\,\cite{GHY}
because the correlation (\ref{eq:Re-dGs-sol}) applies to both of them. But there are large
differences between these two figures, the major reason is that we input the parameter
$\theta_{23}-45^\deg$ in Fig.\,13 according to the oscillation data (Table-1 or Table-2)
and without invoking seesaw, while the $\theta_{23}-45^\deg$ in the Fig.\,2 of Ref.\,\cite{GHY}
was derived as a function of fundamental $\mutau$ and CP breaking parameters in the
seesaw Lagrangian which were scanned within their theoretically allowed ranges.
This also leads to a stronger upper limit of
$\,\theta_{13}\lesssim 6^\deg\,$ in Ref.\,\cite{GHY}.

\vspace*{7mm}
\noindent
\addcontentsline{toc}{section}{Note Added-2\,}
\underline{\bf Note Added-2\,:}
\\[2mm]
After the publication of this paper in Phys.\ Rev.\ D\,84 (2011) 033009,
Daya Bay and RENO collaborations announced new measurements of nonzero $\tac$
on March~8, 2012 \cite{daya2012} and April~8, 2012 \cite{reno2012}, respectively.
Daya Bay experiment made a $5.2\sigma$ discovery of nonzero $\tac$ \cite{daya2012},
$\,\sin^22\tac = 0.092\pm 0.016({\rm stat})\pm 0.005({\rm syst})$;\,
and RENO found a nonzero $\tac$ at $4.9\sigma$ level \cite{reno2012},
$\,\sin^22\tac =0.113\pm 0.013({\rm stat})\pm 0.019({\rm syst})$\,.\,
These give the following $3\sigma$ ranges of nonzero \,$\tac$\,,
\beqs
\beqa
\label{eq:DY2012}
\text{Daya Bay:} && 5.7^\circ < \tac\, (8.8^\deg) <  11.1^\circ \,,
\\
\label{eq:RE2012}
\text{RENO:} && 5.9^\circ < \tac\, (9.8^\deg) < 12.6^\circ \,,
\eeqa
\eeqs
where the numbers in the parentheses
$\,\tac = 8.8^\circ$\, and $\,\tac =9.8^\circ$\,
correspond to the central values.

\begin{figure}[H]
\vspace*{-7mm}
\centering
\includegraphics[width=12.5cm,height=10cm,clip=true]{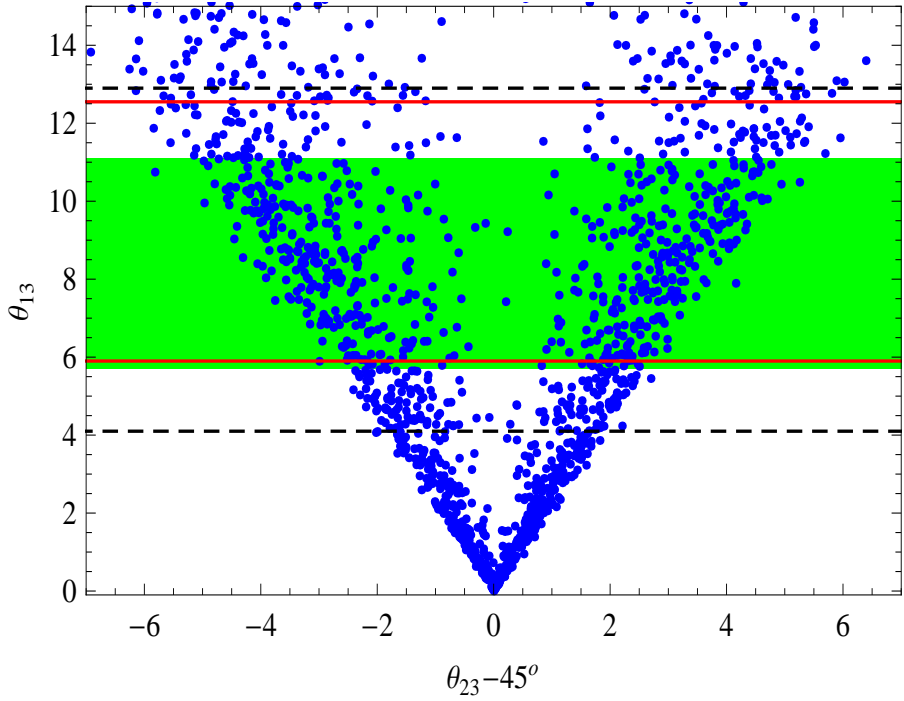}
\vspace*{-2.5mm}
\caption{Update of Fig.\,10, with 2000 samples.
The experimental inputs are scanned within $3\sigma$ ranges.
The $3\sigma$ ranges of the new Daya Bay data \cite{daya2012} are shown as
the green shaded region; and the $3\sigma$ limits of the
new RENO data \cite{reno2012} are depicted by the horizontal red-lines.
The horizontal black dashed-lines denote the $3\sigma$ limits of
the global fit\,\cite{Fogli:2011qn}.
}
\label{fig14:d13-d23-n}
\end{figure}

\begin{figure}[H]
\vspace*{-8mm}
\centering
\includegraphics[width=12.5cm,height=10cm,clip=true]{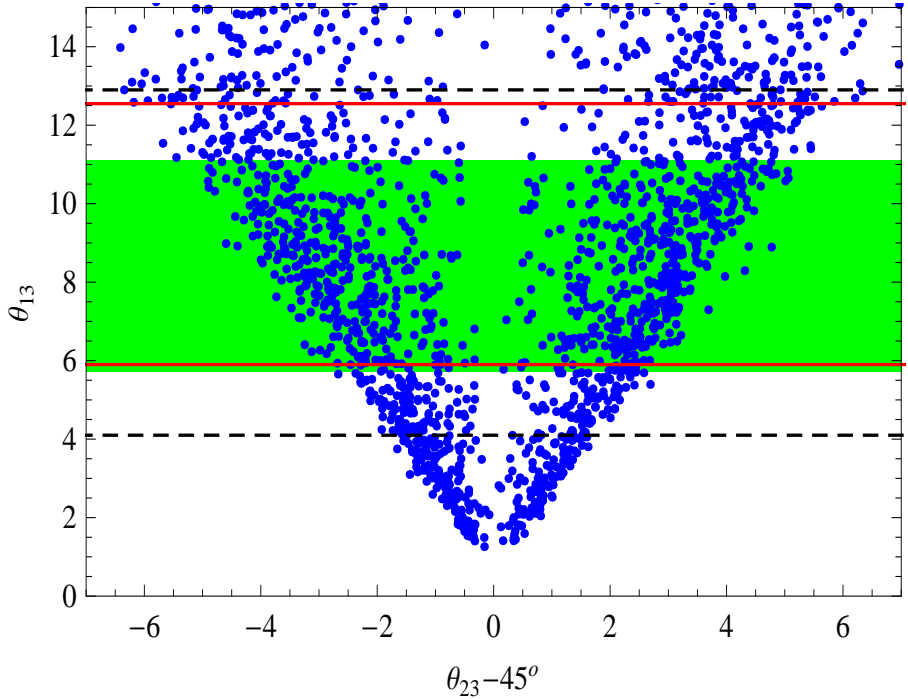}
\vspace*{-2.5mm}
\caption{Update of Fig.\,11, with 2000 samples.
The experimental inputs are scanned within $3\sigma$ ranges.
The $3\sigma$ ranges of the new Daya Bay data \cite{daya2012} are shown as
the green shaded region; and the $3\sigma$ limits of the
new RENO data \cite{reno2012} are depicted by the horizontal red-lines.
The horizontal black dashed-lines denote the $3\sigma$ limits of
the global fit\,\cite{Fogli:2011qn}.
}
\label{fig15:d13-d23-n}
\end{figure}

\begin{figure}[H]
\vspace*{3mm}
\centering
\includegraphics[width=12.5cm,height=10cm]{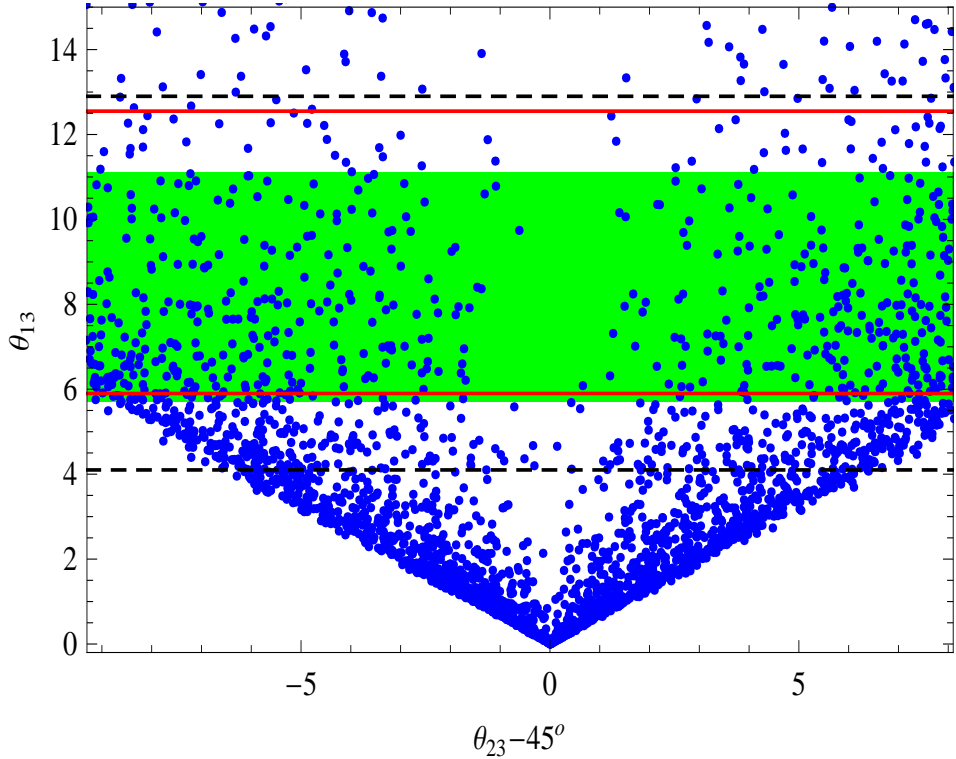}
\caption{Correlation of $\theta_{13}$ and $\theta_{23}-45^\deg$ as predicted
by our Eq.\,(\ref{eq:Re-dGs-sol})  without seesaw and under the assumption of
exact $\,\ZZ_2^s\,$ symmetry, with 2000 samples.
The experimental inputs are scanned within $3\sigma$ ranges.
The $3\sigma$ ranges of the new Daya Bay data \cite{daya2012} are shown as
the green shaded region; and the $3\sigma$ limits of the
new RENO data \cite{reno2012} are depicted by the horizontal red-lines.
The horizontal black dashed-lines denote the $3\sigma$ limits of
the global fit\,\cite{Fogli:2011qn}.
  }
\label{fig16:d13-d23-n}  
\end{figure}

Then, we can re-plot the
Fig.\,\ref{fig10:d13-d23-new}, Fig.\,\ref{fig11:d13-d23-new} and Fig.\,\ref{fig13:t13-t23-p1}(b)
as the new Fig.\,\ref{fig14:d13-d23-n}, Fig.\,\ref{fig15:d13-d23-n} 
and Fig.\,\ref{fig16:d13-d23-n}, respectively.
In these new plots, we have scanned the experimental inputs within $3\sigma$ ranges.
For the successful leptogenesis, we find that the lower bound on the leptogenesis scale
$M_1$ becomes, $\,M_1 > 2 \times 10^{13}\,$,\, at $3\sigma$ level. 
The successful leptogenesis in Fig.\,\ref{fig15:d13-d23-n} further requires, 
$\,\theta_{13}^{} \gtrsim 1^\deg \,$.\,

To compare with our predictions, we have displayed the $3\sigma$ range of \,$\tac$\, from 
the new Daya Bay measurement\,\cite{daya2012} in the green shaded region. 
Furthermore, we show the $3\sigma$ lower and upper limits of \,$\tac$\, from
the new RENO data\,\cite{reno2012} by the horizontal red-lines. 
The horizontal black dashed-lines in each plot denote the $3\sigma$ limits from
the global fit\,\cite{Fogli:2011qn}.
From Figs.\,\ref{fig14:d13-d23-n}-\ref{fig15:d13-d23-n}, we see that the new limits 
from Daya Bay\,\cite{daya2012} and RENO\,\cite{reno2012} experiments 
nicely pick up the central regions of our predicted parameter space of \,$\tac$\,.

Finally, Fig.\,\ref{fig16:d13-d23-n} shows the correlation of 
$\,\theta_{13}^{}\,$ and $\,\theta_{23}-45^\deg$\, as predicted by our 
Eq.\,(\ref{eq:Re-dGs-sol}) [Sec.\,5.2] under the assumption of
an exact $\,\ZZ_2^s\,$ symmetry and without invoking seesaw. 
We have scanned the $3\sigma$ ranges of $\,\tab\,$ in (\ref{eq:Re-dGs-sol}). 
We see that the new data of Daya Bay\,\cite{daya2012} and RENO\,\cite{reno2012} 
pick up the upper parts of our predicted parameter space of \,$\tac$\,.

\vspace*{7mm}
\noindent
\addcontentsline{toc}{section}{Acknowledgments\,}
{\bf Acknowledgments}
 \\[2mm]
 We thank C.\ S.\ Lam, Rabindra N.\ Mohapatra, Werner Rodejohann and Alexei Yu.\ Smirnov
 for useful discussions on this subject, and Eligio Lisi for discussing the
 updated global fit\,\cite{Fogli:2011qn} in connection with the new data from
 T2K\,\cite{Abe:2011sj} and Minos\,\cite{MINOS2011-t13}.
 We are grateful to Yi-Fang Wang, Kam-Biu Luk and Jun Cao
 for discussing Daya Bay experiment\,\cite{DayaBay}.
 We also wish to thank Profs.\ T.\ D.\ Lee and R.\ Friedberg for valuable correspondence
 and discussions on the comparison of $\,\theta_{13}\,$ predictions in
 our study and their work\,\cite{FL2010},
 which we showed at the end of
 Sec.\,4.1 [cf.\ Eqs.\,(\ref{eq:FL})-(\ref{eq:da-dx-expand-ds2})].
 HJH thanks the Center for High Energy Physics at Peking University
 for kind hospitality and support.
 This research was supported by the NSF of China (under grants 10625522 and 10635030),
 the National Basic Research Program of China (under grant 2010CB833000),
 and by Tsinghua University.



\begin{thebibliography}{99}
\bibitem{footnote-3}
 Here the RHS of (\ref{eq:RGE-mj}) also contains flavor-dependent terms
 which are suppressed at least by the square of tau-lepton Yukawa coupling,
 $\,y_\tau^2 =\O(10^{-4})\,$, relative to the universal $\,\ahat$-term,
 and thus completely negligible for the present study.
\bibitem{footnote-6}
 As we will further demonstrate in Sec.\,5.\,3, this consistency
 condition is not always ensured in a given seesaw model, and thus the $\ZZs$ symmetry
 could be partially broken, leading to modified correlation between the $\mutau$ breaking
 observables.
\bibitem{footnote-7}
 For instance, a simple choice of such $\,d_\ell^{(j)}\,$ is,
 $\,d_\ell^{(1)} = \textrm{diag}(1,\,\eta,\,\eta^2)\in \ZZ_3\,$
 with $\,\eta =\exp\(i\f{2\pi}{3}\)\,$,\,  and $\ZZ_3$ is a subgroup of the general
 symmetry $\,\GG_\ell =U(1)\otimes U(1)\,$ of lepton mass-matrix. The invariance of
 lepton mass-matrix $\,M_\ell^{}M_\ell^\dag\,$ under the transformation
 $\,F_1 =d_\ell^{(1)}\,$ automatically ensures left-handed leptons in their mass-diagonal basis.
\bibitem{footnote-8}
 The 3-dimensional representation (\ref{eq:Gs0-k}) of $\,\ZZ_2^s\,$ was derived in
 Eq.\,(6.26) of Ref.\,\cite{GHY}
 for our soft $\mutau$ breaking seesaw model where the group parameter $k$
 (re-denoted as $k'$ here) is related to the current $k$ of (\ref{eq:Gs0-k})
 by a simple notational conversion, $\,k'\equiv -\sqrt{2}/k\,$.
\end{thebibliography}
 \end{document}